\documentclass[fleqn,usenatbib]{mnras}

\usepackage{newtxtext,newtxmath}
\usepackage[T1]{fontenc}

\DeclareRobustCommand{\VAN}[3]{#2}
\let\VANthebibliography\thebibliography
\def\thebibliography{\DeclareRobustCommand{\VAN}[3]{##3}\VANthebibliography}
\usepackage{graphicx}
\usepackage{amsmath}
\usepackage{microtype}
\usepackage[flushleft]{threeparttable}
\usepackage[dvipsnames]{xcolor}
\usepackage{subfig}
\usepackage{subfiles}
\newcommand{\noopsort}[2]{#2}

\title[Apparent dearth of NSRs in the Local Group]{On an apparent\,dearth of recurrent\,nova super-remnants in the Local\,Group}

\author[M. W. Healy-Kalesh et al.]{M. W. Healy-Kalesh,$^{1}$\thanks{E-mail: M.W.HealyKalesh@ljmu.ac.uk (MWH-K)} M. J. Darnley,$^{1}$\thanks{E-mail: M.J.Darnley@ljmu.ac.uk (MJD)} and M. M. Shara$^{2}$
\\
$^{1}$Astrophysics Research Institute, Liverpool John Moores University, IC2 Liverpool Science Park, Liverpool, L3 5RF, UK\\
$^{2}$Department of Astrophysics, American Museum of Natural History, Central Park West at 79th Street, New York, NY 10024, USA\\
}

\date{Accepted 2024 January 08. Received 2023 January 08; in original form 2023 November 20}

\pubyear{2024}

\begin{document}
\label{firstpage}
\pagerange{\pageref{firstpage}--\pageref{lastpage}}
\maketitle

\begin{abstract}
The Andromeda Galaxy is home to the annually erupting recurrent nova (RN) M\,31N 2008-12a (12a); the first nova found to host a nova super-remnant (NSR). A NSR is an immense structure surrounding a RN, created from many millions of eruptions sweeping up material in the local environment to form a shell tens of parsecs across. Theory has demonstrated that NSRs should be found around all RNe, even those systems with long periods between eruptions. Befittingly, the second NSR was found around the Galactic classical (and long suspected recurrent) nova, KT Eridani. In this Paper, we aim to find more of these phenomena through conducting the first ever survey for NSRs in M\,31 and the Large Magellanic Cloud (LMC). We find that the surroundings of fourteen RNe in M\,31 as well as the surroundings of the four RNe in the LMC do not show any evidence of vast parsec-scale structures in narrowband (H$\alpha$ and $[\ion{S}{ii}]$) images, unlike the one clearly seen around 12a, and therefore conclude that observable NSRs are either rare structures, or they are too faint (or small) to be detected in our existing datasets. Yet, the NSR surrounding 12a would also likely to have been overlooked in our study if it were approximately one magnitude fainter. Searches for NSRs around other RNe `masquerading' as classical novae may prove to be fruitful as would whole surveys of other Local Group galaxies.
\end{abstract}


\section{Introduction}

Nova eruptions rank among some of the most luminous stellar transients. At their heart is a white dwarf (WD), accreting hydrogen-rich material from a donor within a close binary configuration \citep{1995cvs..book.....W}. Once a critical mass has accumulated on the WD surface, a thermonuclear runaway ensues that ejects a proportion of the accreted material \citep{1972ApJ...176..169S,1978A&A....62..339P} --- this is the nova eruption itself and it is associated with a substantial increase in the luminosity of the system.

Nova eruptions are inherently recurrent and systems are commonly classified by virtue of their observed inter-eruption cycle. The classical novae (CNe) are those systems with long recurrence periods $P_\mathrm{rec}$, whereas the recurrent novae (RNe) have undergone at least two observed eruptions. Here, there are clear selection effects that limit the known RN population, with many recurrents `masquerading' as their classical siblings \citep{2014ApJ...788..164P}. As an example, the Galactic CN KT Eridani has been observed only once in eruption (in 2009) however, after detailed analysis of system's characteristics, \citet{2022MNRAS.517.3864S} conclude that KT Eridani is a RN with a $P_\mathrm{rec}$ of 40$-$50 years (as do \citealt{2023arXiv231017055S} and \citealt{2023arXiv231017258H}).

Recurrence period is primarily a function of the WD mass and the mass accretion rate ($\dot{M}$). The combination of a high mass WD and high $\dot{M}$ is required to drive $P_\mathrm{rec}$ below, of order, a century \citep[see, for e.g.,][]{2005ApJ...623..398Y} to move a nova into the position of potentially being classed as a RN \citep[see discussion within][]{2021gacv.workE..44D}. Such a high mass WD and likelihood to grow more massive due to the retention of accumulated material \citep[see, for example,][]{2005ApJ...623..398Y,2015ApJ...808...52K,2015MNRAS.446.1924H,2016ApJ...819..168H} place RNe among the front-runners for progenitors of type Ia supernovae \citep[SNe Ia;][]{1973ApJ...186.1007W,1999ApJ...519..314H,1999ApJ...522..487H,2000ARA&A..38..191H}. 

Every nova eruption ejects high velocity material into the surrounding interstellar medium (ISM). This is well evidenced through the recovery of numerous remnants from single nova eruptions such as those around Z Camelopardalis \citep{2007Natur.446..159S}, GK Persei \citep{2004ApJ...600L..63B,2016A&A...595A..64H}, AT Cancri \citep{2012ApJ...758..121S}, DQ Herculis \citep{1978ApJ...224..171W}, HR Delphini \citep{2003MNRAS.344.1219H}, DO Aquilae and V4362 Sagittarii \citep{2020MNRAS.499.2959H}. With recurrence periods for most of the known CNe being longer than 1\,kyr, perhaps up to 1\,Myr, many remnants will have essentially dissipated before the subsequent eruption takes place \citep{2023MNRAS.521.3004H}.
\begin{figure*}
\centering
\subfloat{\includegraphics[width=.35\textwidth]{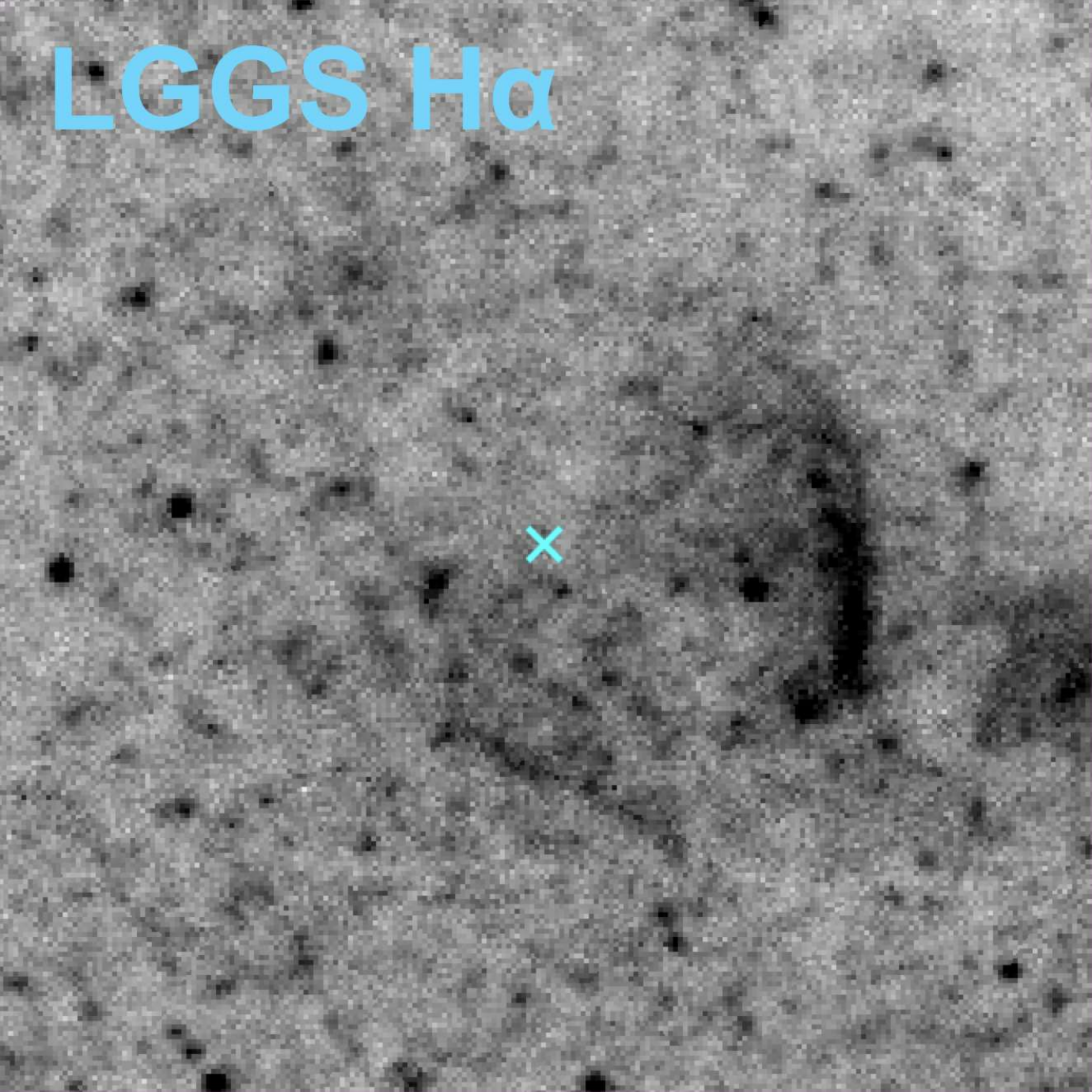}} \quad
\subfloat{\includegraphics[width=.35\textwidth]{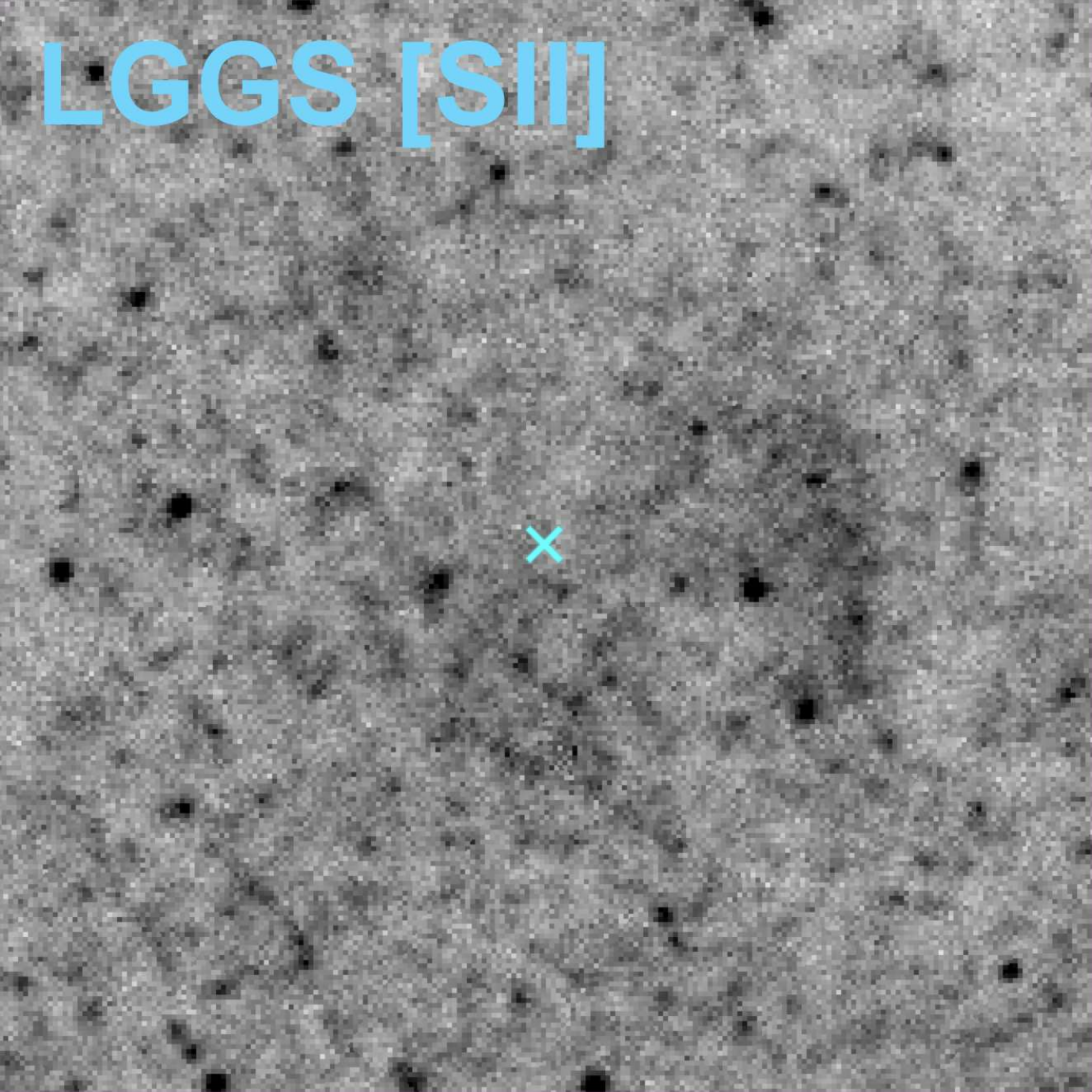}}\\
\subfloat{\includegraphics[width=.35\textwidth]{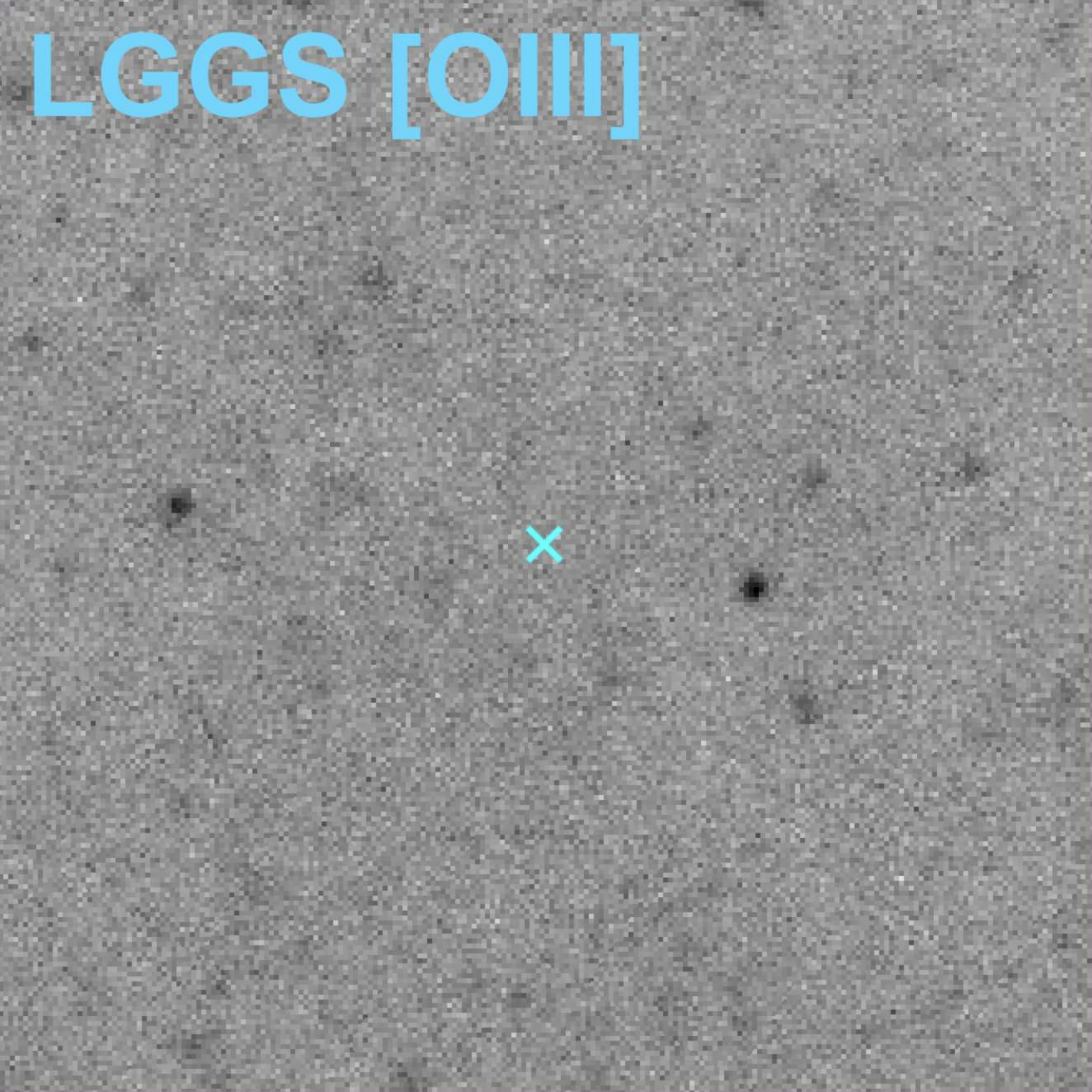}} \quad
\subfloat{\includegraphics[width=.35\textwidth]{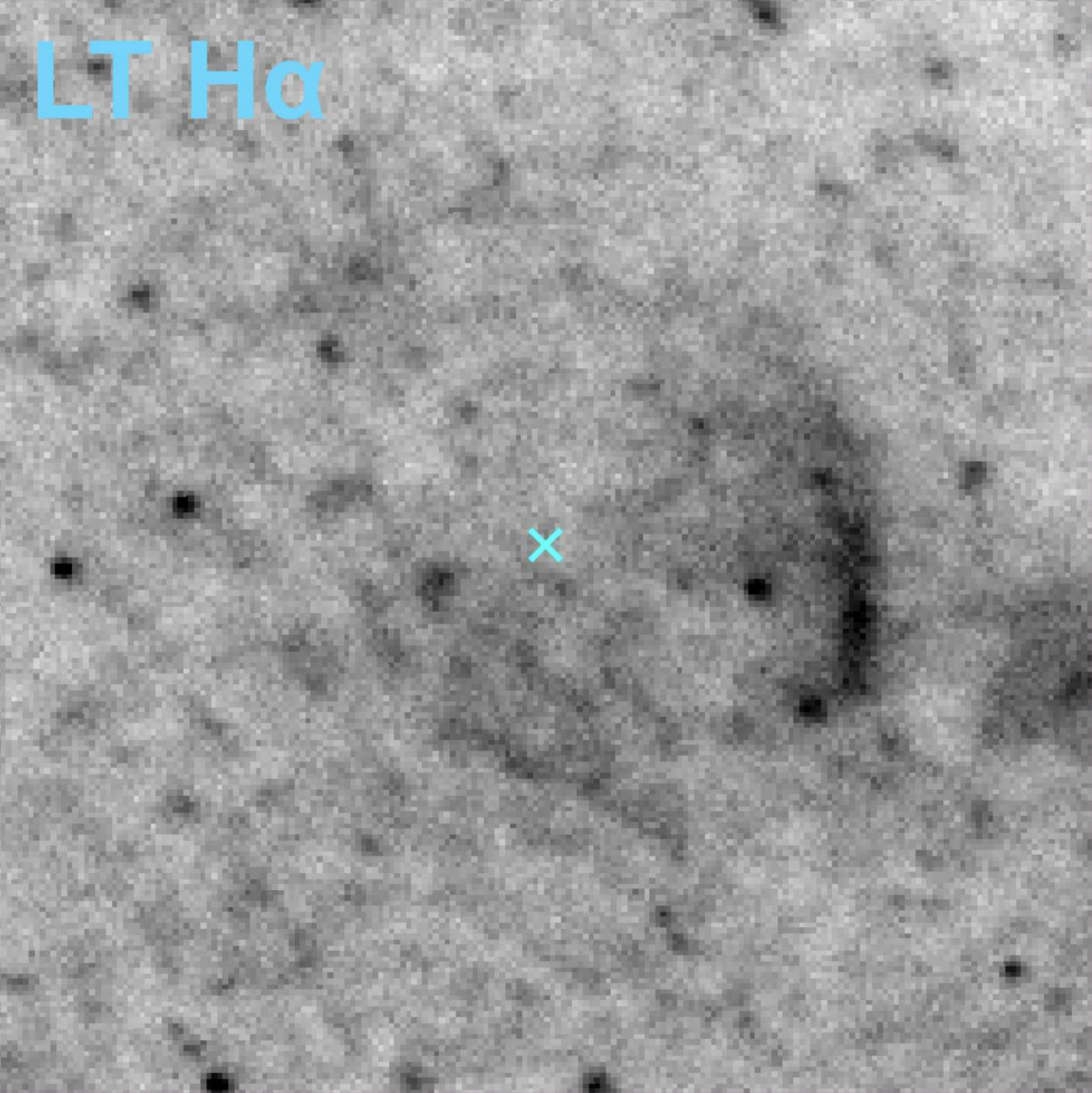}}
\caption{Local Group Galaxies Survey \citep[LGGS;][]{2007AJ....134.2474M} narrow band imaging (from field 2) of the $1^\prime\times1^\prime$ region surrounding M31N\,2008-12a. Top-left:\ LGGS H$\alpha$. Top-right:\ LGGS [\ion{S}{ii}]. Bottom-right:\ LGGS [\ion{O}{iii}]. Bottom-left:\ Liverpool Telescope deeper H$\alpha$ image.}
\label{12a NSR images}
\end{figure*}

However, for RNe, this is not the case. Here, the inter-eruption timescales are short enough ($P_\mathrm{rec}<100$\,yr), that the \textit{next} eruption will occur while the previous ejecta maintain a substantial over-density and before they have had time to cool. A striking example of interacting ejecta from a handful of recent eruptions surrounds the Galactic RN T\,Pyxidis, in the form of over two thousand [$\ion{N}{ii}$]-emitting knots \citep{1997AJ....114..258S,2015ApJ...805..148S}. The knots (and filaments) surrounding T\,Pyxidis are a consequence of Rayleigh-Taylor instabilities brought about by the interaction of six RN ejecta shells with a proposed CN shell \citep{2013ApJ...768...48T,2015ApJ...805..148S}. Three-dimensional simulations mimicking this history \citep{2013ApJ...768...48T} supports the \citet{2010ApJ...708..381S} hypothesis that T\,Pyxidis experienced a CN eruption in 1866 after millennia of hibernation, and followed this up with six RN eruptions driven by a period of elevated mass loss from the donor that is a proposed consequence of the CN eruption. We note that the T\,Pyxidis scenario is not expected to be typical of RNe.

The Andromeda Galaxy (M\,31) RN M31N\,2008-12a (hereafter simply `12a') exhibits the shortest recurrence period currently known:\ $P_\mathrm{rec}=359\pm12$\,days \citep{2014A&A...563L...9D,2016ApJ...833..149D}. The eruptions from 12a are powered by a near-Chandrasekhar mass \citep[$M_\mathrm{WD}\simeq1.38\,\mathrm{M}_\odot$;][]{2015ApJ...808...52K} WD accreting at $1.6\times10^{-7}\lesssim\dot{M}\lesssim1.4\times10^{-6}\,\mathrm{M}_\odot\,\mathrm{yr}^{-1}$ \citep{2017ApJ...849...96D} -- both paramaters are at the extreme high end of values seen in other novae. \citet{2015A&A...580A..45D} reported the discovery of a vast shell-like nebula surrounding 12a, measuring 134\,pc across at its largest extent. Supported by hydrodynamic simulations of $10^{5}$ 12a-like eruptions, \citet{2019Natur.565..460D} proposed that this nebula was the result of ${\sim}6\times10^{6}$ eruptions as the 12a WD has grown (over ${\sim}6\times10^{6}$\,yr) toward the Chandrasekhar mass --- producing the first discovered nova super-remnant (NSR).

Subsequent, and more detailed, hydrodynamic modelling by \citet{2023MNRAS.521.3004H} has shown that vast dynamical NSRs should exist around all RNe, including those systems with shrinking WDs or with eruptions thousands of years apart. In addition, \citet{2023MNRAS.521.3004H} demonstrated that the size of a NSR depends strongly upon $\dot{M}$, the surrounding ISM density, and the observed $P_\mathrm{rec}$ (i.e.\ the instantaneous WD mass) whereas WD temperature and initial mass have less of an impact. Though, if the initial mass is very high, indicative of an oxygen-neon (ONe) WD, the extent of the grown NSR is largely reduced. Yet, it is only systems with high accretion rates that are predicted to have {\it observable} NSRs \citep{2023MNRAS.521.3004H}. Furthermore, as the population of novae surrounded by a NSR likely host a high mass WD, possibly close to the Chandrasekhar limit, then these vast structures may represent indicators for the locations of upcoming or past SNe Ia events \citep{2021gacv.workE..44D}. As such, NSRs may help explore the SN\,Ia population that has taken the single degenerate (nova) pathway.

\begin{table*}
\caption{Details of the 20 known M\,31 recurrent novae and 4 LMC recurrent novae. For M\,31, we state the LGGS fields covering each nova with LGGS field numbers shown in parentheses indicating novae that are contained within an LGGS field, but are too close to an edge to allow sufficient remnant analysis. We also use modelling of H$\alpha$ emission from simulations of NSRs presented in \citet{2023MNRAS.521.3004H} to predict the H$\alpha$ luminosity of each NSR (see Section~\ref{sec:Predicted Ha luminosity} for details and Figure~\ref{fig:Halpha Evolution}). References -- (1) \citet{2015ApJS..216...34S}, (2) \citet{2021gacv.workE..44D}, (3) \citet{2023ATel16361....1S}, (4) \citet{2017ATel10042....1W}, (5) \citet{2022RNAAS...6..241S}, (6) \citet{2022RNAAS...6..214S}, (7) \citet{2021ATel14750....1D}, (8) \citet{2023TNSTR3081....1X}, (9) \citet{2023ATel16354....1S}, (10) \citet{2020MNRAS.491..655K}, (11) \citet{2018ATel11384....1M}, (12) \citet{2016ApJ...818..145B}, (13) \citet{2004IAUC.8424....1B}, (14) \citet{2004IAUC.8424....2M}.}
\label{tab:M31,LMC-RNe}
\begin{center}
\begin{tabular}{llllllll}
\hline\hline
Name & R.A.\ & Decl.\ & Eruptions & $P_\mathrm{rec}$ & LGGS field(s) & Predicted NSR H$\alpha$ \\
& (J2000) & (J2000) & & (years) & & luminosity (erg s$^{-1}$) & References \\
\hline
M31N 2008-12a & 00:45:28.80 & +41:54:10.1 & 19 & $0.996\pm0.030$ & 2, 3 & $1.15 \times 10^{33}$ & 1, 2, 3 \\
M31N 2017-01e & 00:44:10.72 & +41:54:22.1 & 4 & $2.545\pm0.020$ & 3 & $3.63 \times 10^{32}$ & 4, 5 \\
M31N 1926-07c & 00:42:53.37 & +41:15:43.7 & 3 & $\sim3$ & 5, 6 & $3.62 \times 10^{32}$ & 6 \\
M31N 1997-11k & 00:42:39.59 & +41:09:04.0 & 3 & $\sim4$ & 5, 6 & $2.63 \times 10^{32}$ & 1, 2 \\
M31N 1963-09c & 00:42:57.74 & +41:08:12.1 & 4 & $\sim5$ & 5, 6, (7) & $2.23 \times 10^{32}$ & 1, 2\\
M31N 1960-12a & 00:42:55.66 & +41:14:11.7 & 3 & $\sim6$ & 5, 6 & $1.90 \times 10^{32}$ & 1, 2\\
M31N 1984-07a & 00:42:47.24 & +41:16:19.8 & 3 & $\sim8$ & (4), 5, 6 & $1.53 \times 10^{32}$ & 1, 2\\
M31N 2006-11c & 00:41:33.17 & +41:10:12.4 & 2 & $\sim8$ & 6 & $1.53 \times 10^{32}$ & 1, 2\\
M31N 1990-10a & 00:43:04.05 & +41:17:07.5 & 3 & $\sim9$ & 4, 5, 6 & $1.42 \times 10^{32}$ & 1, 2\\
M31N 2007-11f & 00:41:31.52 & +41:07:13.1 & 2 & $\sim9$ & 6, 7 & $1.42 \times 10^{32}$ & 1, 2\\
M31N 1923-12c & 00:42:38.37 & +41:08:45.8 & 2 & $\sim9$ & 5, 6, (7) & $1.42 \times 10^{32}$ & 7 \\
M31N 2013-10c & 00:43:09.32 & +41:15:41.6 & 2 & $\sim10$ & 5, 6 & $1.28 \times 10^{32}$ & 8, 9 \\
M31N 2007-10b & 00:43:29.50 & +41:17:13.0 & 2 & $\sim10$ & 4, 5, 6 & $1.28 \times 10^{32}$ & 1, 2\\
M31N 1982-08b & 00:46:06.60 & +42:03:48.0 & 2 & $\sim14$ & 1, 2, 3 & $9.96 \times 10^{31}$ & 1, 2\\
M31N 1945-09c & 00:41:28.55 & +40:53:14.1 & 2 & $\sim27$ & 6, 7, 8 & $5.64 \times 10^{31}$ & 1, 2\\
M31N 1926-06a & 00:41:41.00 & +41:03:36.7 & 2 & $\sim37$ & 6, 7 & $4.15 \times 10^{31}$ & 1, 2\\
M31N 1966-09e & 00:39:30.80 & +40:29:15.0 & 2 & $\sim41$ & 8, 9 & $3.75 \times 10^{31}$ & 1, 2\\
M31N 1961-11a & 00:42:31.43 & +41:16:22.1 & 2 & $\sim44$ & 5, 6 & $3.46 \times 10^{31}$ & 1, 2\\
M31N 1953-09b & 00:42:20.74 & +41:16:14.1 & 2 & $\sim51$ & 5, 6 & $2.89 \times 10^{31}$ & 1, 2\\
M31N 1919-09a & 00:43:28.76 & +41:21:42.6 & 2 & $\sim79$ & 4, 5, 6 & $2.03 \times 10^{31}$ & 1, 2\\
\hline
LMCN 1968-12a & 05:09:58.40 & --71:39:52.7 & 4 & $6.2\pm1.2$ & -- & $1.85 \times 10^{32}$ & 10 \\
LMCN 1996 & 05:13:30.00 & --68:38:00.0 & 2 & $\sim22$ & -- & $6.89 \times 10^{31}$ & 11 \\
LMCN 1971-08a & 05:40:44.20 & --66:40:11.6 & 2 &$\sim38$ & -- & $4.04 \times 10^{31}$ & 12 \\
YY Doradus & 05:56:42.40 & --68:54:34.5 & 2 & $\sim67$ & -- & $2.22 \times 10^{31}$ & 13, 14 \\
\hline
\end{tabular}
\end{center}
\end{table*}

As predicted by \citet{2023MNRAS.521.3004H}, the on-sky size of any Galactic NSRs would be prohibitively large for facilities with small field-of-views. Indeed, attempts so far to find evidence of associated NSRs around a number of Galactic RNe using conventional small-mid sized facilities has proved futile; this is not surprising as these NSRs will have the same (very low, if not lower) surface brightness as the 12a NSR but cover a much larger proportion of the sky. Significantly though, the $2.3 \times 1.5$\,deg$^2$ FOV and dedicated observing campaign with the new Condor Array Telescope \citep{2023PASP..135a5002L} has led to the discovery of a new NSR: a 50 pc shell surrounding the Galactic nova, KT Eridani, thereby likely confirming its RN status \citep{2023arXiv231017055S,2023arXiv231017258H}. 

M\,31 is the leading laboratory to study nova population statistics \citep{2019enhp.book.....S,2020AdSpR..66.1147D}, due to its high nova rate \citep{2006MNRAS.369..257D}, the uniform, yet close, distance to all novae and the low absorbing column in that direction. Additionally, in M\,31 and the sufficiently distant Large Magellanic Cloud (LMC), NSRs should be of order a few tens of arcseconds across unlike their Galactic counterparts that span many degrees on the sky. As such, M\,31 and the LMC may provide ideal sites to search for NSRs; this is the motivation for the following work.

In this paper we present both the results of a targeted search within the LMC and a survey within M\,31 specifically searching for NSRs. In Section~\ref{sec:Observations and simulations} we outline the data used in this work and follow this with our analysis of the surroundings of each recurrent nova in Section~\ref{sec:Annular Photometry}. We present of results of this analysis in Section~\ref{sec:Results} and discuss implications in Section~\ref{sec:Discussion}, before concluding our paper in Section~\ref{sec:Conclusions}.

\section{Observations and simulations}\label{sec:Observations and simulations}
\subsection{M31N 2008-12a Nova Super-Remnant}
Prior to the discovery of 12a \citep[by][]{2008Nis} and the subsequent realisation of its recurrent nature, its surrounding NSR had already been serendipidously captured by several wide-field surveys \cite[see, for e.g.,][]{1992A&AS...92..625W,2011AJ....142..139A}. However, its potential association with 12a only became apparent following the 2014 eruption \citep[see][and references therein]{2015A&A...580A..45D,2019Natur.565..460D}. The 12a NSR was clearly visible in ground-based narrow band H$\alpha$ and [\ion{S}{ii}] $\lambda$6713, 6731 imaging data (but not [\ion{O}{iii}] $\lambda$5007, due to a lack of such emission) collected as part of the Local Group Galaxies Survey \citep[LGGS;][]{2006AJ....131.2478M,2007AJ....134.2474M}. Figure~\ref{12a NSR images} presents LGGS data in the immediate vicinity of 12a and reveals the extended nebulosity in H$\alpha$ and [\ion{S}{ii}] around the system; the LGGS [\ion{O}{iii}] and a deeper Liverpool Telescope H$\alpha$ image are also provided.
\begin{figure*}
\includegraphics[width=\textwidth]{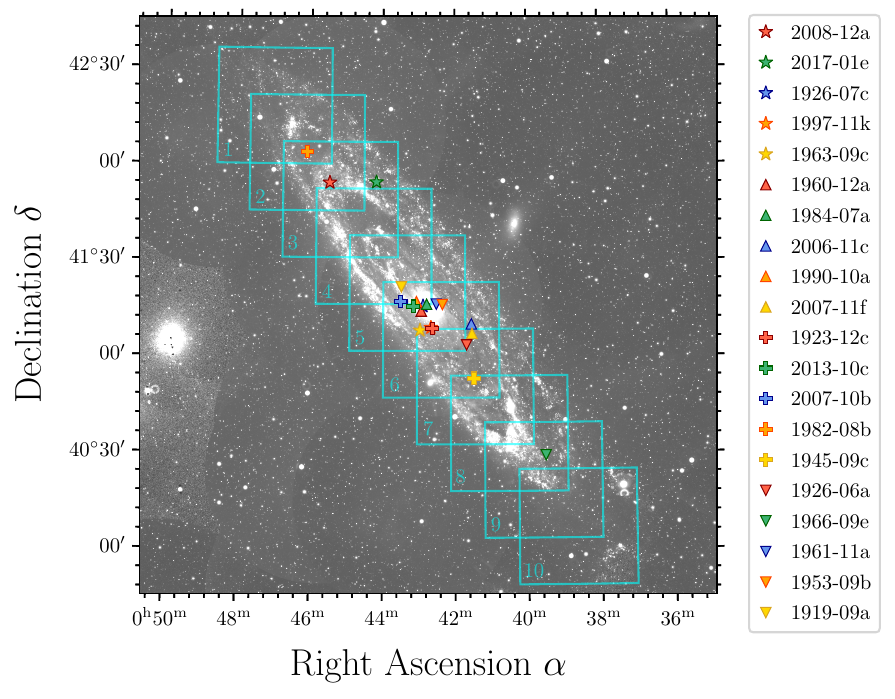}
\caption{A near-UV $3^\circ\times 3^\circ$ mosaic image of M\,31 taken by GALEX. The approximate positions of the ten LGGS fields are overlaid in cyan and the positions of the 20 known M\,31 recurrent novae are provided. In Figure~\ref{fig:M31 bulge RNe}, we show the same near-UV GALEX image but zoomed-in on the bulge of M\,31. \label{fig:LGGS-fields}}
\end{figure*}

\subsection{Local Group Galaxy Survey data and image processing}\label{LGGS data and image processing}
At the time of writing, there are 20 known RNe in M\,31 (see Table~\ref{tab:M31,LMC-RNe}). The majority were recovered through work by \citep{2015ApJS..216...34S}, with M31N 2017-01e confirmed by \citet{2017ATel10042....1W}, M31N 1923-12c by \citet{2021ATel14750....1D} and M31N 2013-10c by \citet{2023ATel16354....1S}.

The positions of all 20 RNe are well covered by at least one field of the LGGS survey, with 13 appearing in two of the ten fields, and five in three fields. In Figure~\ref{fig:LGGS-fields}, we show the approximate placing of the ten LGGS fields and the positions of all 20 known RNe. In this study we utilise the narrowband H$\alpha$, [\ion{S}{ii}] and [\ion{O}{iii}] and broad-band $V$ and $R$ imaging data from LGGS. The LGGS processed science data \citep[see][]{2006AJ....131.2478M,2007AJ....134.2474M} were obtained from the Lowell web pages\footnote{\url{http://www.lowell.edu/users/massey/lgsurvey} given in \citet{2006AJ....131.2478M}.}.

Continuum subtraction of the LGGS images was performed using IRAF \citep{1986SPIE..627..733T,1993ASPC...52..173T} following matching photometry with DAOPHOT \citep{1987PASP...99..191S} between suitable narrow- and broad-band pairs:\ $R$ for H$\alpha$ and [\ion{S}{ii}]; $V$ for [\ion{O}{iii}]. As described in Section~\ref{sec:Annular Photometry}, narrowband fluxes, per square-arcsecond, for H$\alpha$ and [\ion{S}{ii}] were computed following the methodology presented in \citet{2010ApJS..187..275S}. The projected distance from the source is computed assuming an M31 distance of 778\,kpc \citep{1998ApJ...503L.131S}. Figure~\ref{2008-12a surrounding sub images} illustrates the results of this image processing for 12a where the H$\alpha$ and [\ion{S}{ii}] subtracted images show a clear broad emission structure, compared to the [\ion{O}{iii}] image. In the same manner, we illustrate the results of the continuum subtraction for the LGGS H$\alpha$, [\ion{S}{ii}] and [\ion{O}{iii}] images surrounding the other 19 RNe in Figures~\ref{2017-01e surrounding sub images}--\ref{1919-09a surrounding sub images}.
\begin{figure}
\includegraphics[width=\columnwidth]{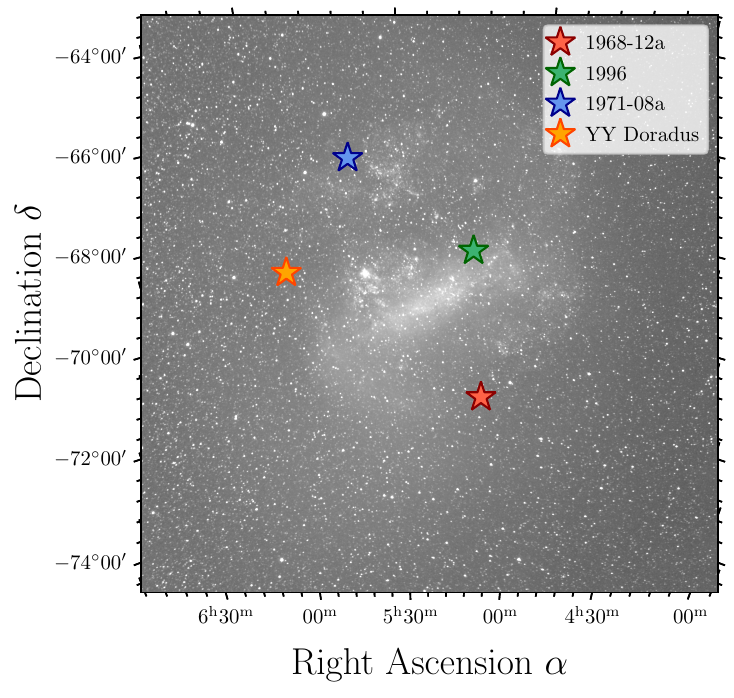}
\caption{An optical $12^\circ\times 12^\circ$ mosaic image of the LMC taken by Axel Mellinger \citep{2009PASP..121.1180M}. The positions of the 4 known LMC recurrent novae are provided.\label{fig:LMC RNe}}
\end{figure}

\subsection{Faulkes Telescope South data and image processing}\label{FTS data and image processing}

In the Large Magellanic Cloud, there are four known RNe (see Table~\ref{tab:M31,LMC-RNe}) -- we show their locations in Figure~\ref{fig:LMC RNe}. As for the potential NSRs in M\,31, the NSRs surrounding each of the RNe in the LMC will subtend a small enough angular size\footnote{Angular radii of $1.4^{\prime}$, $1.0^{\prime}$, $0.8^{\prime}$ and $0.4^{\prime}$ for LMCN\,1968-12a, LMCN\,1996, LMCN\,1971-08a and YY Doradus, respectively.} to be observed within one field of view of many ground-based telescopes.
\begin{table}
\caption{Details of observations of the surroundings of the four RNe in the LMC with the Spectral Imager on the Faulkes Telescope South (exposure and number of exposures).}
\label{tab:LMC FTS}
\begin{center}
\begin{tabular}{llllllll}
\hline\hline
Name & H$\alpha$ & R-band & [\ion{O}{iii}] & V-band \\
\hline
LMCN\,1968-12a & 720s $\times$ 4 & 180s $\times$ 3 & 180s $\times$ 3 & 180s $\times$ 3 \\
LMCN\,1996 & 720s $\times$ 4 & 180s $\times$ 3 & 180s $\times$ 3 & 180s $\times$ 3 \\
LMCN\,1971-08a & 720s $\times$ 3 & 180s $\times$ 3 & 180s $\times$ 3 & 180s $\times$ 3 \\
YY Doradus & 720s $\times$ 4 & 180s $\times$ 3 & 180s $\times$ 3 & 180s $\times$ 3 \\
\hline
\end{tabular}
\end{center}
\end{table}

As such, we observed the surroundings of each of the four RNe in the LMC with the Faulkes Telescope South (FTS)\footnote{ID number: NSF2022B-004} as any existing NSRs will fit comfortably within the $10.5^{\prime} \times 10.5^{\prime}$ field of view of the Spectral Imager camera on this 2-m telescope. We utilised narrowband H$\alpha$ and [\ion{O}{iii}] and broad-band $R$ and $V$ filters on FTS (see Table~\ref{tab:LMC FTS} for details of these data): the H$\alpha$ imaging to detect emission from a NSR shell; the [\ion{O}{iii}] imaging to rule out the possibility of any identified remnants having a different origin (for example, see Section~\ref{MCSNR J0514-6840}) and the $R$ and $V$ imaging for continuum subtraction.

We performed continuum subtraction using the same method as with the M\,31 LGGS data using IRAF \citep{1986SPIE..627..733T,1993ASPC...52..173T} and DAOPHOT \citep{1987PASP...99..191S}. As described in Section~\ref{sec:Annular Photometry}, narrowband fluxes, per square-arcsecond, were then computed in the same manner as for the LGGS data following the methodology presented in \citet{2010ApJS..187..275S} and the projected distance from the source is computed assuming a distance to the LMC of 49.59 kpc \citep{2019Natur.567..200P}. We show the results of this image processing for the FTS H$\alpha$ and [\ion{O}{iii}] images surrounding the four RNe in Figures~\ref{LMCN 1968-12a surrounding sub images}--\ref{YY Doradus surrounding sub images}.

\begin{figure}
\includegraphics[width=\columnwidth]{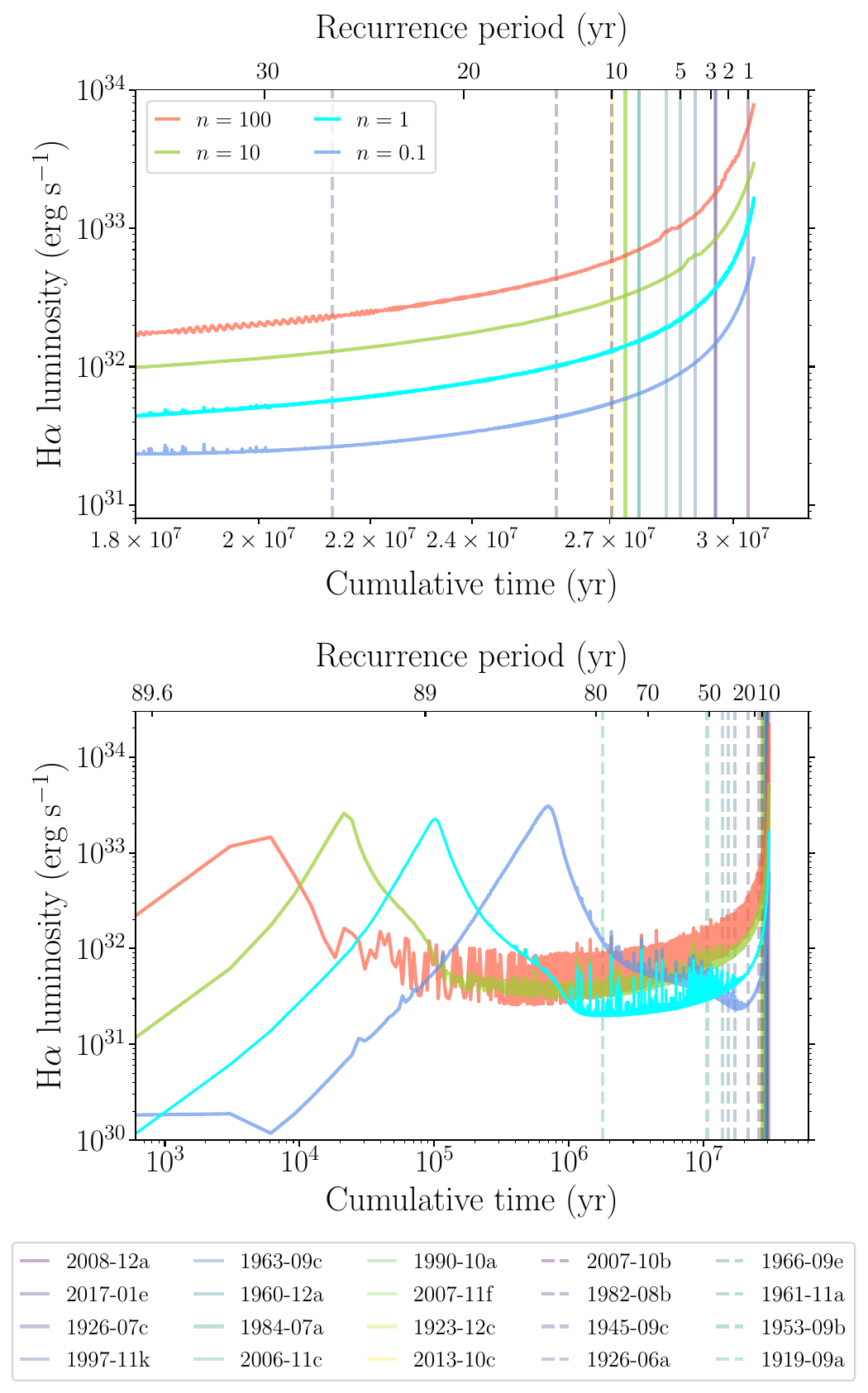}
\caption{The evolution of NSR H$\alpha$ luminosity for the reference simulation ($n=1$) and simulations with difference ISM densities; Run 2 ($n=0.1$), Run 5 ($n=10$) and Run 7 ($n=100$) presented in \citet{2023MNRAS.521.3004H}. The bottom panel shows the NSR H$\alpha$ luminosity throughout the full evolution of the simulations up to ${\sim}3.1 \times 10^7$ years whereas the top panel focusses on a cumulative time of $ 1.8 \times 10^7$ years for clarity. Also, to remove the impact of single eruptions, we re-bin to a lower temporal resolution in the {\it top panel}. The vertical lines indicate the recurrence periods of all 20 RNe in M\,31 and their associated cumulative time from the NSR simulations in \citet{2023MNRAS.521.3004H}.}
\label{fig:Halpha Evolution}
\end{figure}
\subsection{Predicted NSR H\texorpdfstring{\boldmath{$\alpha$}}{alpha} luminosity from simulations}\label{sec:Predicted Ha luminosity}
The simulations presented in \citet{2023MNRAS.521.3004H} replicate the evolution of a range of nova systems and the subsequent creation of each system's NSR. One of these simulations was utilised to model the H$\alpha$ emission from a remnant forming around a $1 \times 10^7$K WD with an accretion rate of $1 \times 10^{-7} \ \text{M}_{\odot} \ \text{yr}^{-1}$ as it grew from $1 \ \text{M}_{\odot}$ to the Chandrasekhar limit \citep[see][their Figure 14]{2023MNRAS.521.3004H}. We reproduce a version of this figure here converted to illustrate H$\alpha$ luminosity evolution in Figure~\ref{fig:Halpha Evolution} for such a system located within an ISM of $1.67 \times 10^{-24} \ \text{g} \ \text{cm}^{-3}$ (1 H atom $\text{cm}^{-3}$), which we will denote as $n=1$ (cyan line), along with the evolution of H$\alpha$ emission from the same system placed within different environments: $1.67 \times 10^{-25} \ \text{g} \ \text{cm}^{-3}$ ($n=0.1$, blue), $1.67 \times 10^{-23} \ \text{g} \ \text{cm}^{-3}$ ($n=10$, green) and $1.67 \times 10^{-22} \ \text{g} \ \text{cm}^{-3}$ ($n=100$, red). We have displayed the 20 RNe in Figure~\ref{fig:Halpha Evolution} using their respective recurrence periods to estimate the H$\alpha$ luminosity from each system. By inspection, the first peak of H$\alpha$ emission in the different environments occurs earlier as the ISM density is increased (${\sim}6 \times 10^3$ years for $n=100$ compared to ${\sim}8 \times 10^5$ for $n=0.1$). However, after this initial peak, the NSR H$\alpha$ luminosity then follows a similar evolutionary trend for each of different ISM densities. Beyond this, the NSR located within the highest density ISM ($n=100$) continues to be the brightest in H$\alpha$, ending with $\text{L}_{\text{H}\alpha}\simeq8.6 \times 10^{33}$ erg\,s$^{-1}$ compared $\text{L}_{\text{H}\alpha}\simeq5.7 \times 10^{32}$ erg\,s$^{-1}$ in the $n=0.1$ surroundings.

As a first-order approximation for the total H$\alpha$ luminosity from each potential NSR in M\,31, we assumed an ISM density of $1.67 \times 10^{-24}$ g cm$^{-3}$ and computed the corresponding H$\alpha$ luminosity at the system's current recurrence period. The predicted total H$\alpha$ luminosity from a NSR for each of the 20 M\,31 RNe and 4 LMC RNe are provided in Table~\ref{tab:M31,LMC-RNe}.

The H$\alpha$ luminosity from the 12a NSR is larger than all other predicted NSRs  (see Table~\ref{tab:M31,LMC-RNe}) due to its higher mass WD and therefore longer evolution. It therefore experiences highly energetic eruptions at late times that travel through the already established NSR ejecta pile-up region before colliding with the formed high density shell, leading to collisional excitation and high levels of recombination \citep{2023MNRAS.521.3004H}. From this modelling, we conclude that 12a should have the NSR with the highest total H$\alpha$ luminosity out of the 20 RNe in M\,31 and the 4 RNe in the LMC.

\section{Annular Photometry}\label{sec:Annular Photometry}
To search for excess, extended, emission around each RN, we perform annular photometry at the position of each M\,31 nova in all of the LGGS fields in which they are found and for the LMC novae in each image, out to 295$^{\prime\prime}$ in all frames. 

The annular photometry consisted of 500 circular annuli (970 for FTS data) with radii logarithmically distributed from 0 to 135 arcseconds (295$^{\prime\prime}$ for FTS data). To remain consistent between different data sets, for both the LGGS data and the FTS data, we used the methodology provided in \citet{2010ApJS..187..275S} and photometry from {\it Gaia} DR2 \citep{2016A&A...595A...1G,2018A&A...616A...1G} to convert the images from counts s$^{-1}$ to flux per count in units of ergs s$^{-1}$ cm$^{-2}$ \AA$^{-1}$, before applying a scaling of the flux to account for differences between the $G_{\text{RP}}$ filter \citep{2018A&A...616A...1G,2018A&A...617A.138W} and the LGGS H$\alpha$ filter and [\ion{S}{ii}] filter \citep{2007AJ....134.2474M} and the FTS H$\alpha$ filter\footnote{\url{https://lco.global/observatory/instruments/filters/}}. Additionally, we applied pixel masking to remove clear artefacts at large radii -- the origins of which were bright stars coincident with the surroundings of the novae.

The results of our continuum-subtracted, pixel-masked H$\alpha$ and [\ion{S}{ii}] luminosity radial profiles for the M\,31 data are shown in Figure~\ref{fig:NSR Ha luminosity} and Figure~\ref{fig:NSR SII luminosity}, respectively. We have removed 1926-07c, 1960-12a, 1984-07a, 1990-10a and 1961-11a from our analysis as they are all too close to the bulge of M\,31 (see Figure~\ref{fig:M31 bulge RNe}) to allow any form of sufficient data processing, after continuum subtraction, in the same manner as other regions of the galaxy. The equivalent for the H$\alpha$ luminosity radial profiles for the LMC data is shown in Figure~\ref{fig:NSR Ha luminosity LMC}. We have not included the [\ion{O}{iii}] luminosity radial profiles for either the M\,31 or LMC data as all profiles are very noisy and are consistent with zero excess flux for all RN, including 12a.

\section{Results}\label{sec:Results}
The aim of this survey of M\,31 and targeted search within the LMC was to discover more NSRs to place alongside the NSRs around M\,31N 2008-12a \citep{2019Natur.565..460D,2023MNRAS.521.3004H} and KT Eridani \citep{2023arXiv231017055S,2023arXiv231017258H}. As the emission from the NSR around 12a can be clearly seen in $1^{\prime} \times 1^{\prime}$ H$\alpha$ and $[\ion{S}{ii}]$ LGGS images centred on the nova (shown in Figure~\ref{12a NSR images} and Figure~\ref{2008-12a surrounding sub images}), we would also expect to see similar emission in the same vicinity of other RNe within M\,31 and the LMC if they were to also host a {\it visible} NSR.

\subsection{M31}\label{sec:Results M31}
We can clearly see by looking at the continuum subtracted narrowband images in Figures~\ref{2017-01e surrounding sub images}--\ref{1919-09a surrounding sub images} that there is an absence of evidence for such emission around all other RNe in M\,31. Furthermore, the results of annular photometry performed on the H$\alpha$ and [\ion{S}{ii}] LGGS images for each of the other RNe, as described in Section~\ref{sec:Annular Photometry} and provided in Figure~\ref{fig:NSR Ha luminosity} and Figure~\ref{fig:NSR SII luminosity}, respectively, quantify this lack of observable emission in comparison to 12a. For the five RNe that are too close to the bulge of M\,31, we cannot speculate about whether these systems host an observable NSR. 

\subsubsection{H\texorpdfstring{\boldmath{$\alpha$}}{alpha} luminosity}\label{M31 Ha luminosity}
\begin{figure}
\includegraphics[width=\columnwidth]{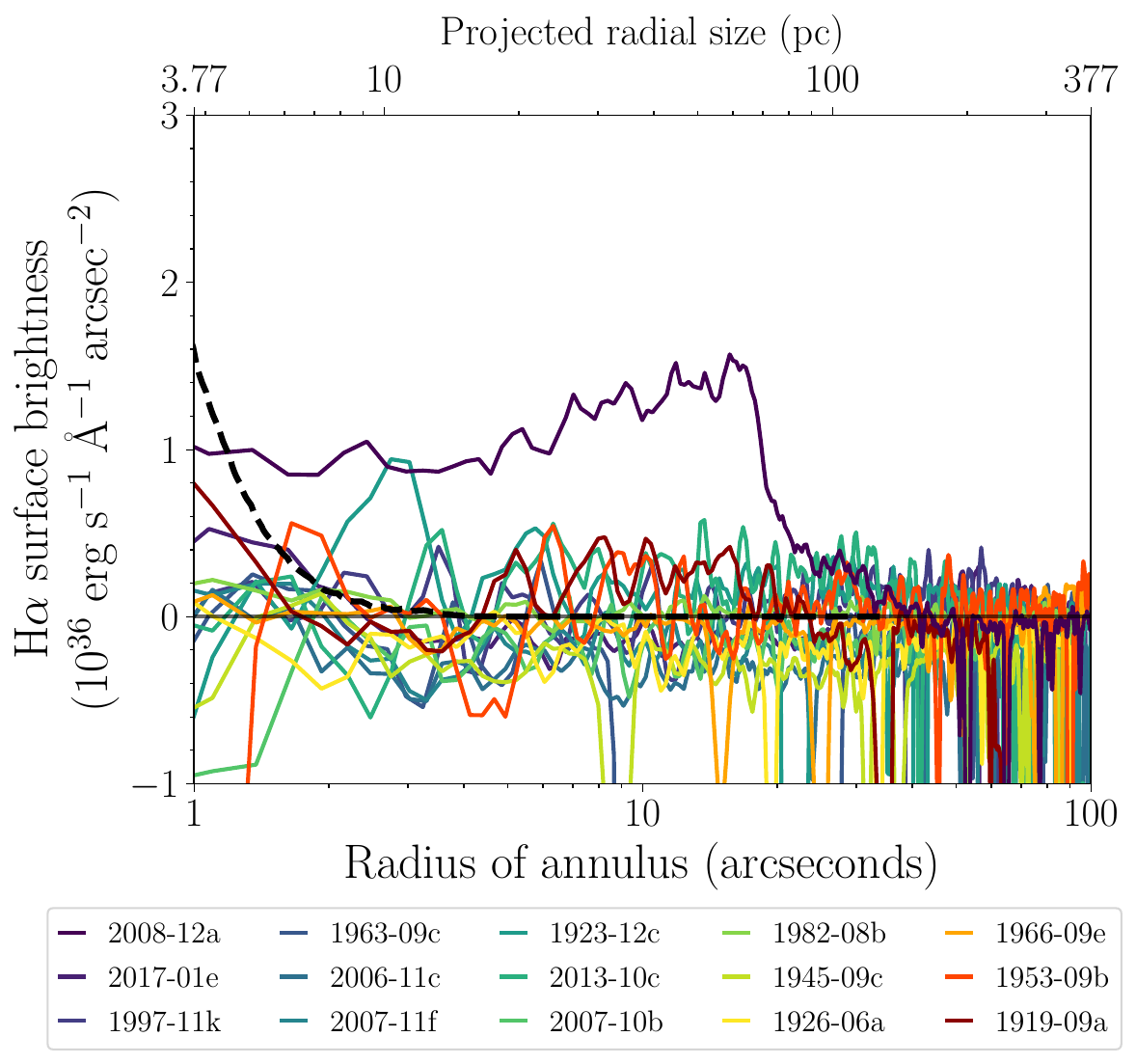}
\caption{The H$\alpha$ luminosity surface brightness of the surroundings of fiftee
novae in M\,31 out to 100$^{\prime \prime}$ (with a projected size of $\sim$ 377 parsecs at a distance of M\,31) -- the luminosity for the other five novae were omitted as their close proximity to the bulge of M\,31 prevented sufficient data processing in comparison to the other novae (see Figure~\ref{fig:M31 bulge RNe}). This is emission determined from annular photometry after continuum subtraction and pixel masking of the bright sources in the images. We also show the PSF for the LGGS data (black dashed line).}
\label{fig:NSR Ha luminosity}
\end{figure}

The characteristic shell surrounding M\,31N 2008-12a is trivially recoverable in the H$\alpha$ LGGS image. This is reflected in Figure~\ref{fig:NSR Ha luminosity}, with the surplus of H$\alpha$ emission from the nova out to approximately 35 arcseconds. At the distance of M\,31, this is equivalent to ${\sim}130$ parsecs -- consistent with the 12a NSR projected size of 134 parsecs \citep{2019Natur.565..460D}. The peak emission is approximately $1.6 \times 10^{36}$ erg s$^{-1}$ \AA$^{-1}$ arcsec$^{-2}$ at ${\sim}$17 arcseconds (projected size of ${\sim}$64 parsecs).

Clearly, the predicted H$\alpha$ emission of 12a from simulations as described in Section~\ref{sec:Predicted Ha luminosity} ($1.15 \times 10^{33}$ erg s$^{-1}$ in Table~\ref{tab:M31,LMC-RNe}) does not match the total integrated H$\alpha$ luminosity from the LGGS observations (${\sim} 6 \times 10^{37}$ erg s$^{-1}$). The discrepancy may arise from the models placing the NSR in an ISM with a much lower density than it is actually situated. From Figure~\ref{fig:Halpha Evolution}, we show that this has a large effect on the H$\alpha$ flux of a NSR. Other factors potentially playing a role in this discrepancy include the cooling package employed in the \citet{2023MNRAS.521.3004H} simulations not being scaled for lower densities of ISM; assuming a pure-hydrogen ISM or assuming the ejecta and ISM are spherically symmetric \citep{2023arXiv231017258H}.

\begin{figure}
\includegraphics[width=\columnwidth]{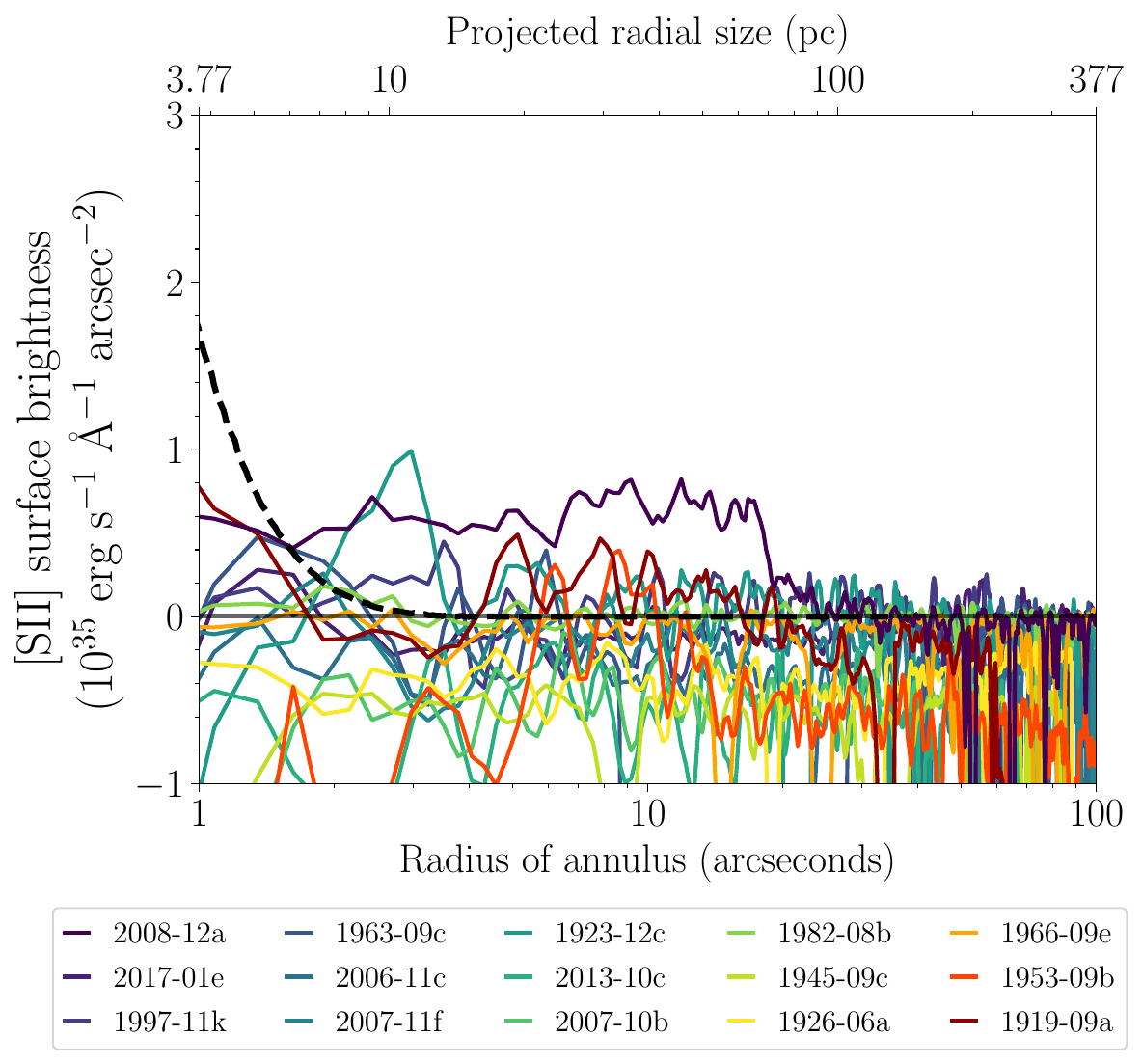}
\caption{As in Figure~\ref{fig:NSR Ha luminosity} but with the [$\ion{S}{ii}$] luminosity surface brightness surrounding each of the fifteen M\,31 RN we analysed.}
\label{fig:NSR SII luminosity}
\end{figure}

\begin{figure}
\includegraphics[width=\columnwidth]{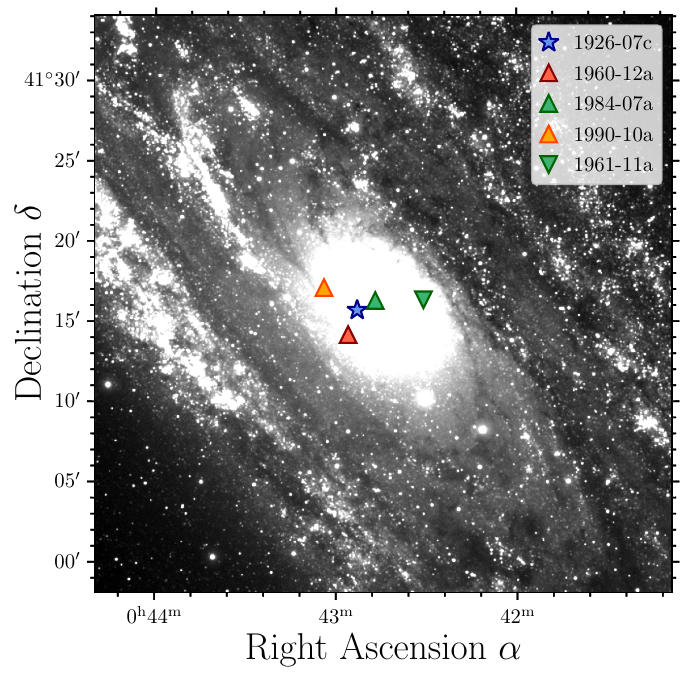}
\caption{A zoom-in of the bulge of M\,31 from the near-UV GALEX image shown in Figure~\ref{fig:LGGS-fields}. Here we show the locations of the five novae that we cannot analyse.\label{fig:M31 bulge RNe}}
\end{figure}

The other fourteen H$\alpha$ flux radial profiles shown in Figure~\ref{fig:NSR Ha luminosity} resemble noise that is consistent with zero excess flux; we have provided a negative section of the y axis to illustrate this. For example, 1923-12c looks to have a peak at ${\sim}2 - 3$ arcseconds however, at radii $<2$ arcseconds the flux drops below zero by approximately the same amount, indicative of noise. Moreover, by inspection, this peak is likely from an artefact created during continuum subtraction of a very nearby source. 1953-09b has a peak of {$\sim$}$6 \times 10^{35}$ erg s$^{-1}$ \AA$^{-1}$ arcsec$^{-2}$ at ${\sim}2$ arcseconds, but again, the flux either side of this region drops below zero by a similar factor. 

Both 2017-01e and 1919-09a have small bumps at the lowest radius (between 1 and 2 arcseconds) followed by a flat, albeit noisy, profile. However, as shown by the scaled point source function (PSF) for the H$\alpha$ LGGS data in Figure~\ref{fig:NSR Ha luminosity}, this initial peak in both profiles can be attributed to the PSF of the central object -- which may indeed be the quiescent nova system, bright in H$\alpha$ due to accretion.

Potentially, some of the RNe without a structure may have a faint NSR only just below the detection limit of LGGS. If this was the case, then combining all of the surroundings centred on each RNe we analysed may reveal a faint glow out to approximately 30$^{\prime \prime}$. To test this we took the $1^{\prime} \times 1^{\prime}$ surroundings of the fourteen M\,31 RNe we analysed, aligned them by the location of the nova and then co-added the images. However, the fully combined image did not display excess emission.

\subsubsection{\texorpdfstring{\boldmath{$[\ion{S}{ii}]$}}{$[\ion{S}{ii}]$} luminosity}\label{M31 [SII] luminosity}
As with the extended emission for 2008-12a derived from the LGGS H$\alpha$ data, we see a similarly extended profile of [$\ion{S}{ii}$] emission. As we would expect, this emission is fainter but is still able to quantify the NSR structure surrounding 12a. The extended emission reaches out to approximately 22 arcseconds which is equivalent to a projected size of ${\sim}$83 parsecs. Whilst this doesn't match the furthest extent of the H$\alpha$ profile of 12a (see Section~\ref{M31 Ha luminosity}), it is likely that the faintness of the [$\ion{S}{ii}$] emission compared to H$\alpha$ contributes to the lack of detection at the outer regions.

The [$\ion{S}{ii}$] luminosity radial profiles for the other M\,31 RNe somewhat mimic their H$\alpha$ counterparts. 1923-12c again has an evident peak between ${\sim}2 - 3$ arcseconds but can be attributed to a similar continuum subtraction artefact of a nearby source and is consistent with noise. There is a bump near the origin ($< 3$ arcseconds) in the luminosity profile for 1963-09c but this fits with the scaled PSF for this data. Similarly, the surroundings of 1919-09a reveal a bump near the centre of the profile ($< 2$ arcseconds) that is consistent with the PSF of the image. Other than these peaks, all of the radial luminosity profiles are relatively flat (accounting for the noise).

\subsubsection{\texorpdfstring{\boldmath{$[\ion{O}{iii}]$}}{$[\ion{O}{iii}]$} luminosity}\label{M31 [OIII] luminosity}
As seen in the LGGS imaging, there is a total absence of [\ion{O}{iii}] from the NSR surrounding M31N 2008-12a, and so we find that this is the case when we quantify the emission; a very noisy luminosity profile averaging zero. The fourteen recurrent novae which we were able to analyse all show the same lack of extended [\ion{O}{iii}] emission (consistent with noisy data averaging zero) therefore we have chosen to exclude the [\ion{O}{iii}] luminosity radial profile.

\subsection{LMC}\label{sec:Results LMC}
\begin{figure}
\includegraphics[width=\columnwidth]{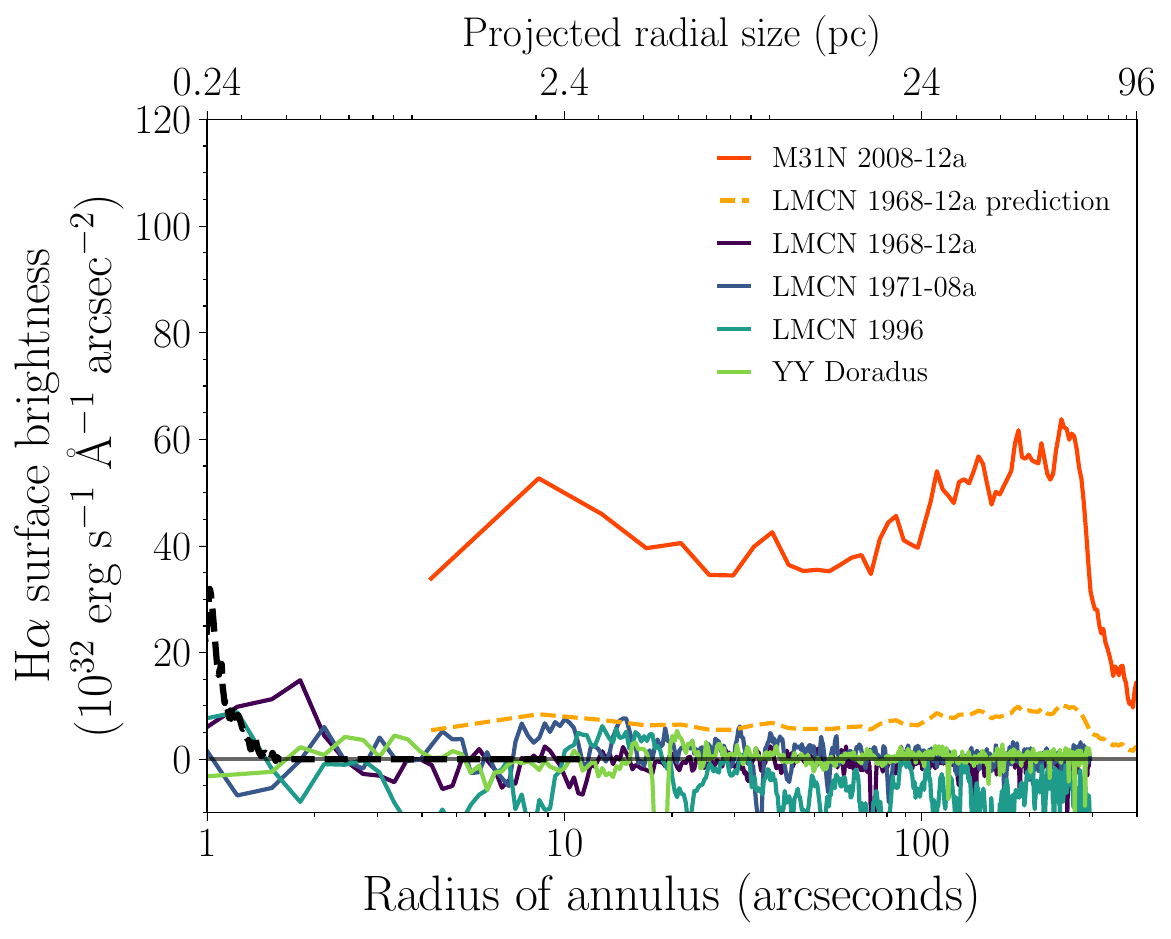}
\caption{The H$\alpha$ luminosity surface brightness of the surroundings of four novae in the LMC out to ${\sim}295^{\prime \prime}$ (with a projected size of $\sim$71 parsecs at a distance of the LMC). This is emission determined from annular photometry after continuum subtraction and pixel masking of the bright sources in the images. As in Figures~\ref{fig:NSR Ha luminosity} and \ref{fig:NSR SII luminosity}, we show the PSF for the FTS data (black dashed line). For comparison, we also include the 12a NSR H$\alpha$ emission (from Figure~\ref{fig:NSR Ha luminosity}; the red line) but scaled to the distance of the LMC. The orange dashed line is a first-order approximation of the NSR emission profile for LMCN 1968-12a (see Section~\ref{LMC Ha luminosity}).}
\label{fig:NSR Ha luminosity LMC}
\end{figure}

The surroundings of the four RNe shown in the continuum subtracted images (Figures~\ref{LMCN 1968-12a surrounding sub images} - \ref{YY Doradus surrounding sub images}) do not unveil any type of structure centred on the central nova. This absence of extended emission is reflected in the results of the annular photometry analysis described in Section~\ref{sec:Annular Photometry} and shown in Figure~\ref{fig:NSR Ha luminosity LMC}.

\subsubsection{H\texorpdfstring{\boldmath{$\alpha$}}{alpha} luminosity}\label{LMC Ha luminosity}
The lack of a discernible structure reminiscent of a NSR surrounding any of the four RNe in the LMC is quantified in Figure~\ref{fig:NSR Ha luminosity LMC}. While noisy, each of the radial luminosity profiles are consistent with zero emission.

The only exception to the flat (noisy) profiles is the bump between ${\sim}1 - 2.5$ arcseconds associated with LMCN 1968-12a. The point spread function for the Faulkes Telescope South data is much smaller (see the black dashed line in Figure~\ref{fig:NSR Ha luminosity LMC}) than in the LGGS data so may not necessarily be the reason behind the small peak. By inspection, within the continuum subtracted FTS H$\alpha$ image containing LMCN 1968-12a, the nova is located between two other stars therefore this apparent bump is likely to be contributions from the residuals of these two sources. The residuals are then likely to be compounded at these small radii, whereby the smallest apertures applied are more sensitive to (potentially extreme) pixel-to-pixel fluctuations.

In Figure~\ref{fig:NSR Ha luminosity LMC}, we show the H$\alpha$ emission for the M\,31 2008-12a NSR but as it would appear if the nova was in the LMC and, clearly, this NSR would detectable in our data. As the NSR for LMCN 1968-12a is predicted to be approximately the same size as the NSR surrounding M\,31 2008-12a, we have scaled the M\,31 2008-12a NSR emission to match the difference in predicted flux for both NSRs (from Table~\ref{tab:M31,LMC-RNe}) as a first-order approximation for the NSR emission profile of LMCN 1968-12a. This is shown as the orange dashed line in Figure~\ref{fig:NSR Ha luminosity LMC} and demonstrates that, given the LMC H$\alpha$ background and the extent of the FTS data, this NSR would be within the noise and therefore not detectable within our FTS data.

\subsubsection{\texorpdfstring{\boldmath{$[\ion{O}{iii}]$}}{$[\ion{O}{iii}]$} luminosity}\label{LMC [OIII] luminosity}
In a same way that the LGGS [\ion{O}{iii}] imaging had no trace of extended emission around any RNe in M\,31 (including 2008-12a as expected), the FTS [\ion{O}{iii}] imaging is also devoid of emission around the four RNe in the LMC. As such, the [\ion{O}{iii}] luminosity radial profiles for the four RNe quantified this absence with a noise profile consistent with zero flux. Therefore, we again omit this.

\subsubsection{Recovered supernova remnant -- MCSNR J0514-6840}\label{MCSNR J0514-6840}
\begin{figure}
\centering
\subfloat{\includegraphics[width=.25\textwidth]{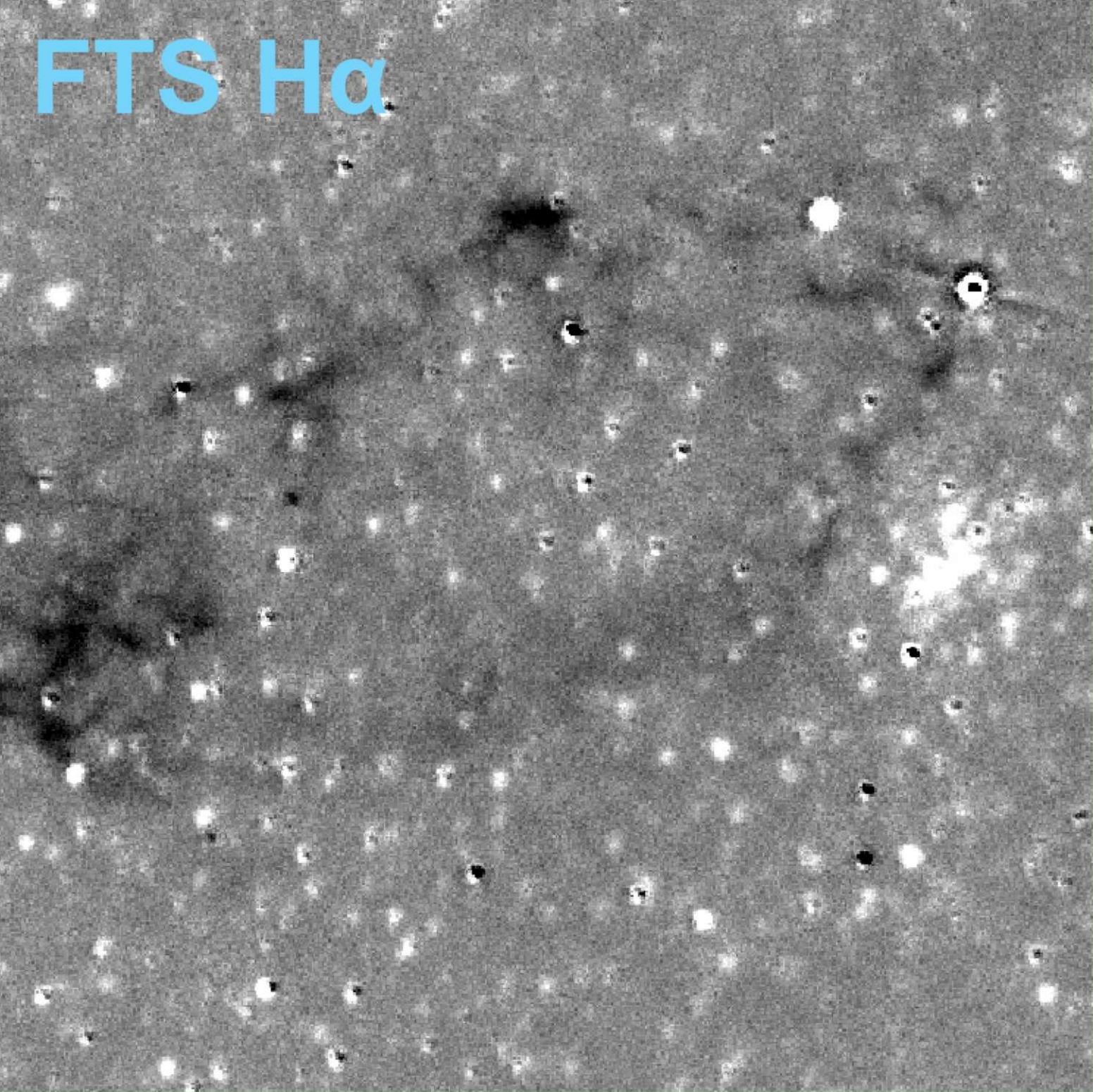}} \\
\subfloat{\includegraphics[width=.25\textwidth]{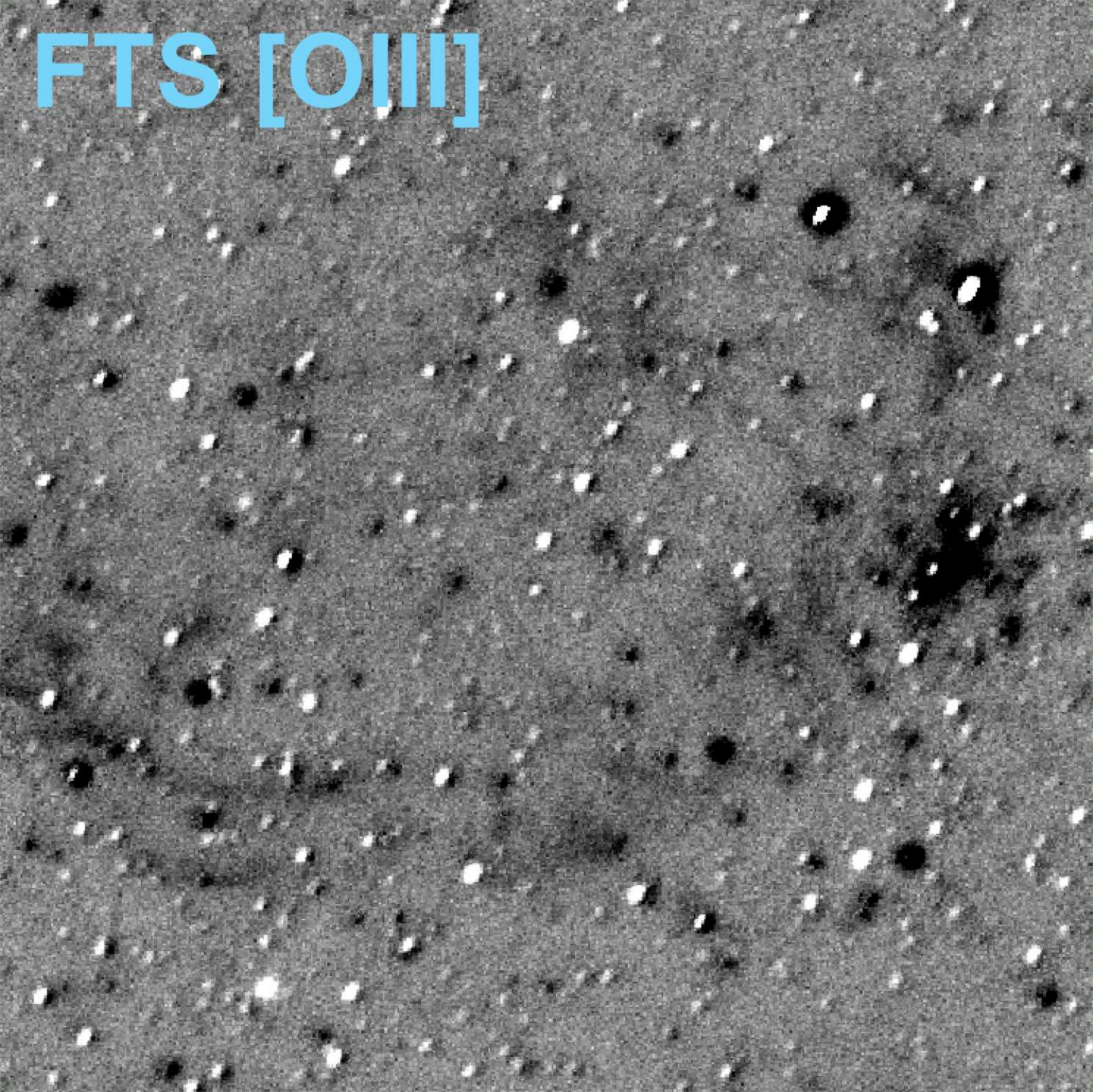}}
\caption{A $2.6^\prime\times2.6^\prime$ region of Faulkes Telescope South data showing the supernova remnant MCSNR J0514-6840. It was found in the same FTS field of view as the RNe LMCN\,1996 however note that the location of this nova is not within the above images. Top:\ Faulkes Telescope South H$\alpha$ image. Bottom:\ Faulkes Telescope South [\ion{O}{iii}] image.}
\label{SNR}
\end{figure}
In the south-west portion of the $10.5^{\prime} \times 10.5^{\prime}$ FTS field centred on LMCN\,1996, we do see extended H$\alpha$ emission from a known supernova remnant -- MCSNR J0514-6840 \citep[for example, see Figure 3 in][]{2014A&A...561A..76M} -- in the form of a shell, which we show in Figure~\ref{SNR}. Likewise, we see the supernova remnant shell, albeit with a different shape in our [\ion{O}{iii}] imaging of the same field (also shown in Figure~\ref{SNR}). In the H$\alpha$ image, the nearest edge of MCSNR J0514-6840 is approximately 3.9 arcminutes from the location of the nova.

While not relevant to the objectives of our study, the detection of this supernova remnant in the LMC does illustrate the suitability of our data. It also demonstrates the requirement for [\ion{O}{iii}] imaging to differentiate NSRs from similar-looking astrophysical phenomena such as supernova remnants.

\subsection{When would the 12a NSR have been missed?}\label{sec:Add noise 12a NSR}
As can be seen clearly in Figure~\ref{fig:NSR Ha luminosity} and Figure~\ref{2008-12a surrounding sub images}, the NSR surrounding M\,31N 2008-12a is well recovered in LGGS images. However, we will now look at how much fainter the structure associated with 12a would have to have been for it have been missed; this could be analogous to the lack of emission seen around the other systems.

We explored this idea by taking the continuum-subtracted LGGS image containing the NSR surrounding 12a and setting the instrumental magnitude of the background to be the zero point. We then added artificial noise to the image using the IRAF package \texttt{mknoise}, taking account of the change in magnitude of the background, until the NSR was no longer visible. By artificially brightening the background, we can determine how many magnitudes fainter the 12a NSR would have to be to evade discovery.

\begin{figure}
\centering
\includegraphics[width=\columnwidth]{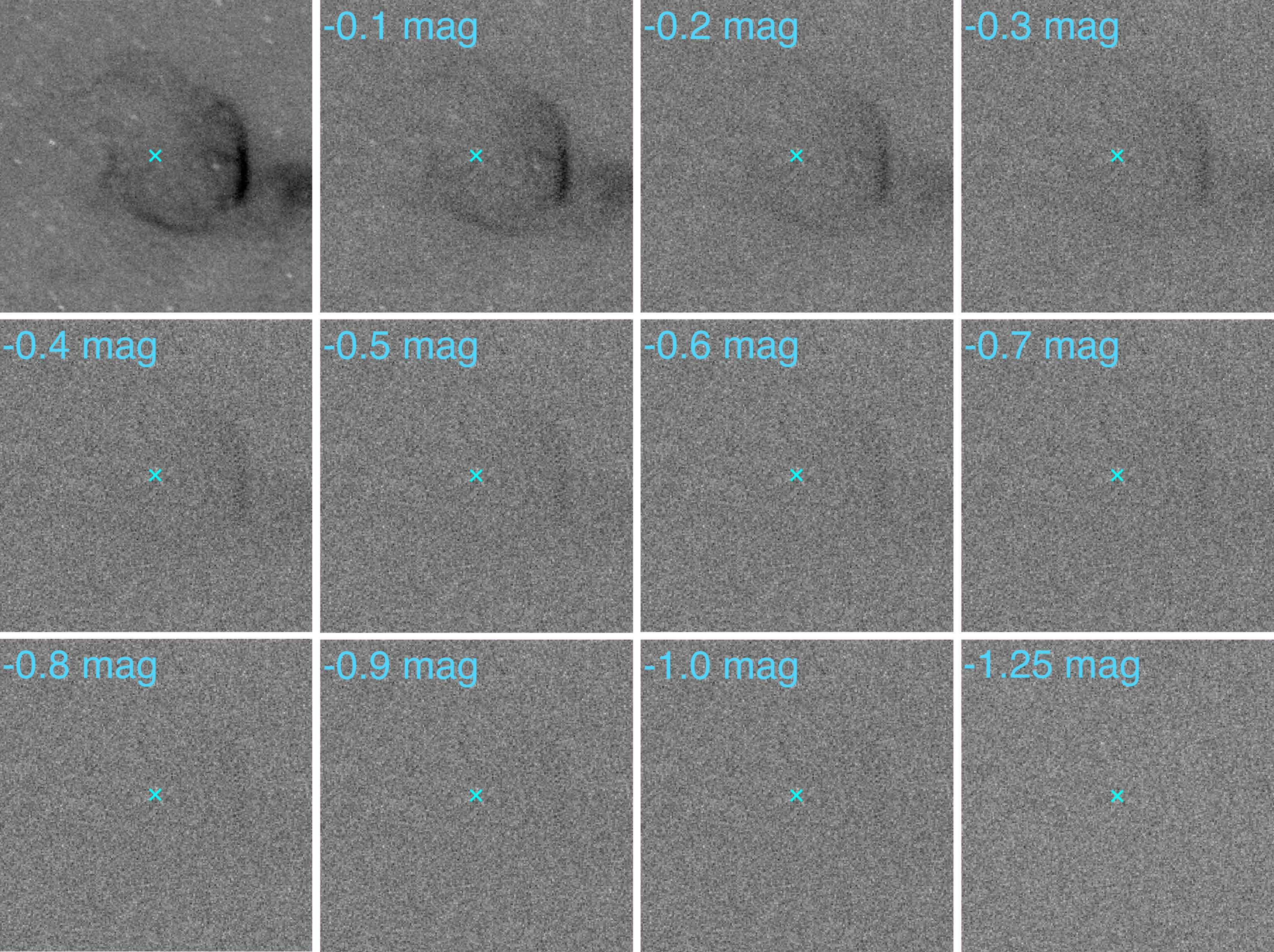}
\caption{Adding artificial noise to the region where the NSR surrounding 12a is located. Left: Each panel shows the $1^{\prime} \times 1^{\prime}$ region surrounding 12a with the level of noise added. The top left panel is the continuum-subtracted LGGS of the 12a NSR (no artificial noise added).}
\label{fig:12a add noise}
\end{figure}

The outcome of this test can be seen in the left panels of Figure~\ref{fig:12a add noise}. In the top left panel we simply show the $1^{\prime} \times 1^{\prime}$ region surrounding 12a (as in Figure~\ref{2008-12a surrounding sub images}). The other eleven panels show the same region but with increasing levels of added Poisson noise to represent worsening signal-to-noise alongside the brightening of the background.

The bottom right panel of Figure~\ref{fig:12a add noise} shows the surroundings of 12a with 1.25 magnitudes of noise added to the image (analogous to the remnant being 1.25 mag fainter). This reflects the difference in luminosity between the 12a NSR and the next brightest predicted NSR associated with M\,31N 2017-01e (see Table~\ref{tab:M31,LMC-RNe}). As can be seen, the 12a NSR is difficult to detect once the background has brightened by approximately one magnitude: this is equivalent to the remnant surrounding 12a not being detected if it were one magnitude fainter. Therefore, the brightest NSR in M\,31 (around 12a) is only on the cusp of visibility. Furthermore, the 2017-01e NSR (and other NSRs in M\,31) should not be detectable in LGGS data.

\section{Discussion}\label{sec:Discussion}
We applied our method of annular photometry to nineteen of the twenty-four known RNe in M\,31 and the LMC (five M\,31 RNe are located too close the bulge of that galaxy for suitable analysis). From the surroundings of these nineteen novae, we only recovered the known nova super-remnant associated with M\,31N 2008-12a. The environs of the other RNe displayed a distance lack of extended emission across the narrowband filters we investigated, even when co-adding the images.

One possible explanation of this NSR dearth could be related to the evolutionary phase of each system. As shown in Section~\ref{sec:Predicted Ha luminosity}, NSRs will be brightest when the responsible nova is at an extreme stage of its evolution. In other words, systems with the most massive (carbon-oxygen) WDs and highest accretion rates (as is the case with 12a), and therefore the shortest recurrence periods, are likely to host the brightest NSRs. Yet, from simulations \citep{2023MNRAS.521.3004H}, even the RN with the next shortest recurrence period (${\sim}$2.5 years for M\,31N 2017-01e) is predicted to have approximately one-quarter of the H$\alpha$ flux as the equivalent simulation with a one-year recurrence period (see the cyan line in Figure~\ref{fig:Halpha Evolution}). Therefore, the NSRs associated with all of the other RNe in M\,31 and the LMC will simply be intrinsically fainter and thus more difficult to detect (as we explored in Section~\ref{sec:Add noise 12a NSR}).

Another potential reason behind the absence of NSRs could be the local ISM density. \citet{2023MNRAS.521.3004H} illustrated that the ISM density has a significant influence on the size of the evolving NSR. As such, a very low ISM density environment would permit the growth of a substantially large remnant. As with a longer recurrence period, this would result in an intrinsically fainter NSR.

An additional possibility for the lack of other NSRs in M\,31 and the LMC may relate to the initial mass of the WD. It was demonstrated in \citet{2023MNRAS.521.3004H} that the initial mass of the WD has a lesser impact on NSR growth in comparison to other system parameters (e.g., accretion rate). However, this small influence applies to WDs with initial masses that sit in the CO regime, whereby early in their evolution, the WDs are not massive enough to power rapidly recurring novae \citep[and so {\it are} able to grow a large dynamical NSR over many eruptions;][]{2023MNRAS.521.3004H}. On the other hand, if the RN originates from a WD initially formed as an ONe WD, as may be the case for a large number of RNe we have considered in this work, then it will have a short recurrence period from birth and consequently will not have been able to grow a large dynamical remnant. This smaller NSR size will thus hamper efforts of detection.

Lastly, we are considering that NSRs exist around all other RNe and so are proposing the reasoning behind not detecting these structures. But we might not be able to observe a 12a-like NSR around the other systems in M\,31 and the LMC because these other novae do not host a NSR. While the simplest explanation for the lack of other remnants, it does not provide a satisfactory narrative in light of the NSRs around 12a and KT Eri being so prominent. Specifically, \citet{2019Natur.565..460D} used hydrodynamical simulations to show that a structure reminiscent of the 12a NSR could be grown from frequent nova eruptions; this prediction was reinforced for 12a by \citet{2023MNRAS.521.3004H}, shown to apply to all RNe systems (whether the WD was growing or shrinking) and recently demonstrated for KT Eri \citep{2023arXiv231017258H}. Simulations correctly predicting the existence of the 12a NSR then not applying to other RNe is dubious -- this lends support to the NSRs in M\,31 and the LMC being too faint (or too small) to currently detect.

\section{Conclusions}\label{sec:Conclusions}
In this paper, we have conducted the first ever survey for nova super-remnants within M\,31 using the imaging data collected as part of the Local Group Galaxies Survey. Additionally, we present our targeted search for nova super-remnants within the LMC using data collected with the Faulkes Telescope South. Here, we summarise the key results of the paper:

\begin{enumerate}
\item The prototypical nova super-remnant, surrounding the rapidly recurring nova M\,31N 2008-12a, is recovered in the LGGS narrowband H$\alpha$ and [$\ion{S}{ii}$] images and absent in the [$\ion{O}{iii}$], as expected.
\item There is no H$\alpha$ or [$\ion{S}{ii}$] emission apparent around the other fourteen recurrent novae in M\,31 which we are able quantify with annular photometry (five novae were omitted from our analysis due to close proximity with the bulge of M\,31).
\item There is no H$\alpha$ emission apparent around the four recurrent novae in the LMC which we are able quantify with annular photometry.
\item The 12a NSR would likely to have been missed if it were approximately one magnitude fainter.
\item The distinct lack of a NSR around the RNe in M\,31 and the LMC could possibly be explained by longer recurrence periods or very low density ISM contributing to their intrinsic faintness.
\item The dearth of NSRs may also be resolved through considering that RNe with ONe WDs have a short recurrence period at birth thus will not host a large NSR, complicating their detection.
\item Alternatively, NSRs around the other RNe we considered may not exist and the 12a NSR is an extremely rare phenomenon, however this seems to be contrived based on the recent discovery of a NSR surrounding KT Eri.
\end{enumerate}

Regardless of the underlying reason, we were not able to recover any other NSR in the vicinity of the majority of RNe in M\,31 and the RNe in the LMC. Moving forward, deeper narrowband observations should be focussed on the surroundings of the novae with the shortest recurrence periods.

\section*{Acknowledgements}
The authors would like to thank our anonymous referee for their time spent reviewing our manuscript. MWH-K acknowledges a PDRA position funded by the UK Science and Technology Facilities Council (STFC). MWH-K and MJD receive funding from STFC grant number ST/S505559/1. MMS acknowledges the support of NSF award 2108234. This work was only possible with data from the Local Group Galaxy Survey. We would also like to thank Emily Manne-Nicholas for their generous help in coordinating our FTS observations for the LMC RNe data. This work has made use of data from the European Space Agency (ESA) mission {\it Gaia} (\url{https://www.cosmos.esa.int/gaia}), processed by the {\it Gaia} Data Processing and Analysis Consortium (DPAC, \url{https://www.cosmos.esa.int/web/gaia/dpac/consortium}). Funding for the DPAC has been provided by national institutions, in particular the institutions participating in the {\it Gaia} Multilateral Agreement. 

\section*{Data Availability}
All data/image analysis was carried out using IRAF \citep{1986SPIE..627..733T,1993ASPC...52..173T}, DAOPHOT \citep{1987PASP...99..191S}, {\tt astropy} \citep[v4.0;][]{2018AJ....156..123T} and {\tt photutils} \citep[v0.7.2;][]{larry_bradley_2019_3568287}. The analysis in this work made use of the Python libraries: Numpy \citep{harris2020array} and Matplotlib \citep{Hunter:2007}.

\bibliographystyle{mnras}
\bibliography{bibliography}

\appendix

\section{Surroundings of recurrent novae}
Here, we provide the Local Group Galaxies Survey \citep[LGGS;][]{2007AJ....134.2474M} narrowband imaging (H$\alpha$, [$\ion{S}{ii}$] and [$\ion{O}{iii}$]) of the $1^\prime\times1^\prime$ region surrounding nineteen of the twenty recurrent novae within M\,31 (the surroundings of M\,31N 1984-07a have been omitted due to its extreme proximity to the centre of M\,31). We also provide Faulkes Telescope South narrowband imaging (H$\alpha$ and [$\ion{O}{iii}$]) of the $1.5^\prime\times1.5^\prime$ region surrounding each of the four recurrent novae within the LMC.

\begin{figure*}
\centering
\subfloat{\includegraphics[width=.32\textwidth]{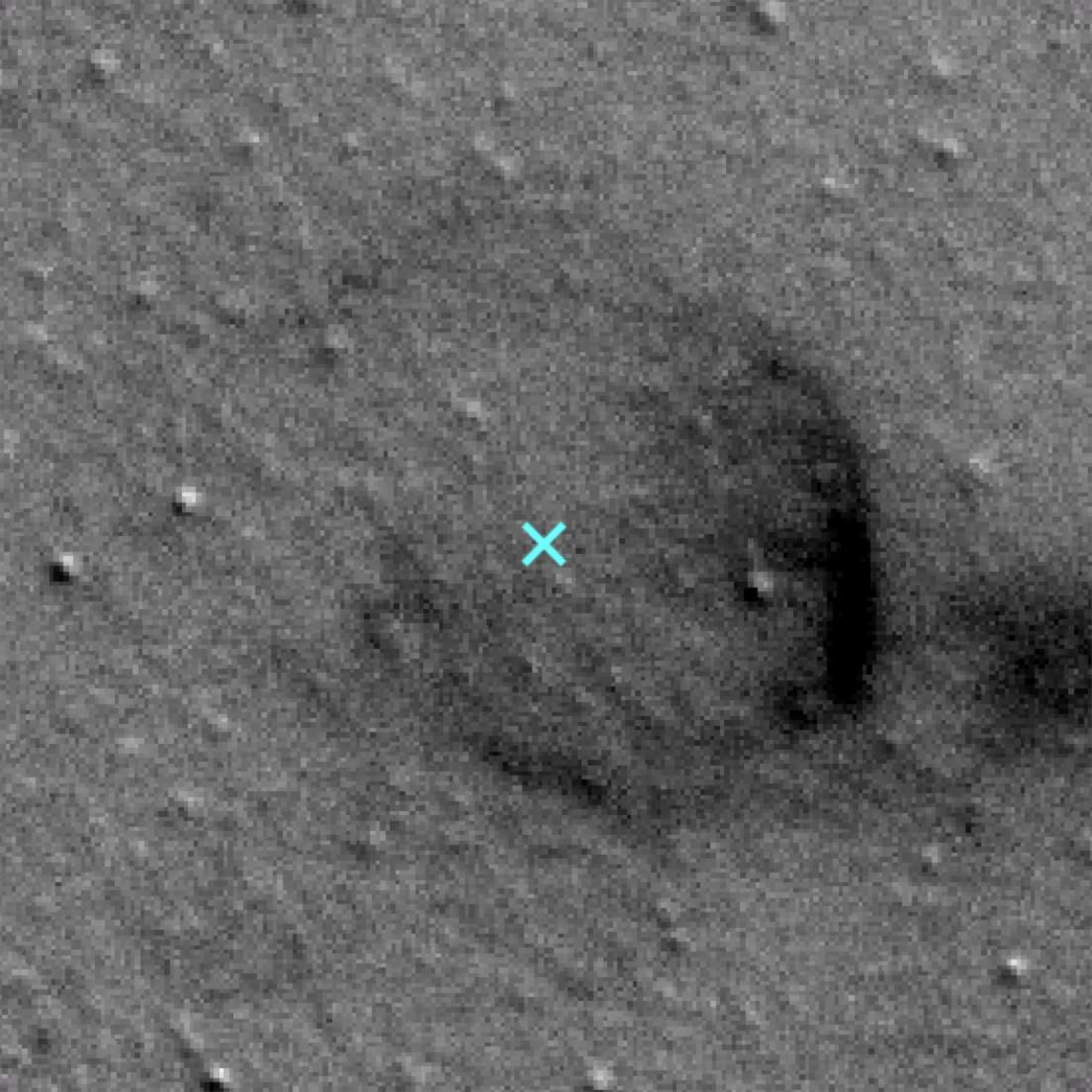}} \quad
\subfloat{\includegraphics[width=.32\textwidth]{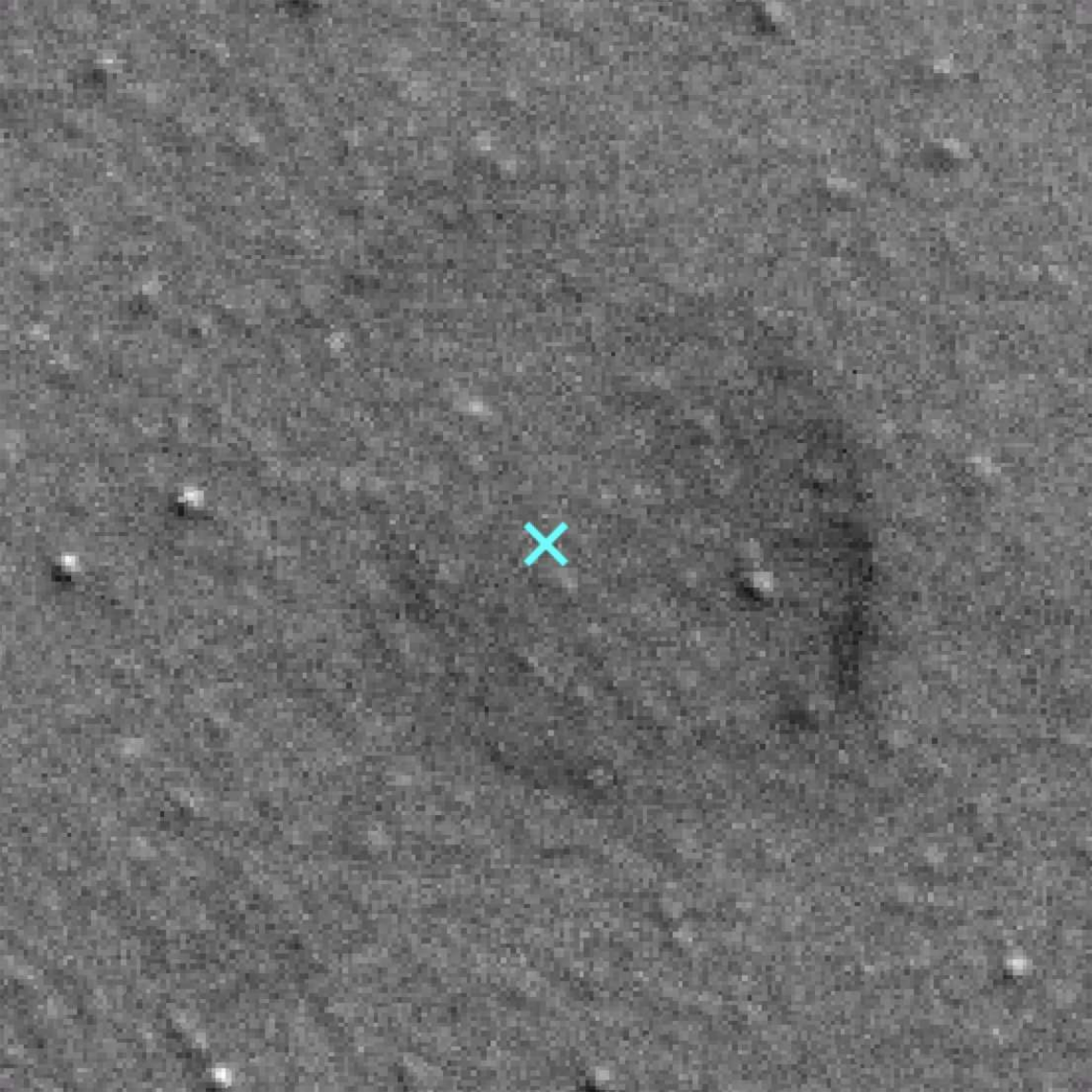}} \quad
\subfloat{\includegraphics[width=.32\textwidth]{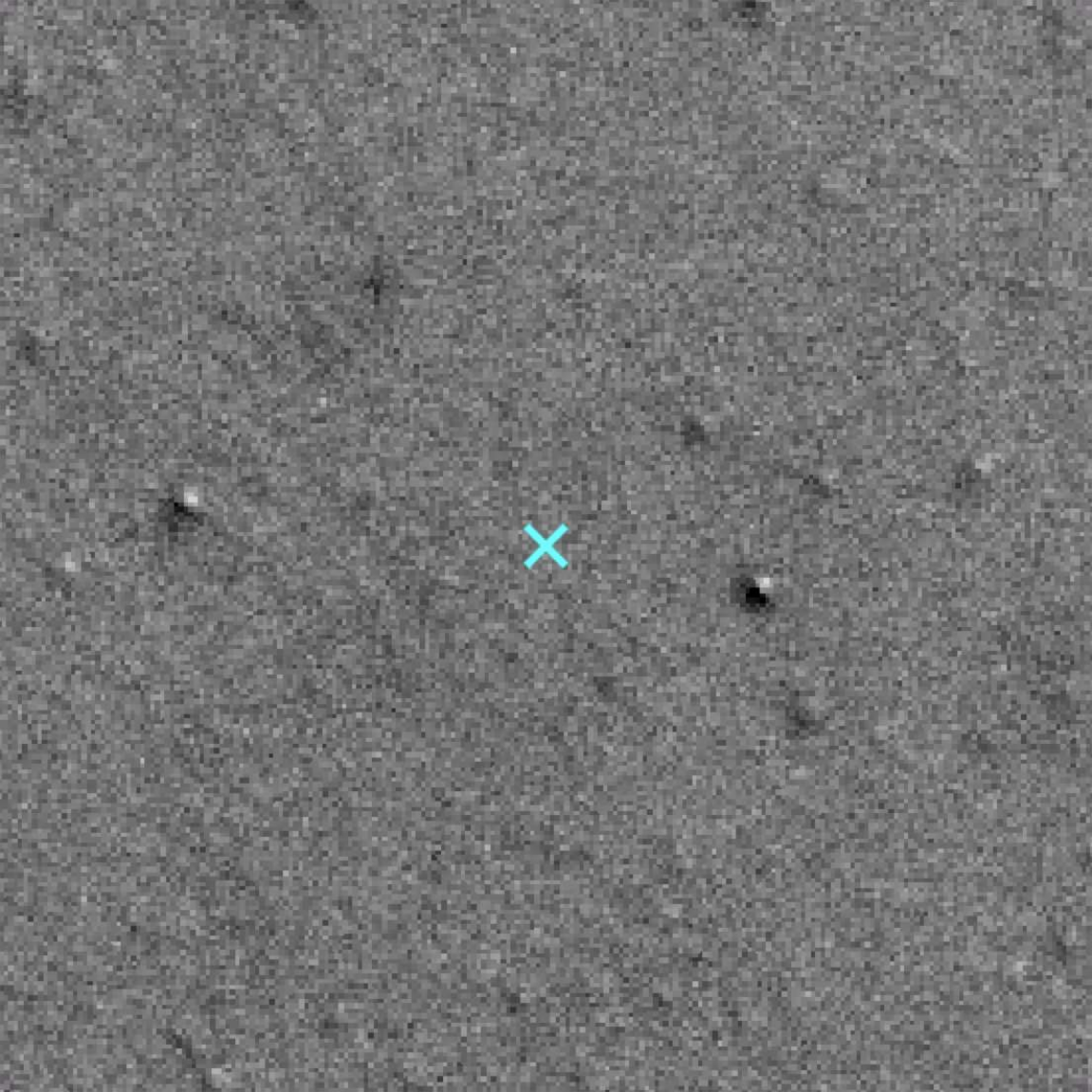}}
\caption{{\bf -- M\,31N 2008-12a}. The location of the nova (in field 3) is indicated by the blue cross. Left:\ Continuum subtracted LGGS H$\alpha$. Middle:\ Continuum subtracted LGGS [\ion{S}{ii}]. Right:\ Continuum subtracted LGGS [\ion{O}{iii}].}
\label{2008-12a surrounding sub images}
\end{figure*}

\begin{figure*}
\centering
\subfloat{\includegraphics[width=.32\textwidth]{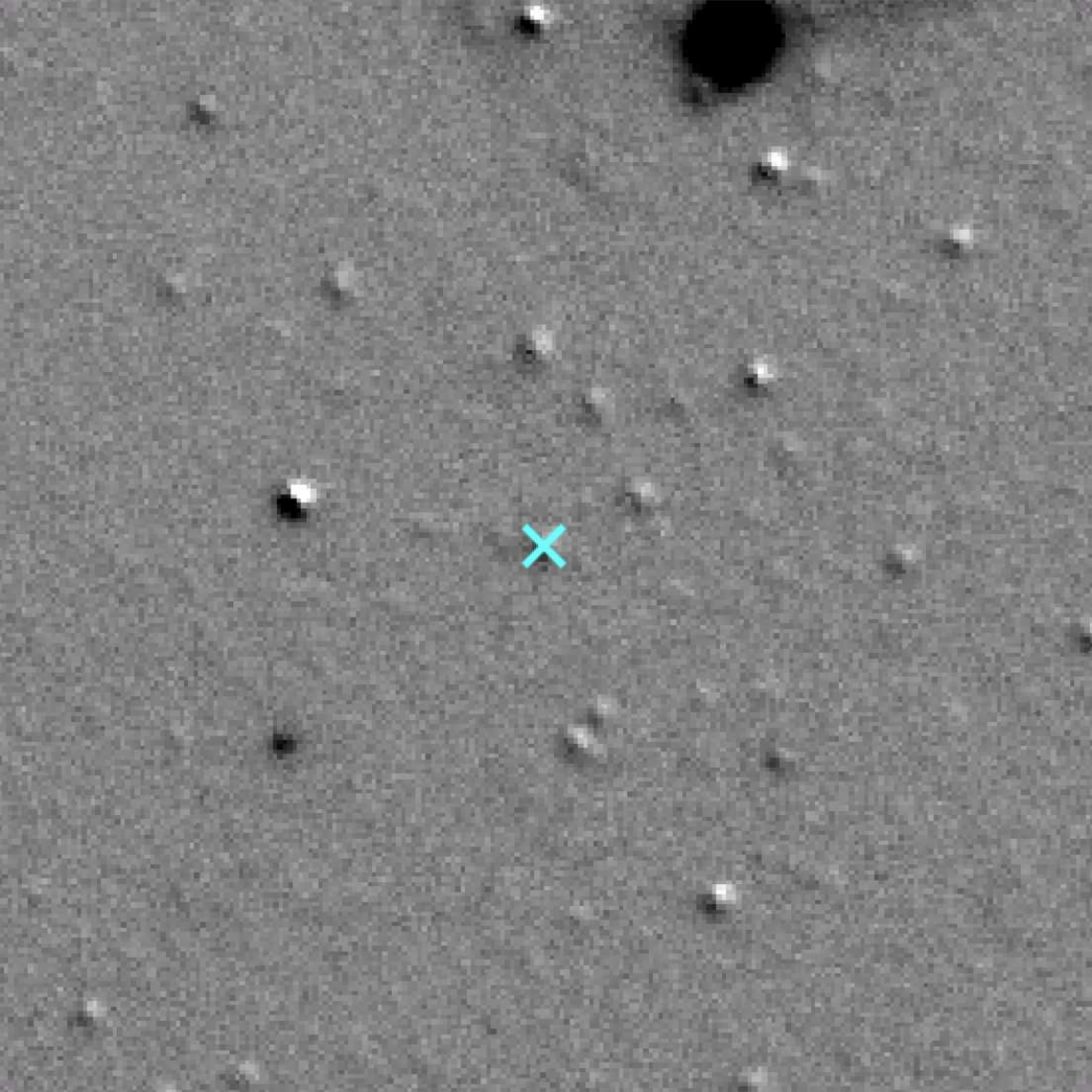}} \quad
\subfloat{\includegraphics[width=.32\textwidth]{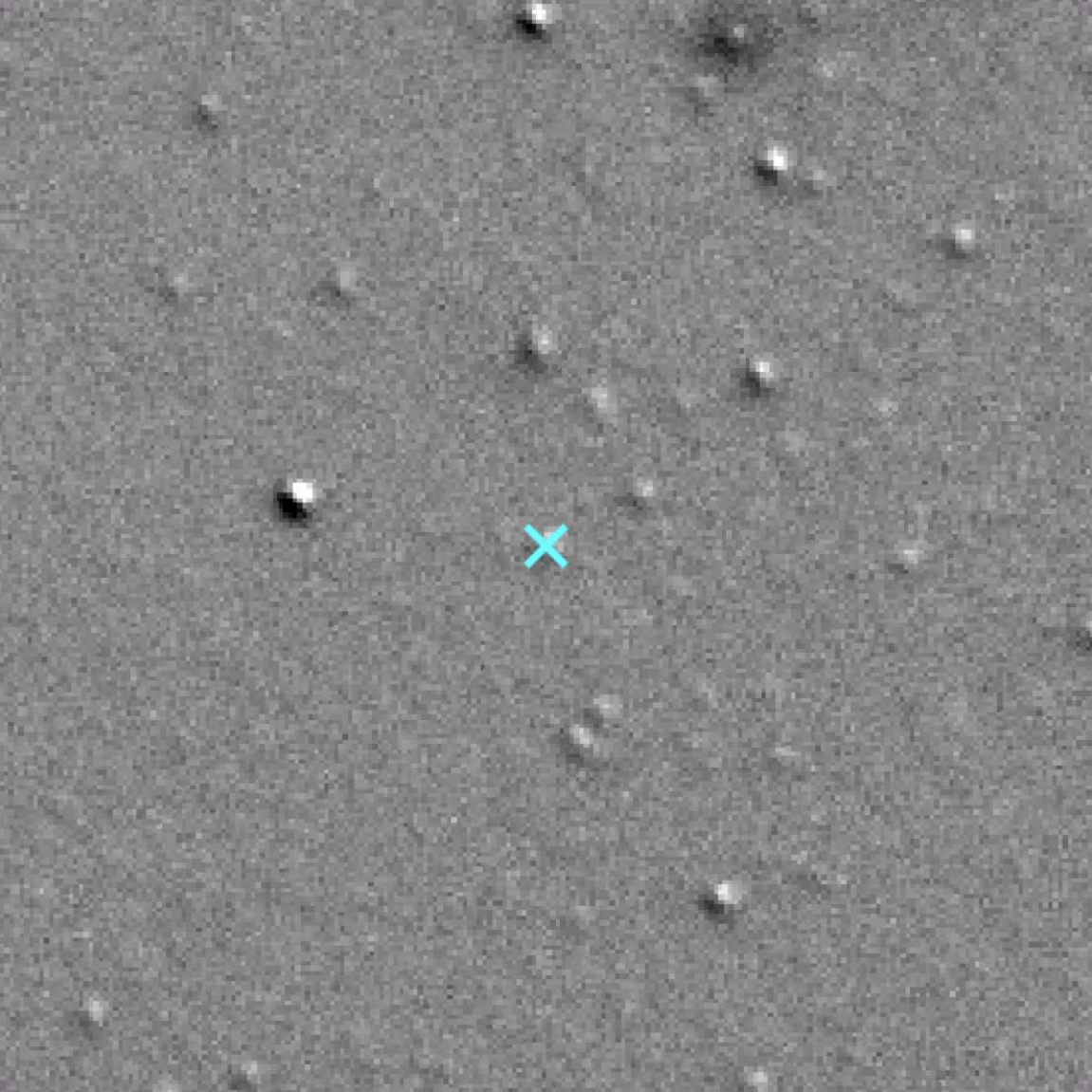}} \quad
\subfloat{\includegraphics[width=.32\textwidth]{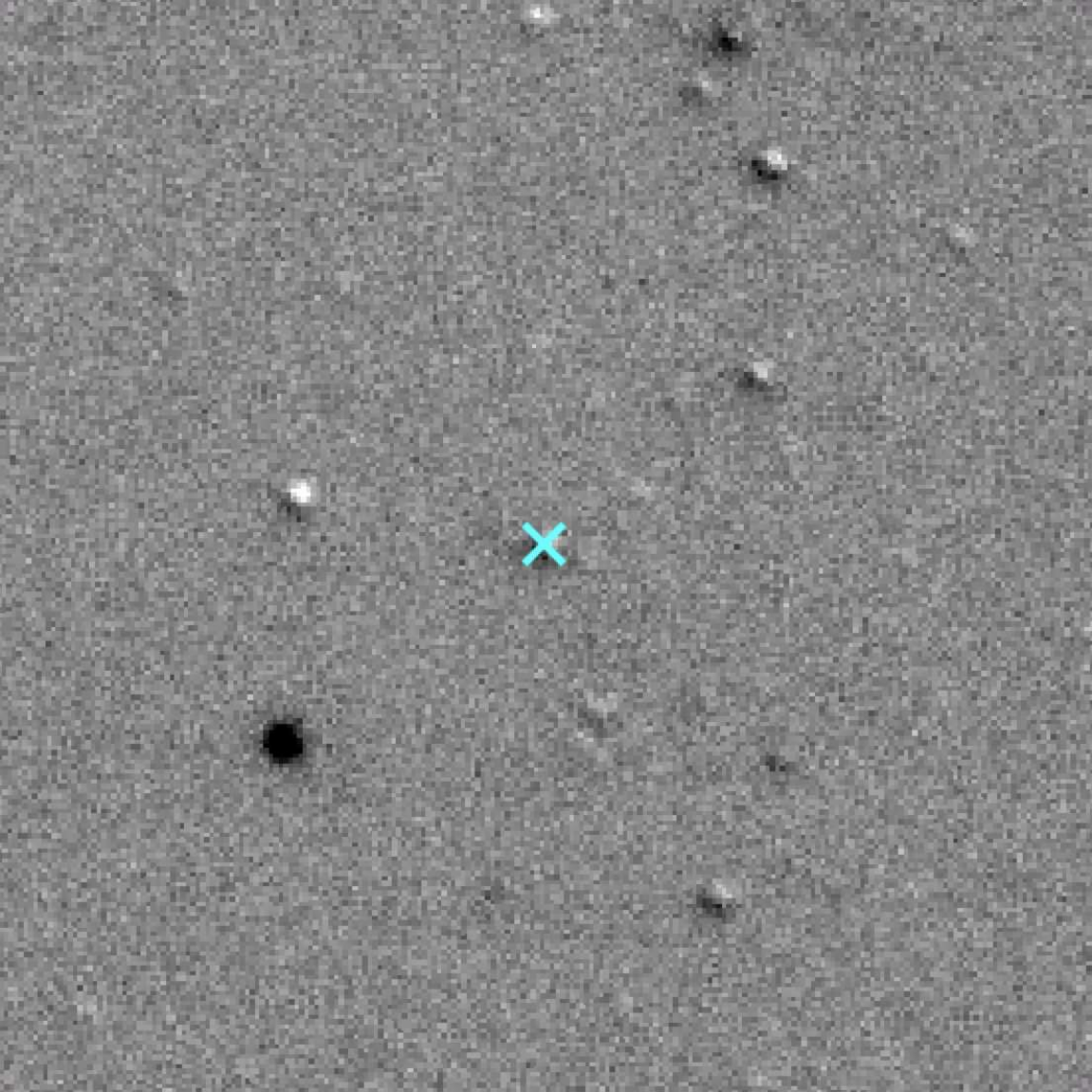}}
\caption{{\bf -- M\,31N 2017-01e}. The location of the nova (in field 3) is indicated by the blue cross. Left:\ Continuum subtracted LGGS H$\alpha$. Middle:\ Continuum subtracted LGGS [\ion{S}{ii}]. Right:\ Continuum subtracted LGGS [\ion{O}{iii}].}
\label{2017-01e surrounding sub images}
\end{figure*}

\begin{figure*}
\centering
\subfloat{\includegraphics[width=.32\textwidth]{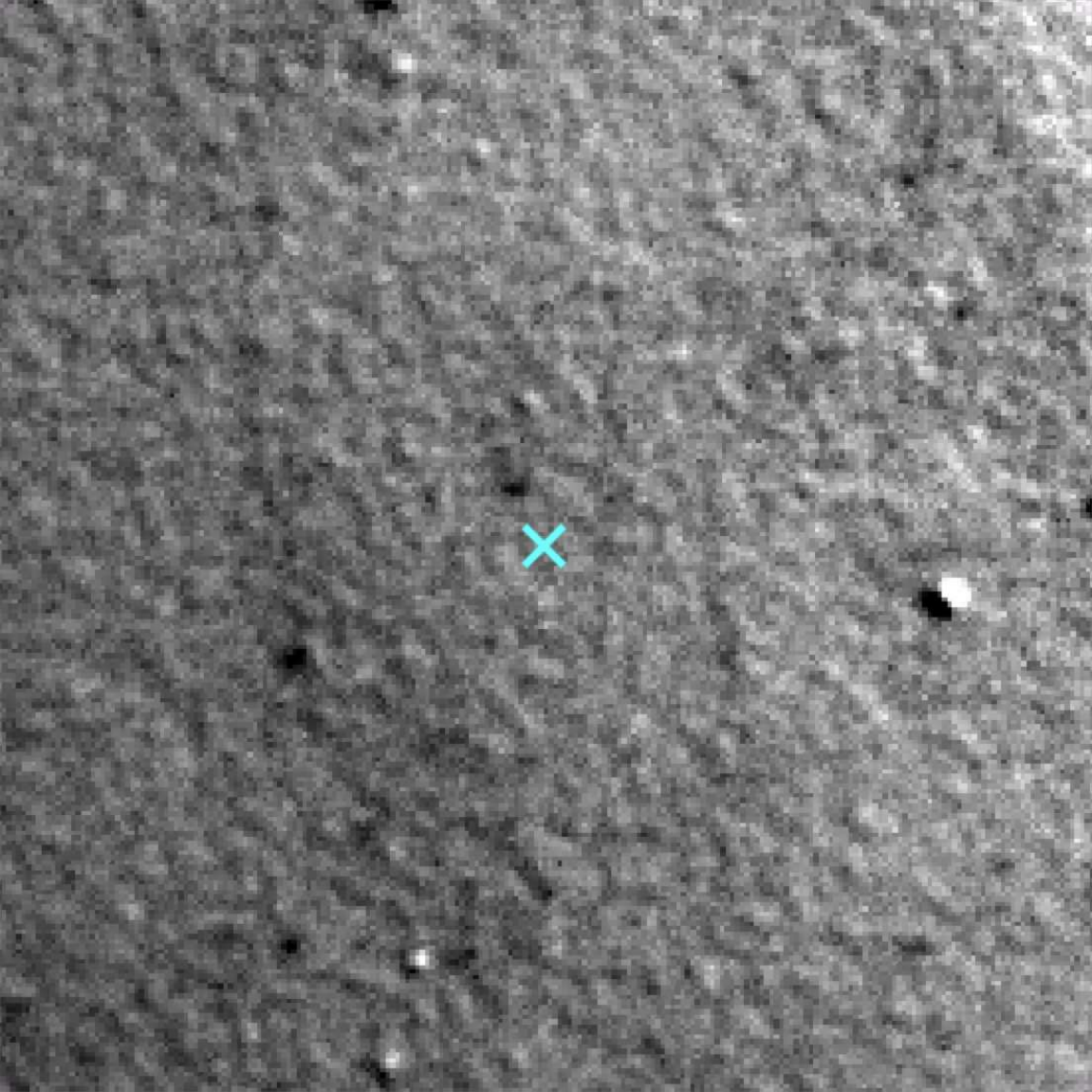}} \quad
\subfloat{\includegraphics[width=.32\textwidth]{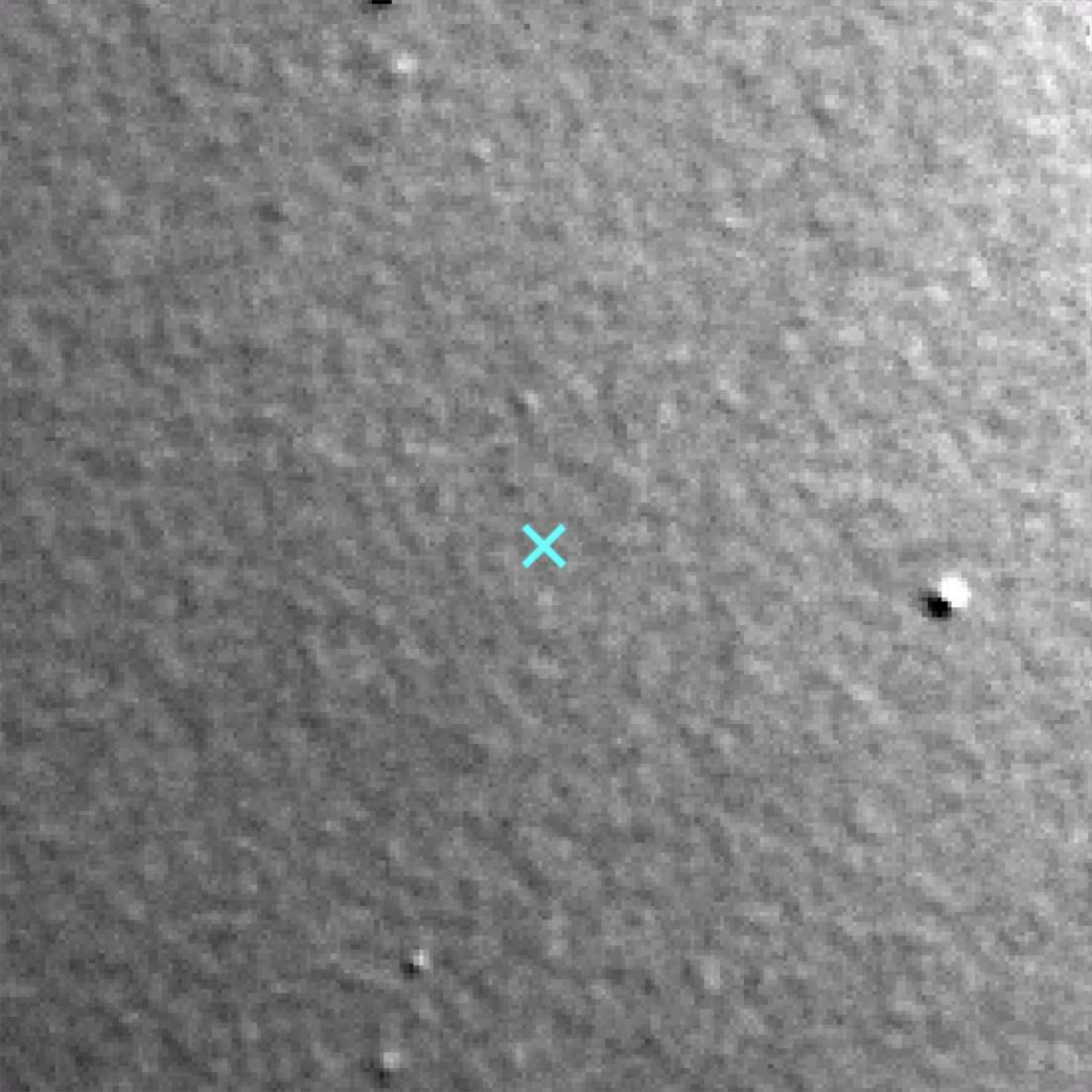}} \quad
\subfloat{\includegraphics[width=.32\textwidth]{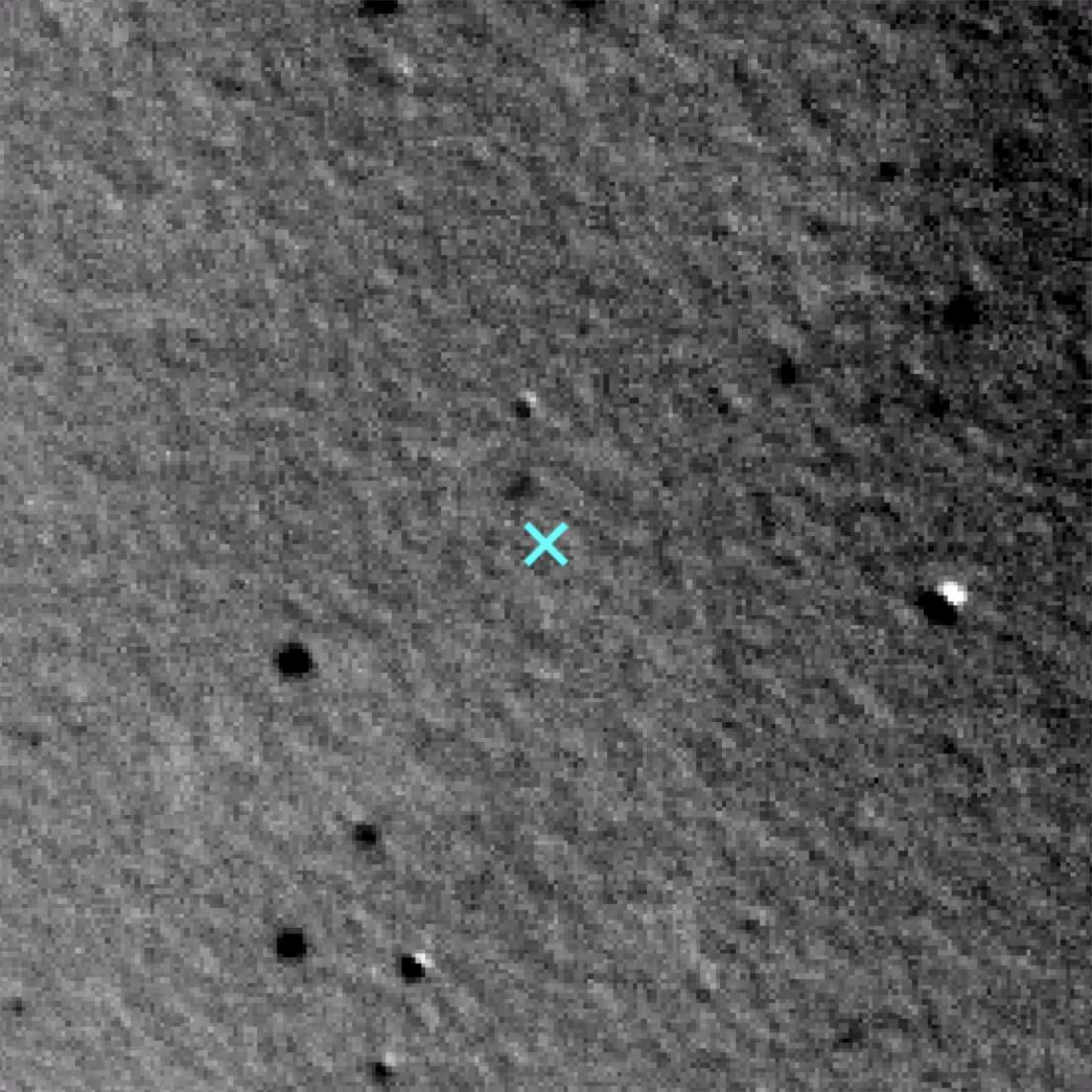}}
\caption{{\bf -- M\,31N 1926-07c}. The location of the nova (in field 6) is indicated by the blue cross. Left:\ Continuum subtracted LGGS H$\alpha$. Middle:\ Continuum subtracted LGGS [\ion{S}{ii}]. Right:\ Continuum subtracted LGGS [\ion{O}{iii}].}
\label{1926-07c surrounding sub images}
\end{figure*}

\begin{figure*}
\centering
\subfloat{\includegraphics[width=.32\textwidth]{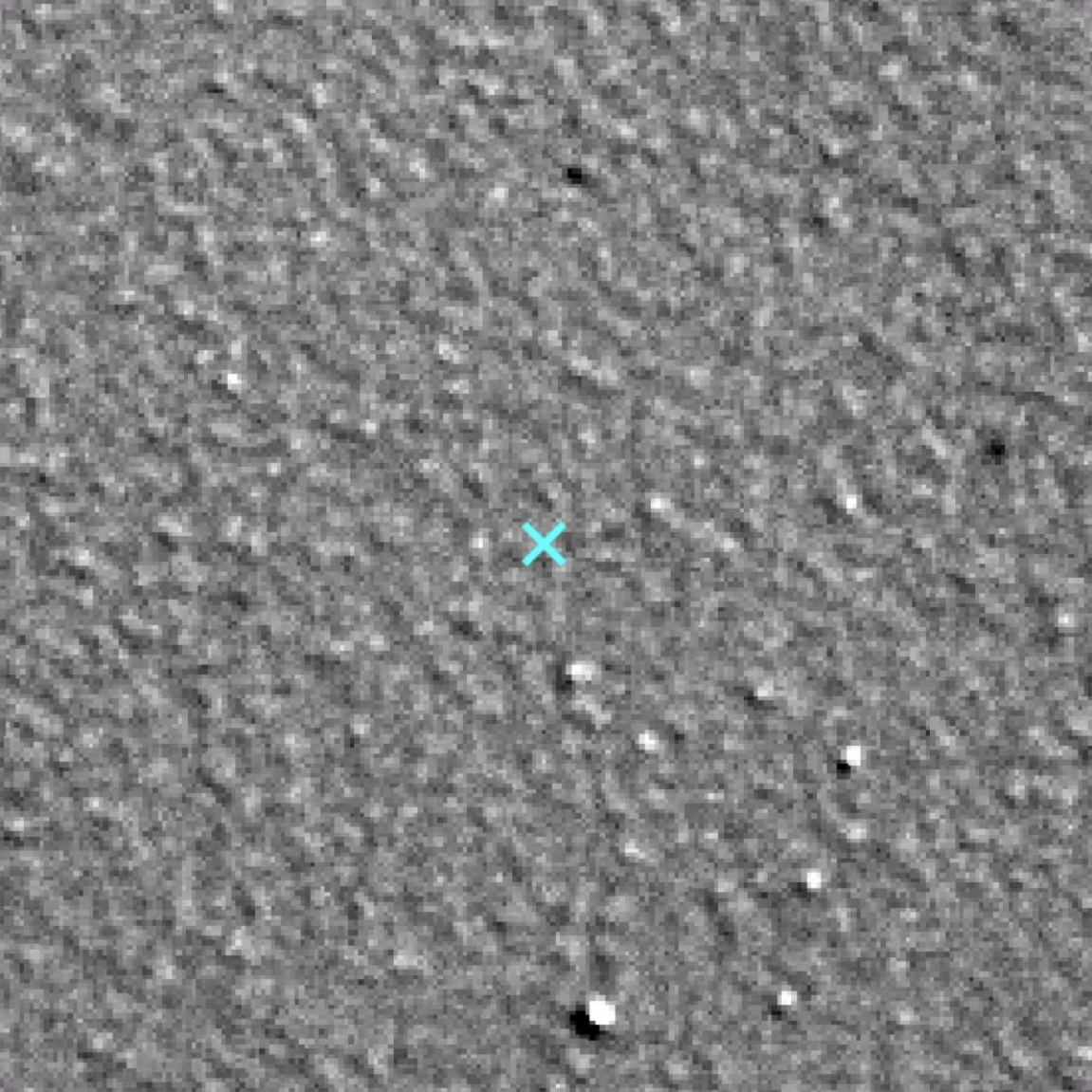}} \quad
\subfloat{\includegraphics[width=.32\textwidth]{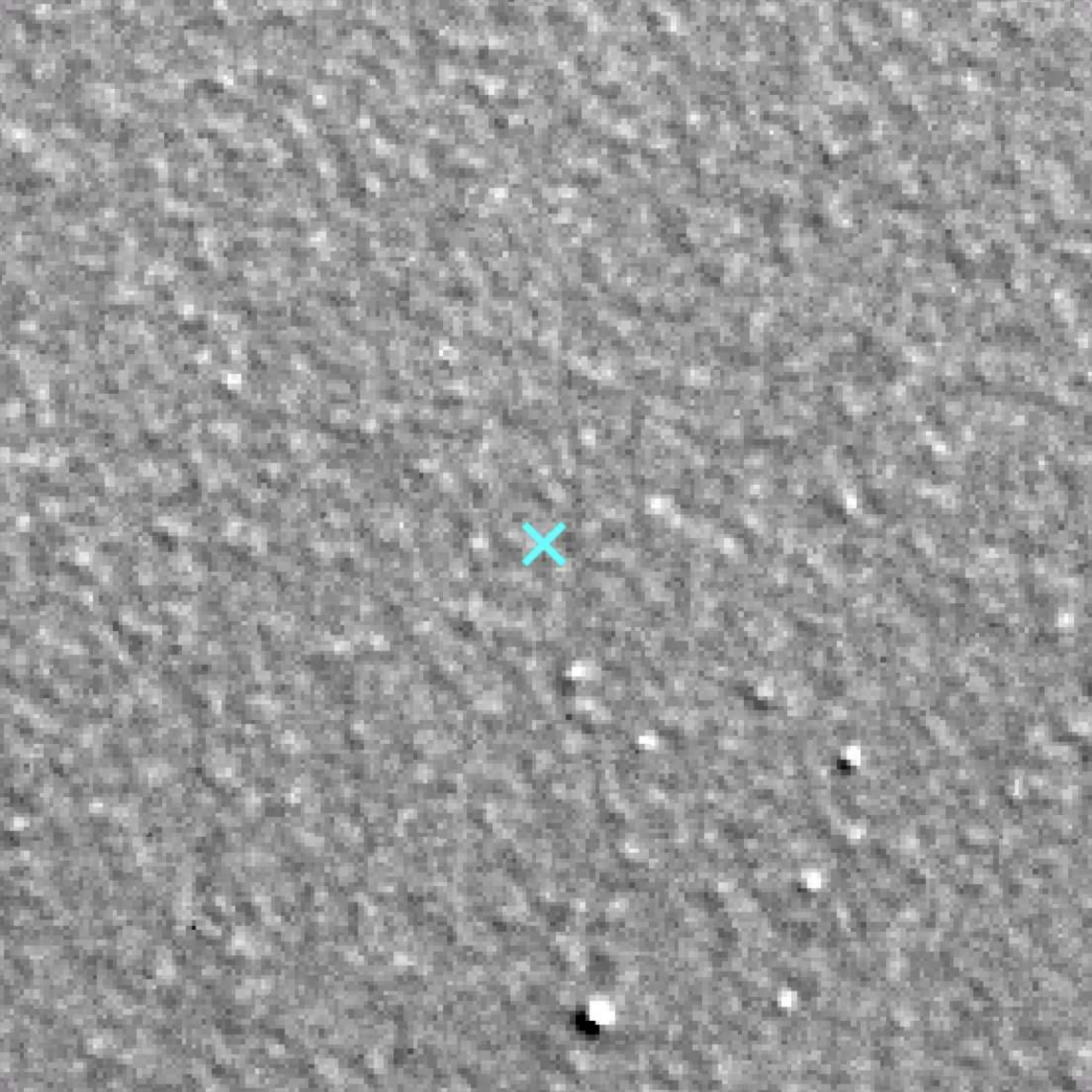}} \quad
\subfloat{\includegraphics[width=.32\textwidth]{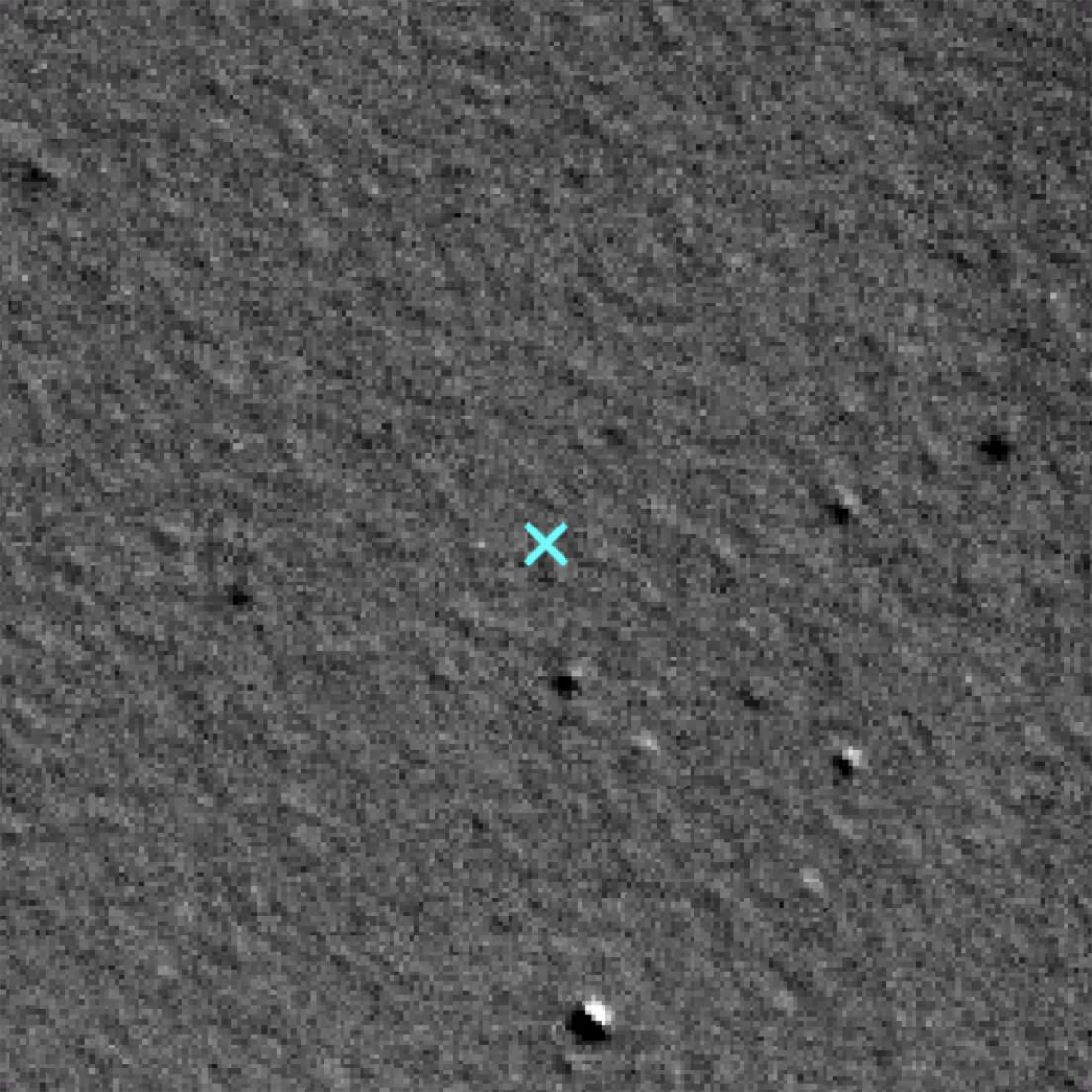}}
\caption{{\bf -- M\,31N 1997-11k}. The location of the nova (in field 6) is indicated by the blue cross. Left:\ Continuum subtracted LGGS H$\alpha$. Middle:\ Continuum subtracted LGGS [\ion{S}{ii}]. Right:\ Continuum subtracted LGGS [\ion{O}{iii}].}
\label{1997-11k surrounding sub images}
\end{figure*}

\begin{figure*}
\centering
\subfloat{\includegraphics[width=.32\textwidth]{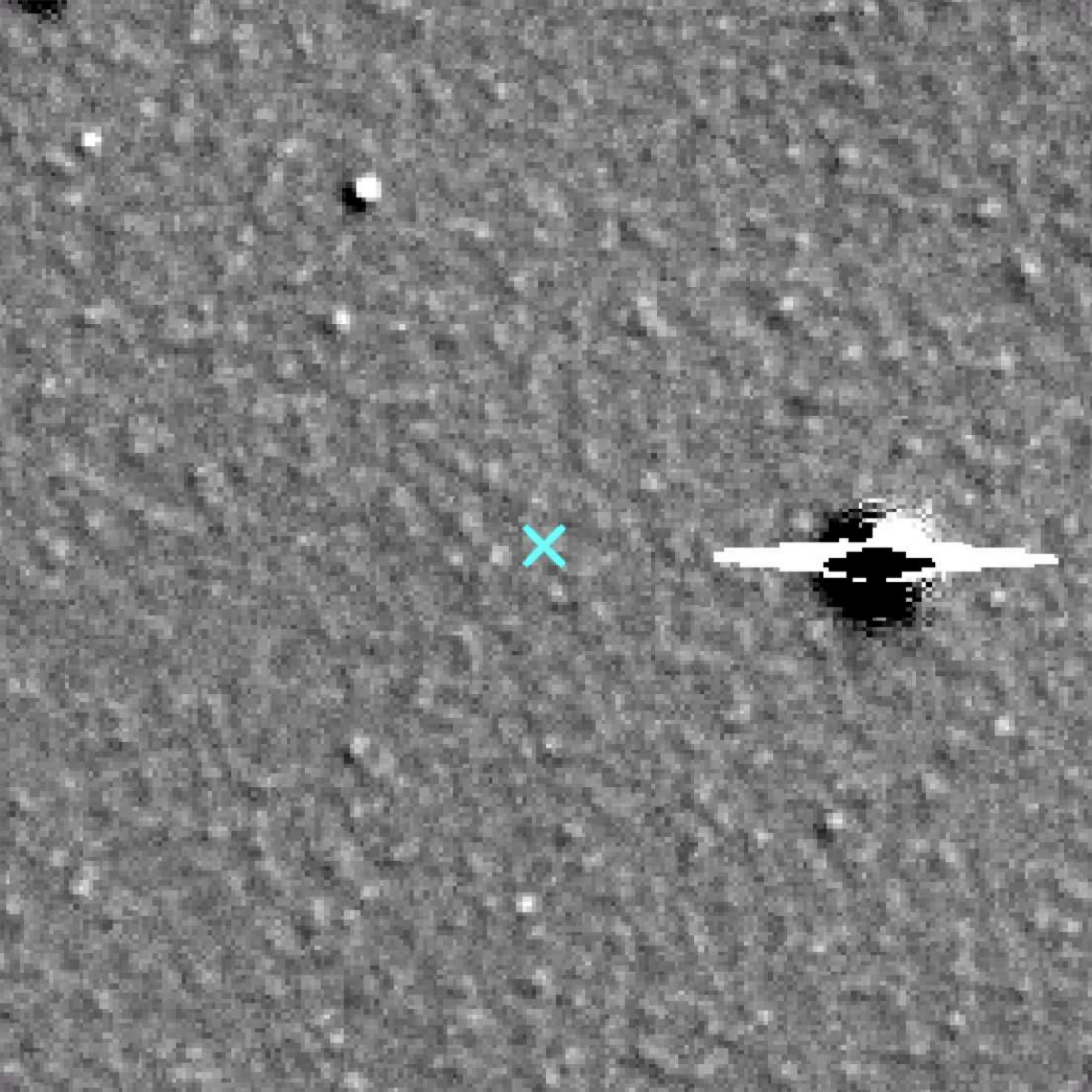}} \quad
\subfloat{\includegraphics[width=.32\textwidth]{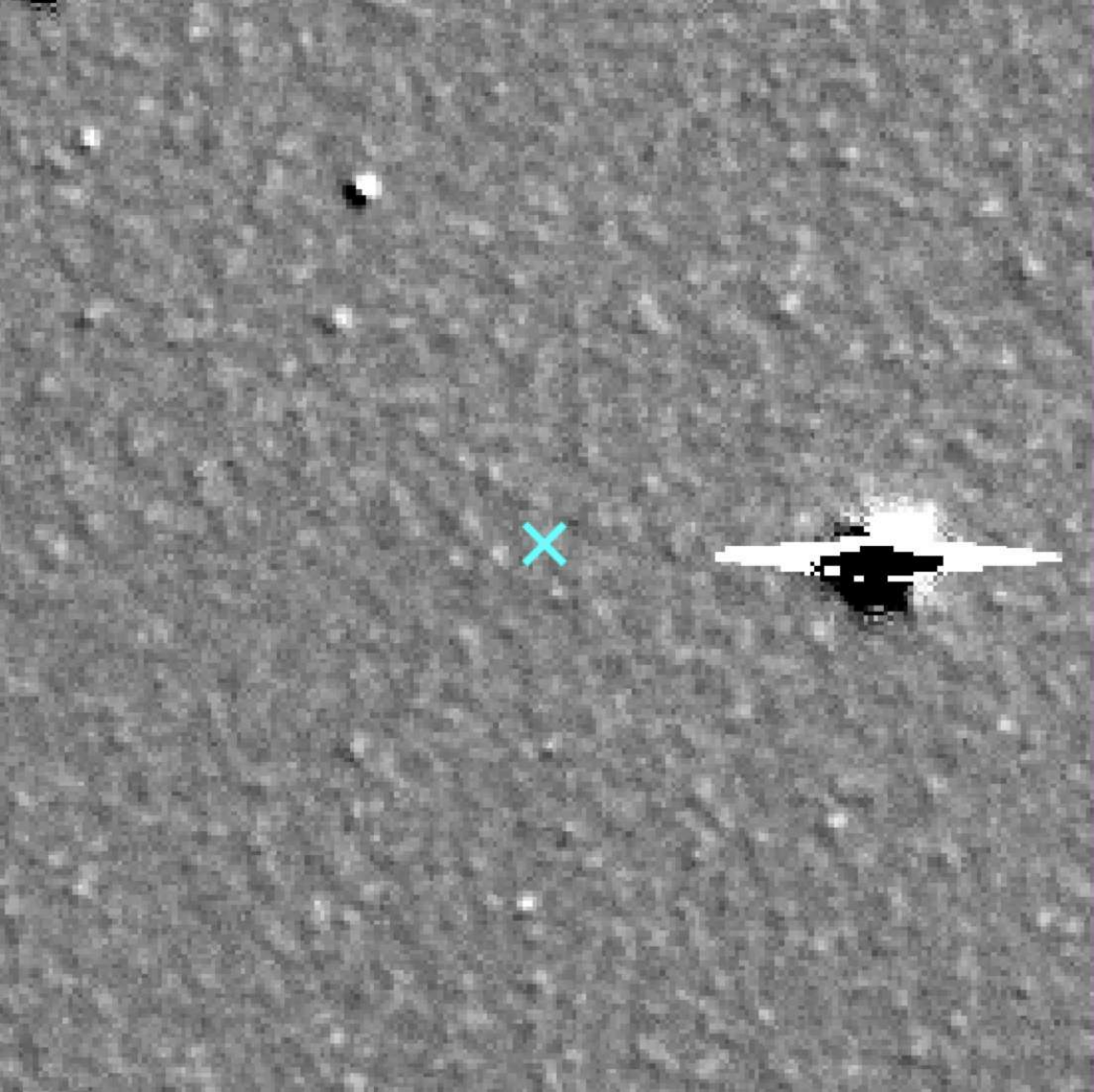}} \quad
\subfloat{\includegraphics[width=.32\textwidth]{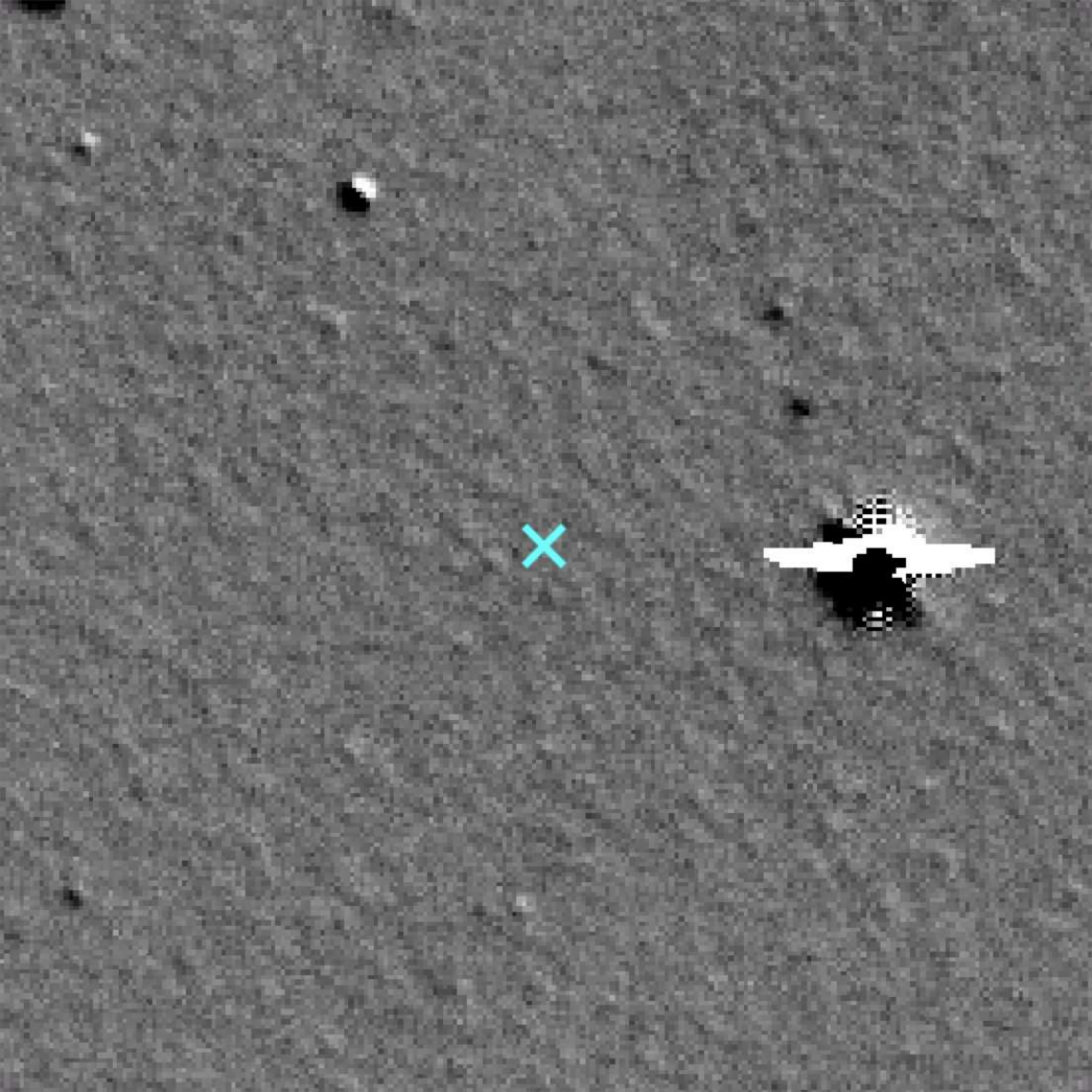}}
\caption{{\bf -- M\,31N 1963-09c}. The location of the nova (in field 6) is indicated by the blue cross. Left:\ Continuum subtracted LGGS H$\alpha$. Middle:\ Continuum subtracted LGGS [\ion{S}{ii}]. Right:\ Continuum subtracted LGGS [\ion{O}{iii}].}
\label{1963-09c surrounding sub images}
\end{figure*}

\begin{figure*}
\centering
\subfloat{\includegraphics[width=.32\textwidth]{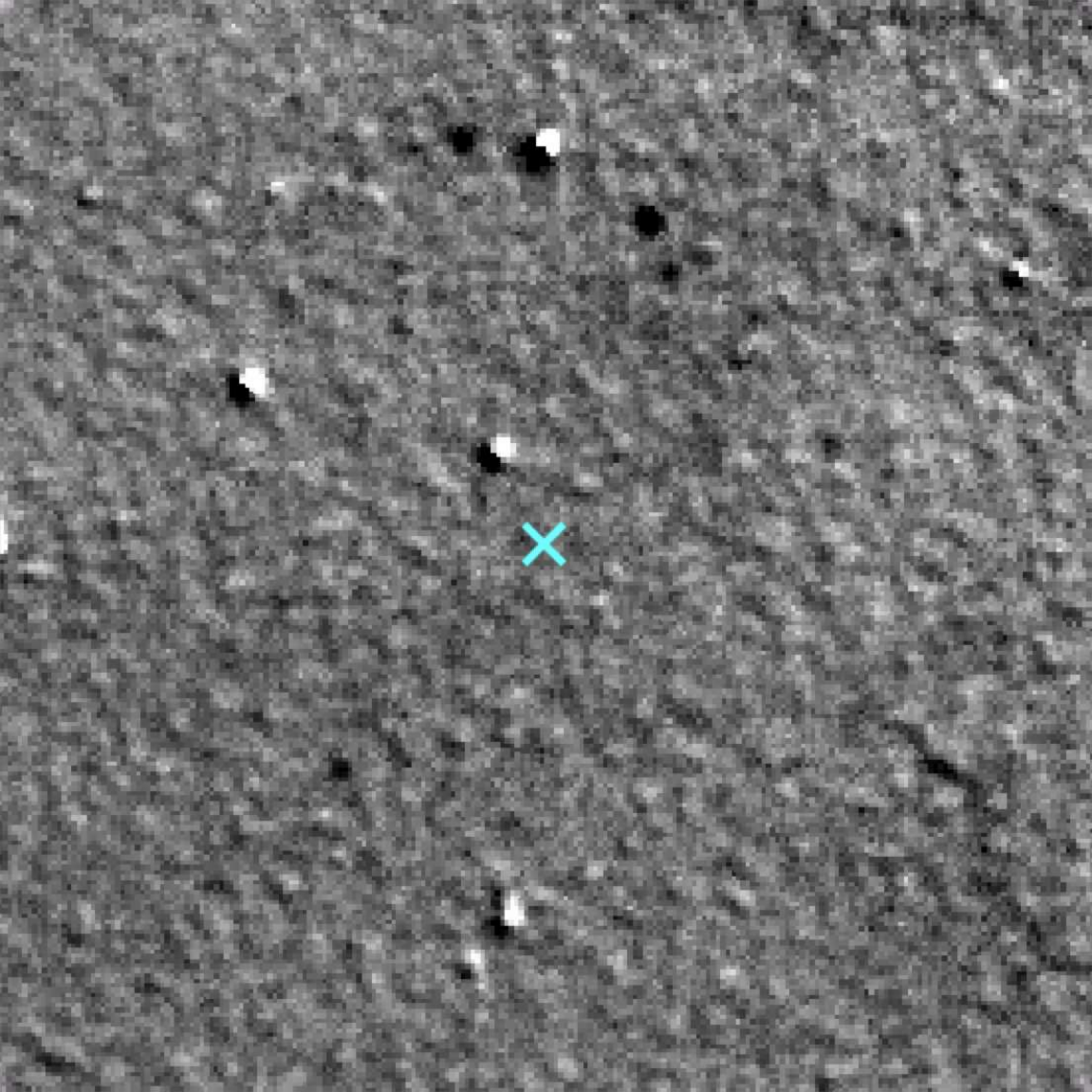}} \quad
\subfloat{\includegraphics[width=.32\textwidth]{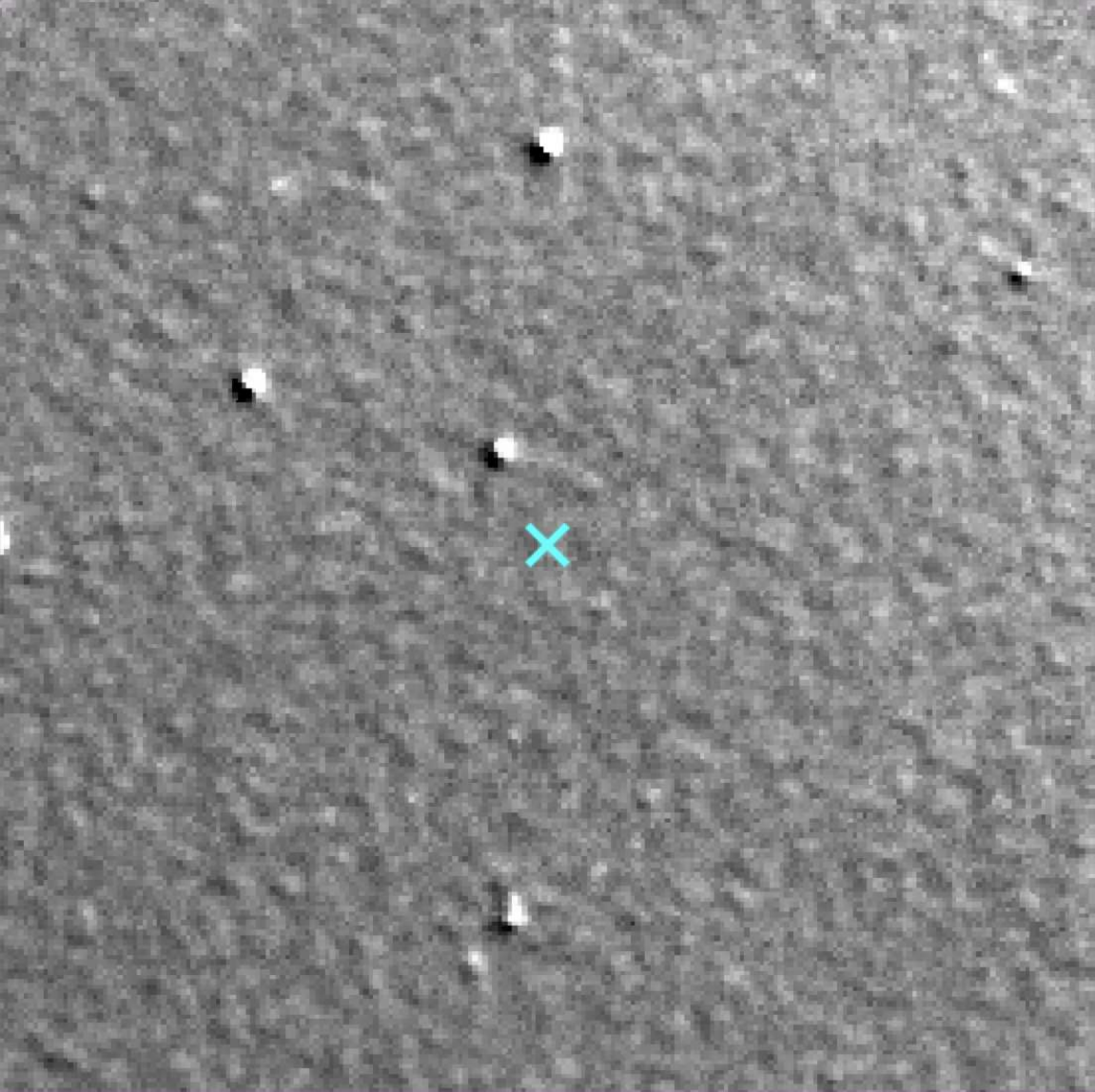}} \quad
\subfloat{\includegraphics[width=.32\textwidth]{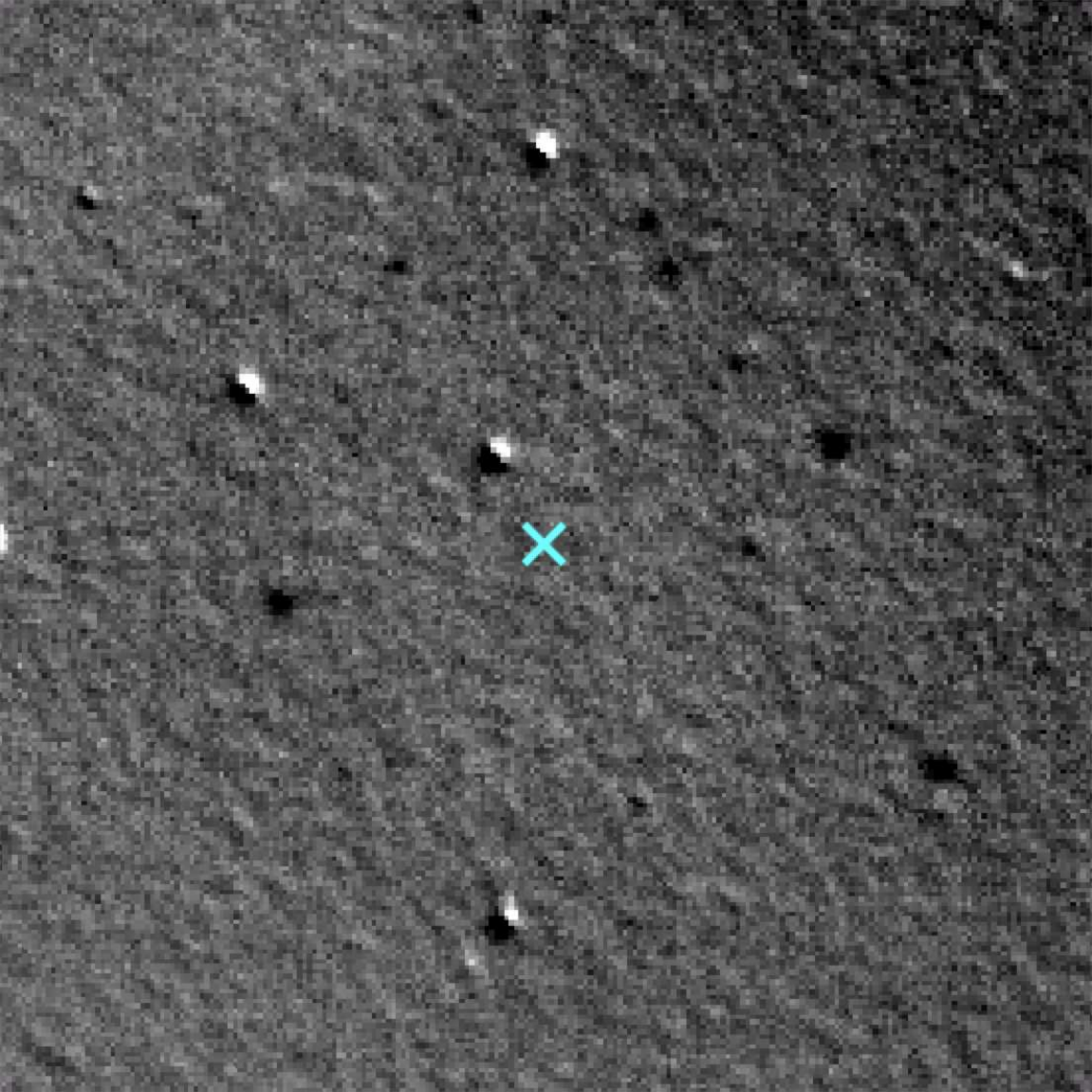}}
\caption{{\bf -- M\,31N 1960-12a}. The location of the nova (in field 6) is indicated by the blue cross. Left:\ Continuum subtracted LGGS H$\alpha$. Middle:\ Continuum subtracted LGGS [\ion{S}{ii}]. Right:\ Continuum subtracted LGGS [\ion{O}{iii}].}
\label{1960-12a surrounding sub images}
\end{figure*}

\begin{figure*}
\centering
\subfloat{\includegraphics[width=.32\textwidth]{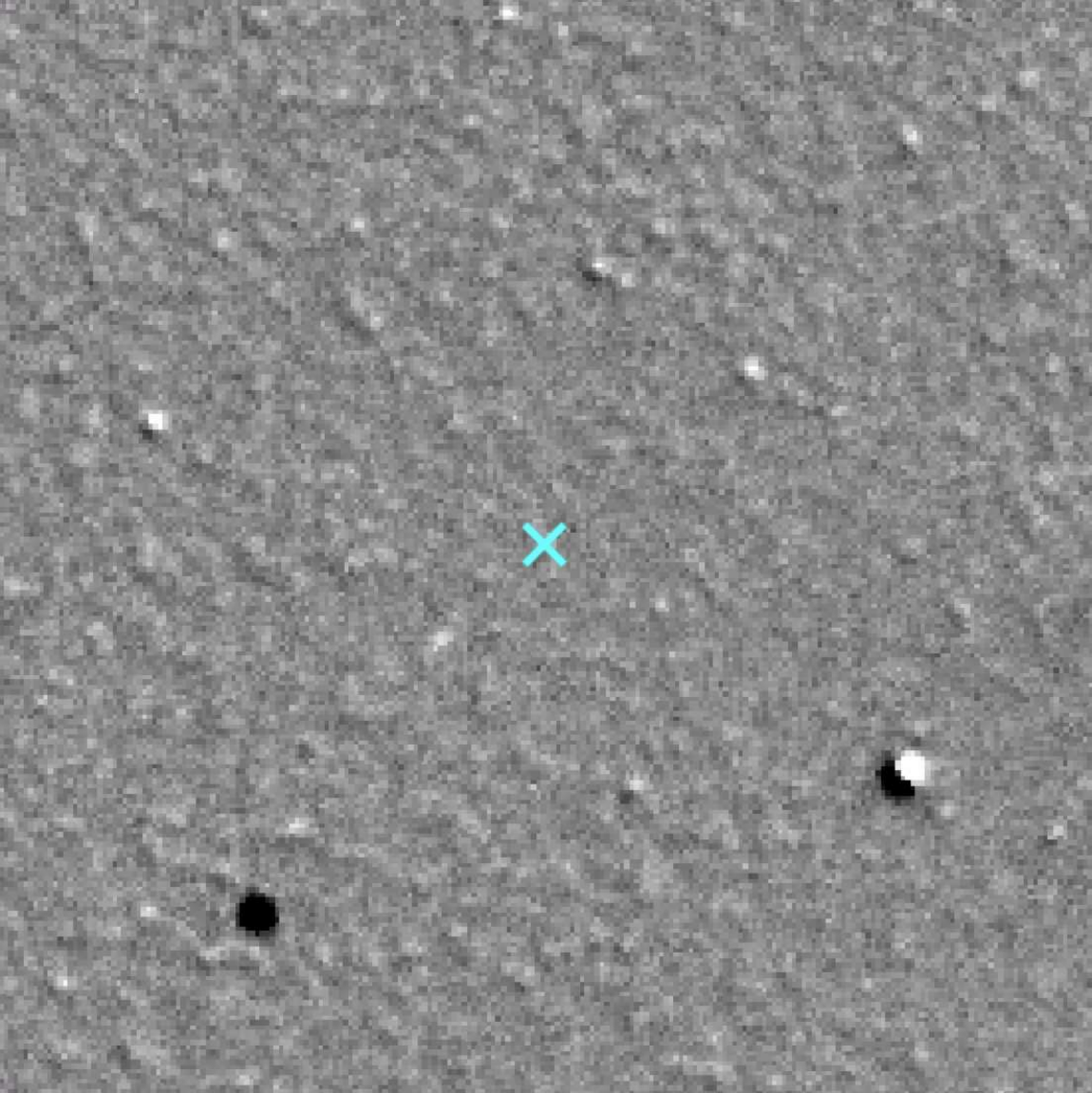}} \quad
\subfloat{\includegraphics[width=.32\textwidth]{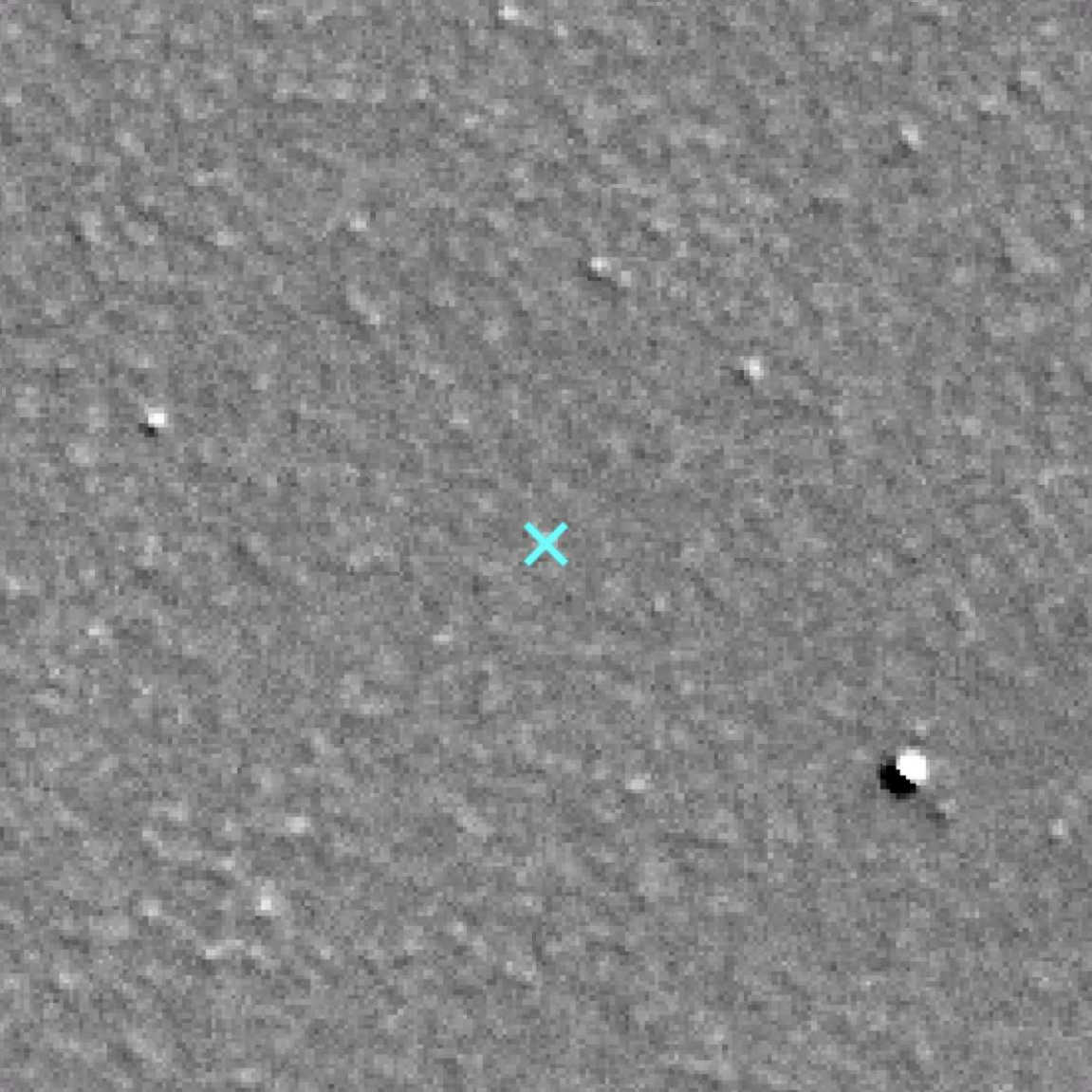}} \quad
\subfloat{\includegraphics[width=.32\textwidth]{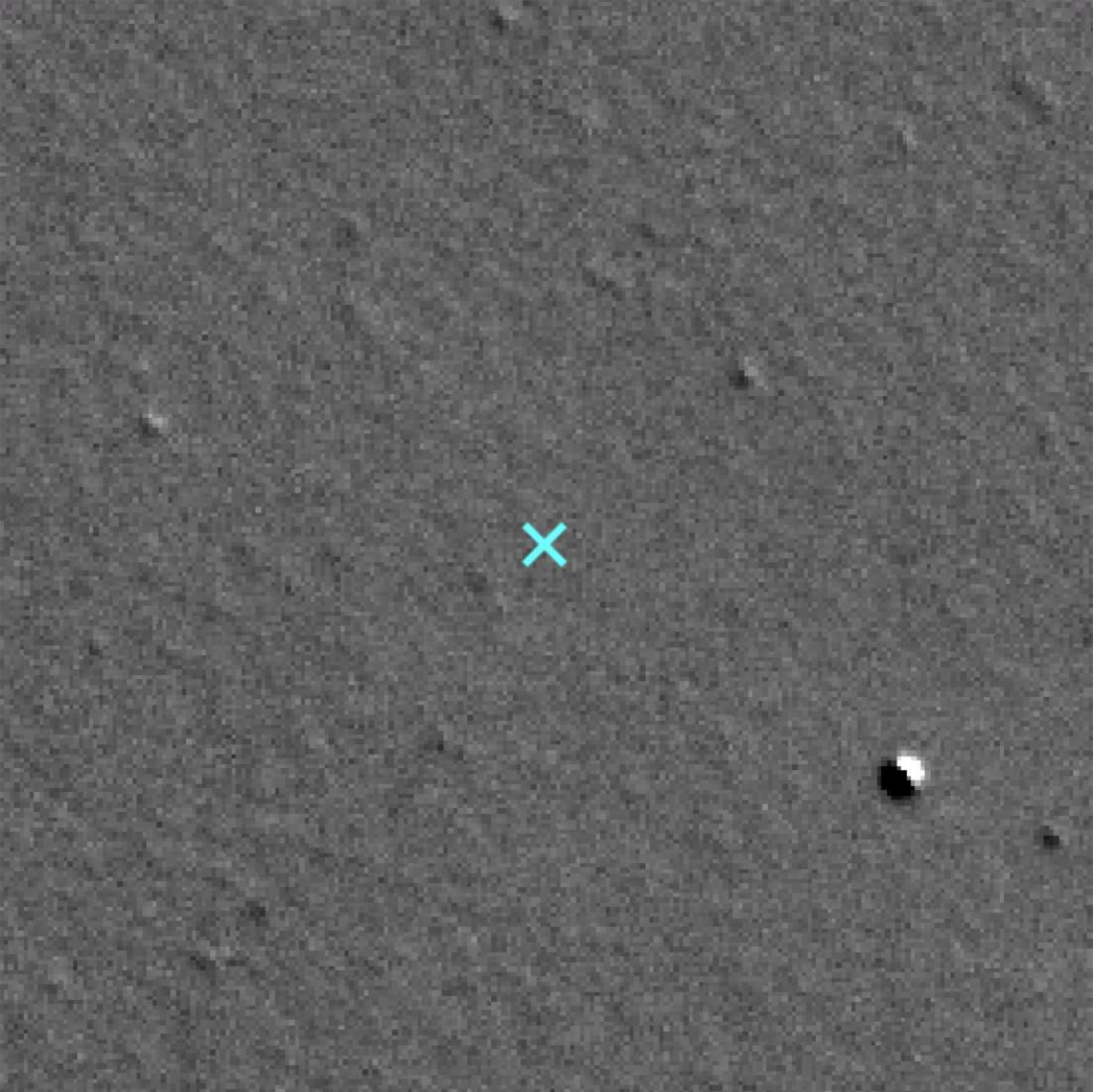}}
\caption{{\bf -- M\,31N 2006-11c}. The location of the nova (in field 6) is indicated by the blue cross. Left:\ Continuum subtracted LGGS H$\alpha$. Middle:\ Continuum subtracted LGGS [\ion{S}{ii}]. Right:\ Continuum subtracted LGGS [\ion{O}{iii}].}
\label{2006-11c surrounding sub images}
\end{figure*}

\begin{figure*}
\centering
\subfloat{\includegraphics[width=.32\textwidth]{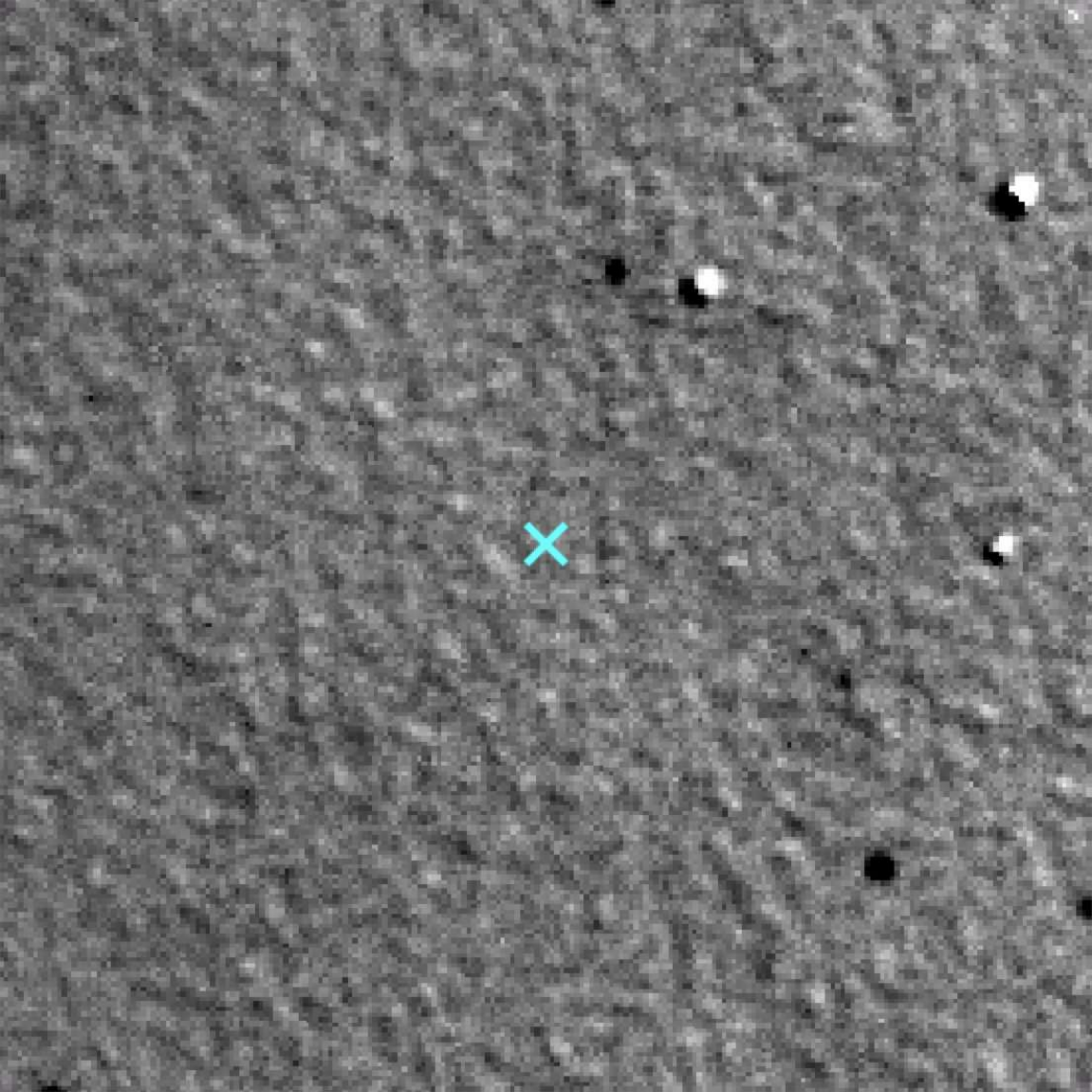}} \quad
\subfloat{\includegraphics[width=.32\textwidth]{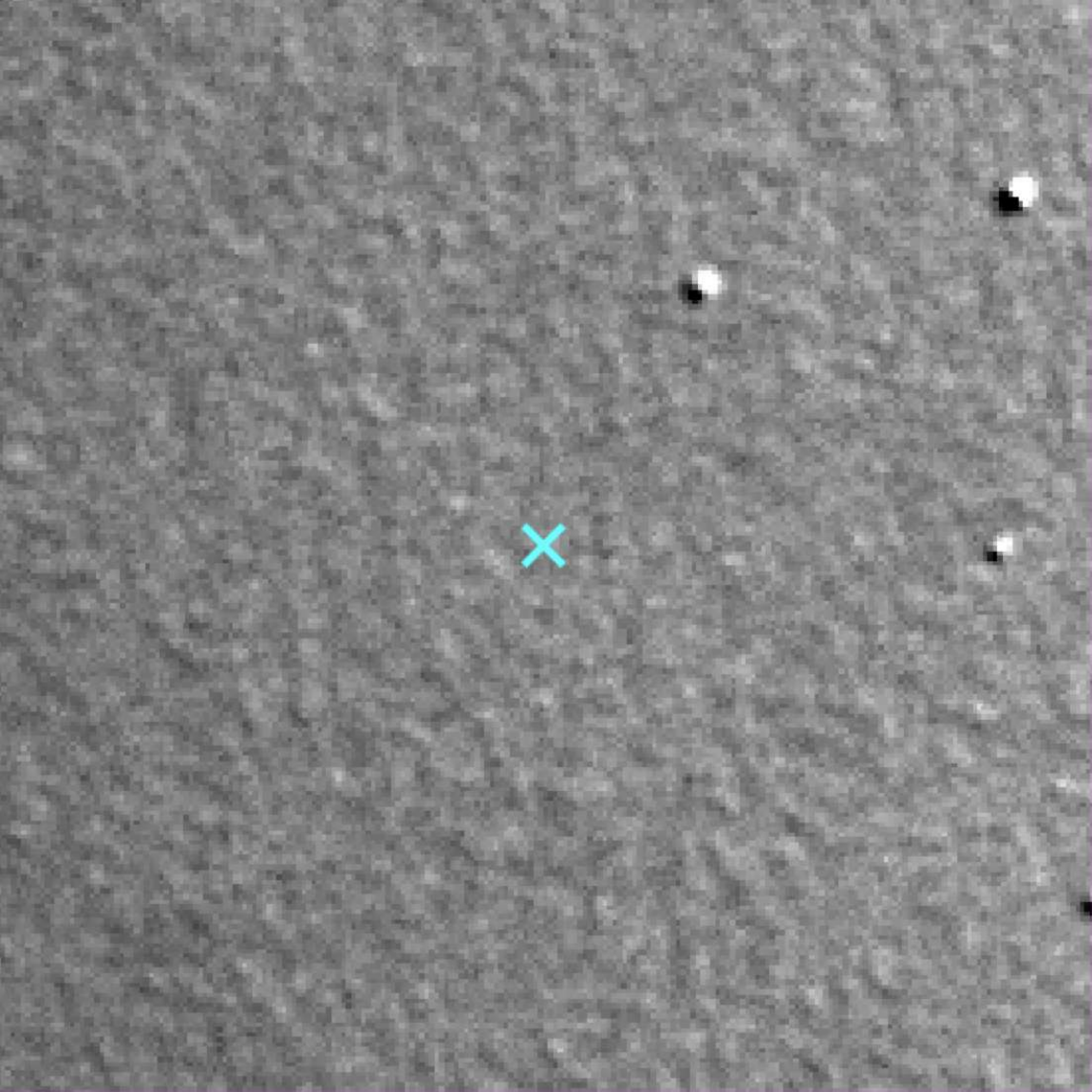}} \quad
\subfloat{\includegraphics[width=.32\textwidth]{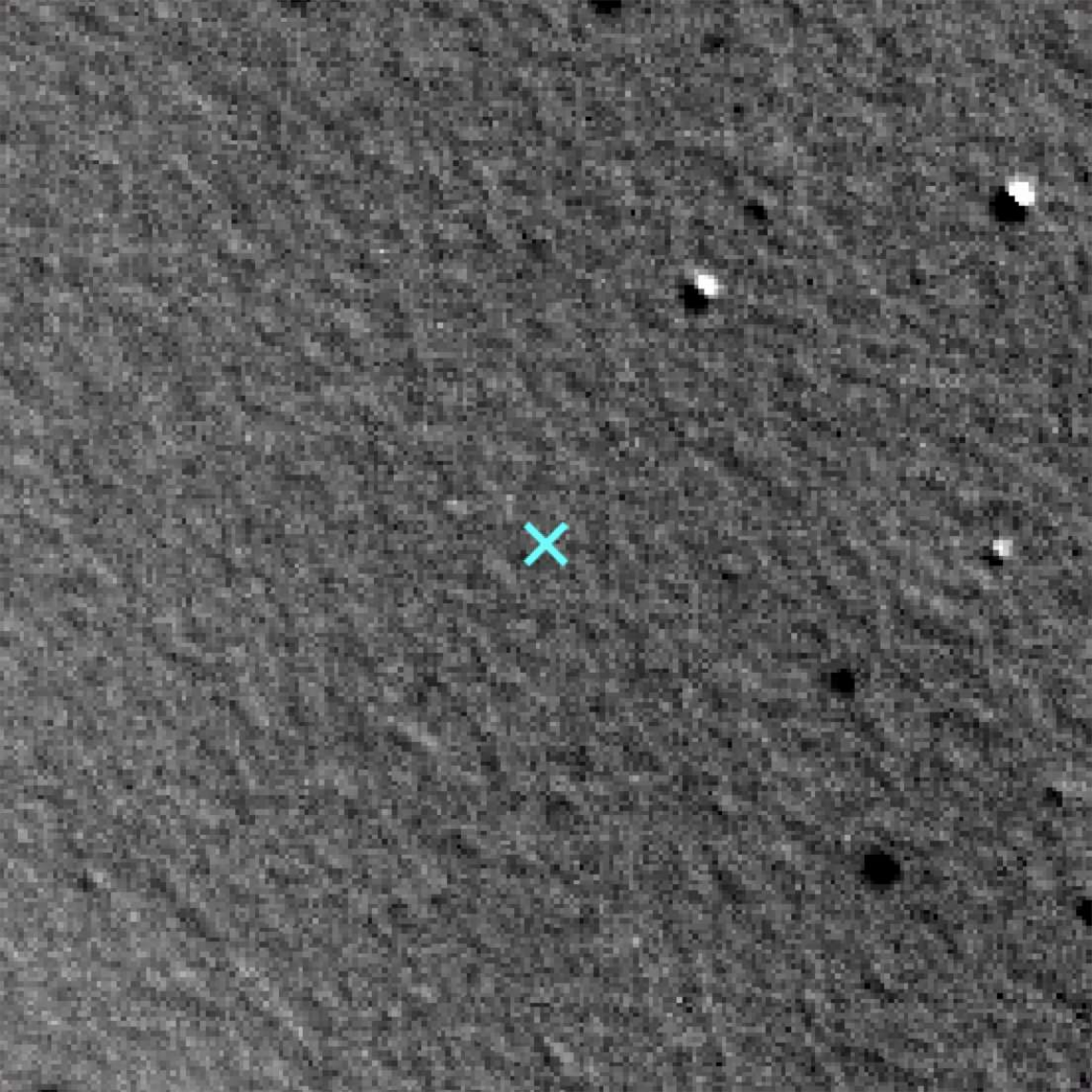}}
\caption{{\bf -- M\,31N 1990-10a}. The location of the nova (in field 6) is indicated by the blue cross. Left:\ Continuum subtracted LGGS H$\alpha$. Middle:\ Continuum subtracted LGGS [\ion{S}{ii}]. Right:\ Continuum subtracted LGGS [\ion{O}{iii}].}
\label{1990-10a surrounding sub images}
\end{figure*}

\begin{figure*}
\centering
\subfloat{\includegraphics[width=.32\textwidth]{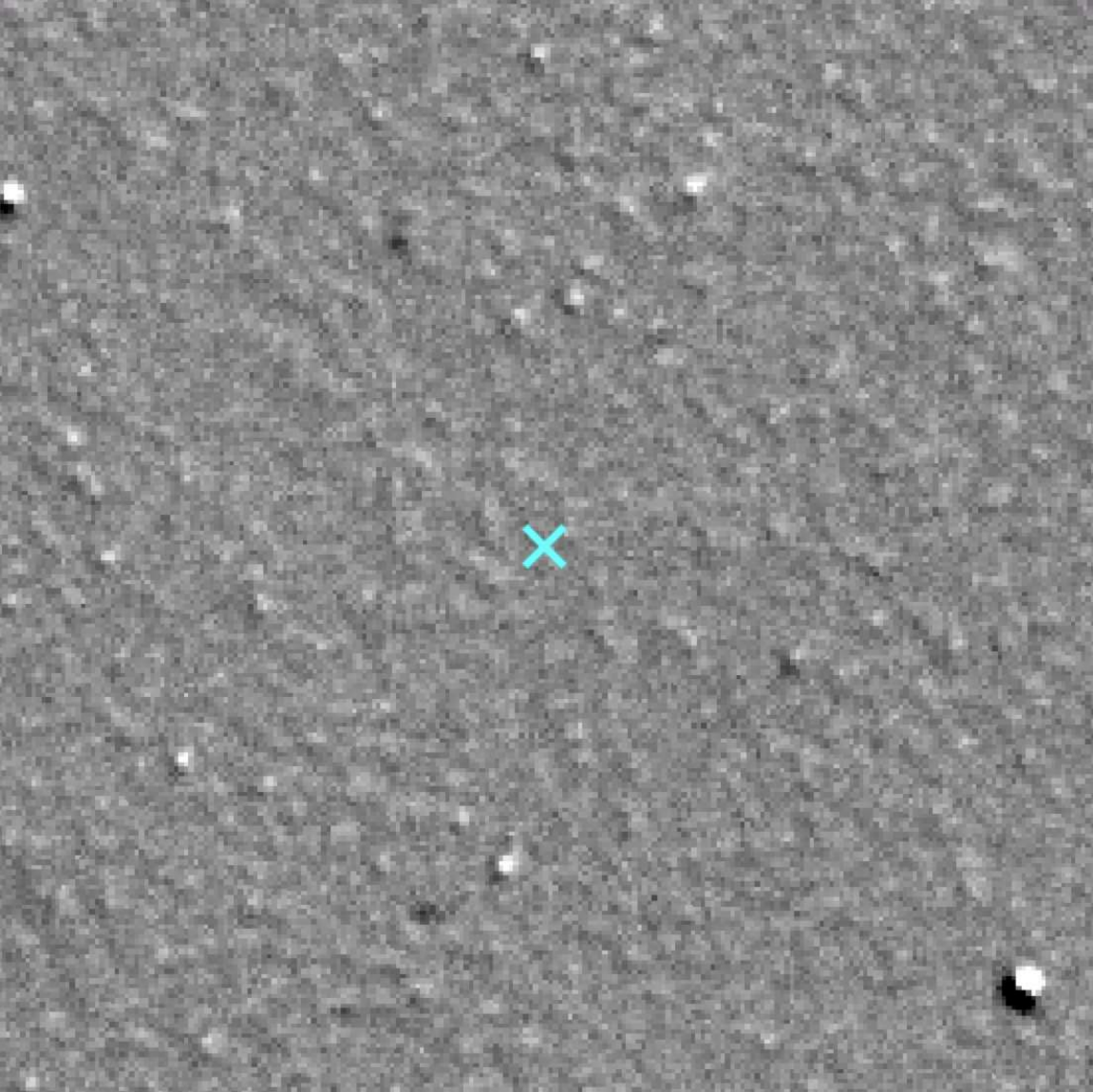}} \quad
\subfloat{\includegraphics[width=.32\textwidth]{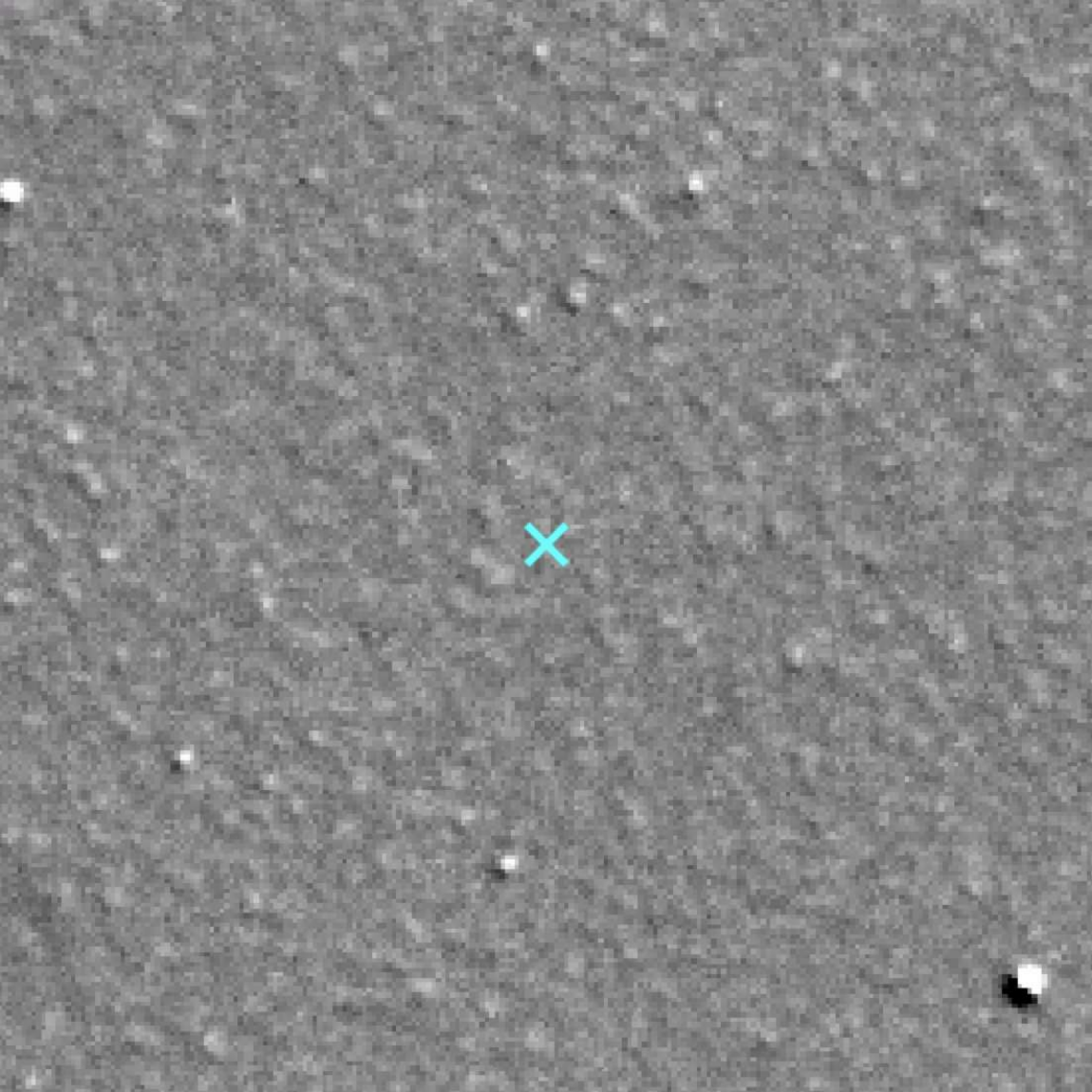}} \quad
\subfloat{\includegraphics[width=.32\textwidth]{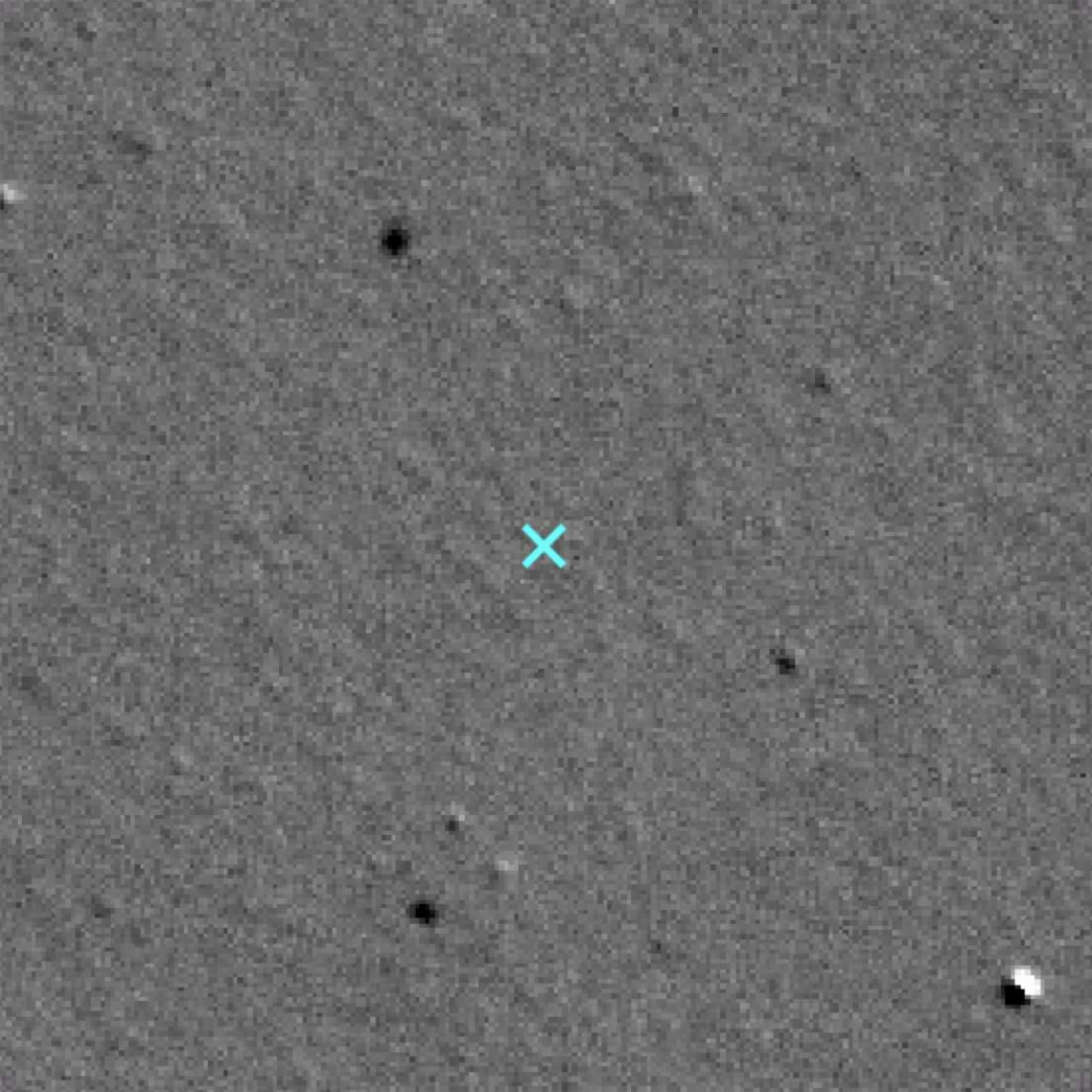}}
\caption{{\bf -- M\,31N 2007-11f}. The location of the nova (in field 6) is indicated by the blue cross. Left:\ Continuum subtracted LGGS H$\alpha$. Middle:\ Continuum subtracted LGGS [\ion{S}{ii}]. Right:\ Continuum subtracted LGGS [\ion{O}{iii}].}
\label{2007-11f surrounding sub images}
\end{figure*}

\begin{figure*}
\centering
\subfloat{\includegraphics[width=.32\textwidth]{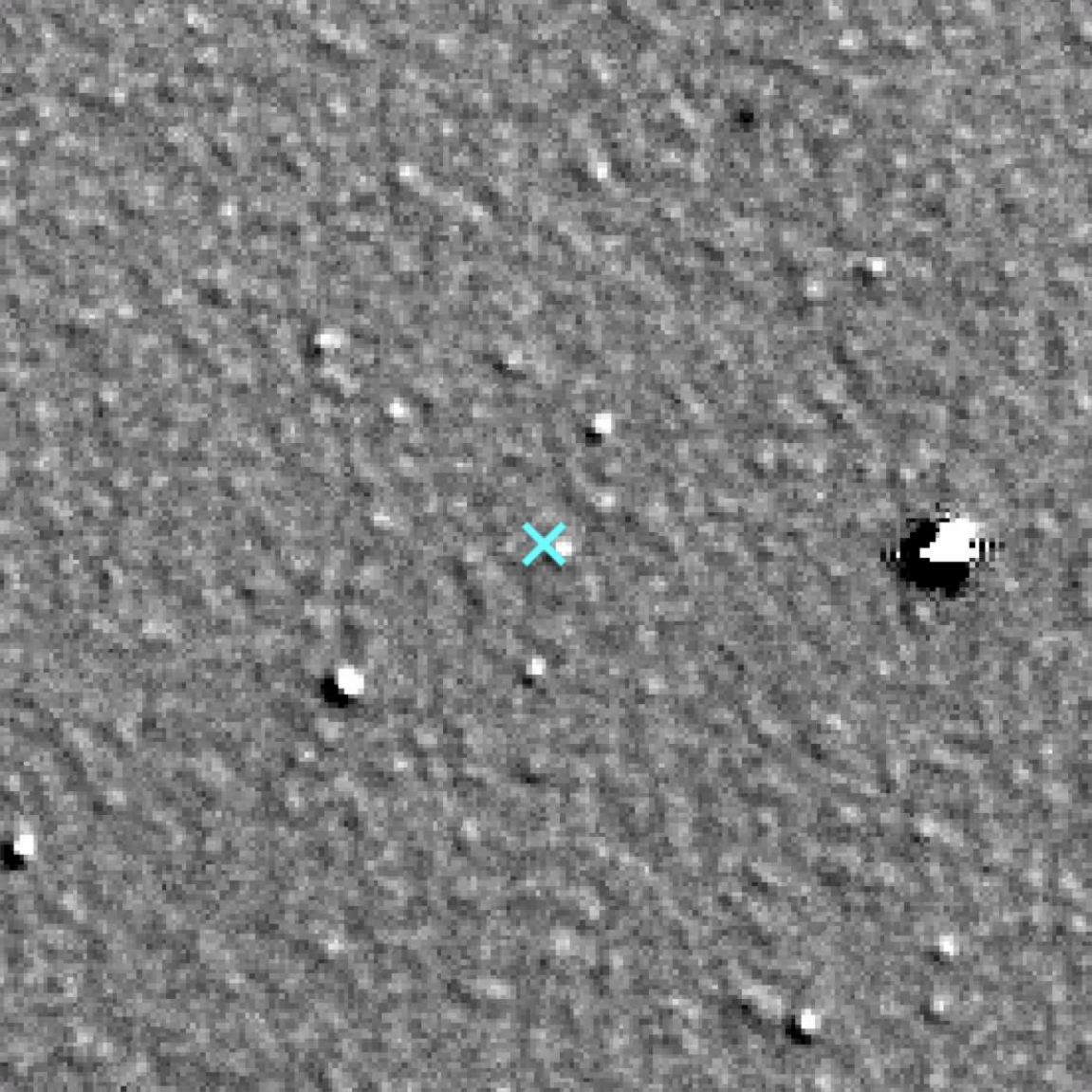}} \quad
\subfloat{\includegraphics[width=.32\textwidth]{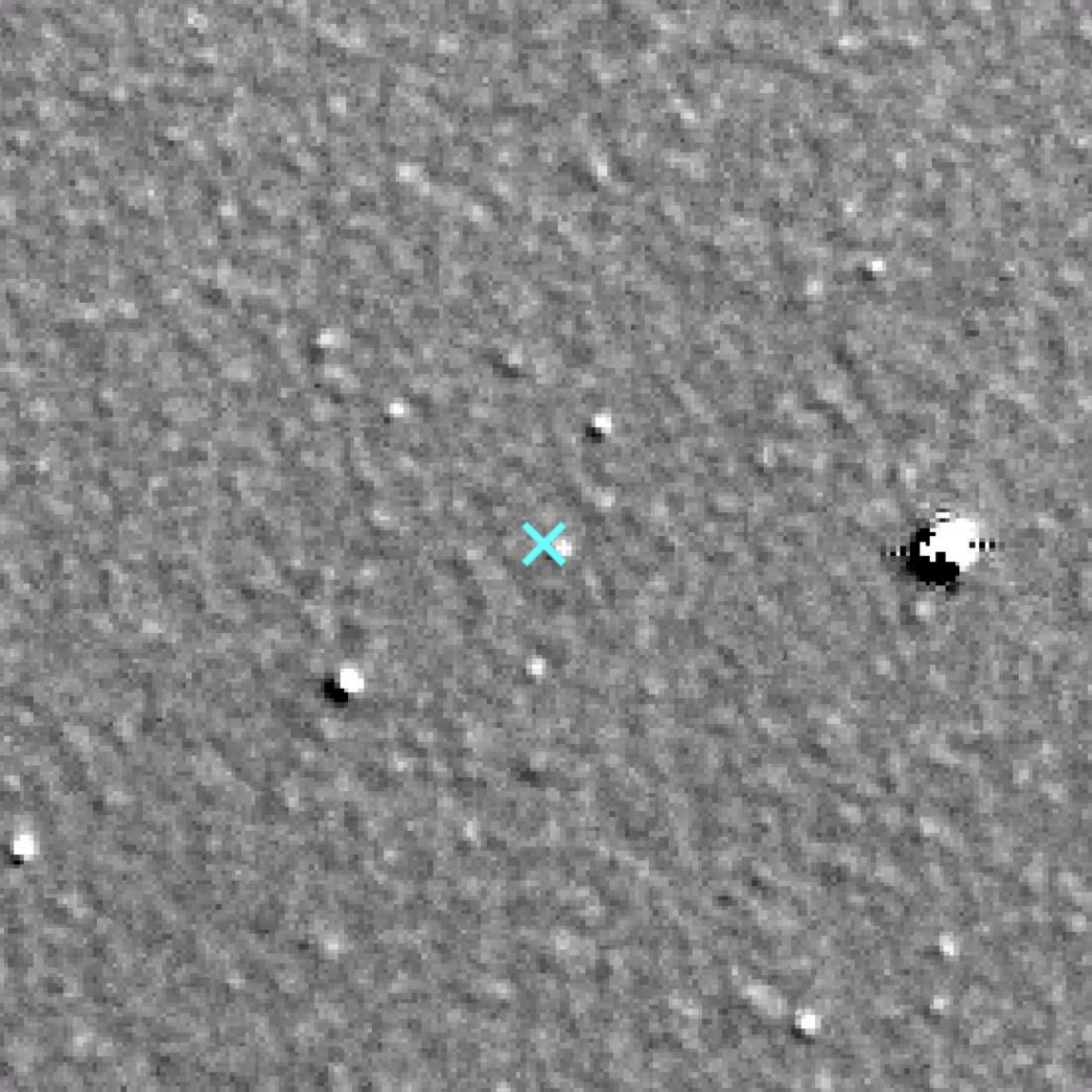}} \quad
\subfloat{\includegraphics[width=.32\textwidth]{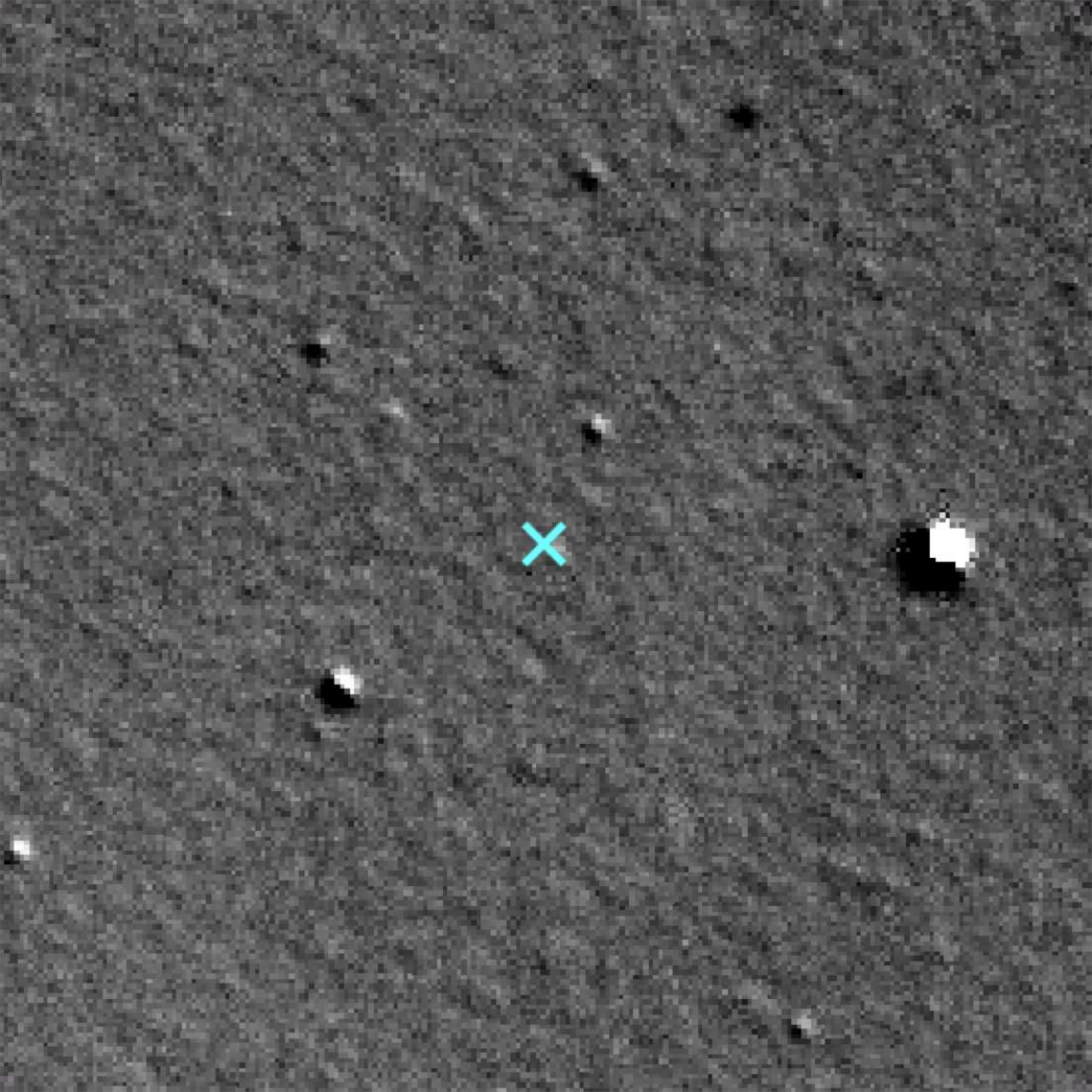}}
\caption{{\bf -- M\,31N 1923-12c}. The location of the nova (in field 6) is indicated by the blue cross. Left:\ Continuum subtracted LGGS H$\alpha$. Middle:\ Continuum subtracted LGGS [\ion{S}{ii}]. Right:\ Continuum subtracted LGGS [\ion{O}{iii}].}
\label{1923-12c surrounding sub images}
\end{figure*}

\begin{figure*}
\centering
\subfloat{\includegraphics[width=.32\textwidth]{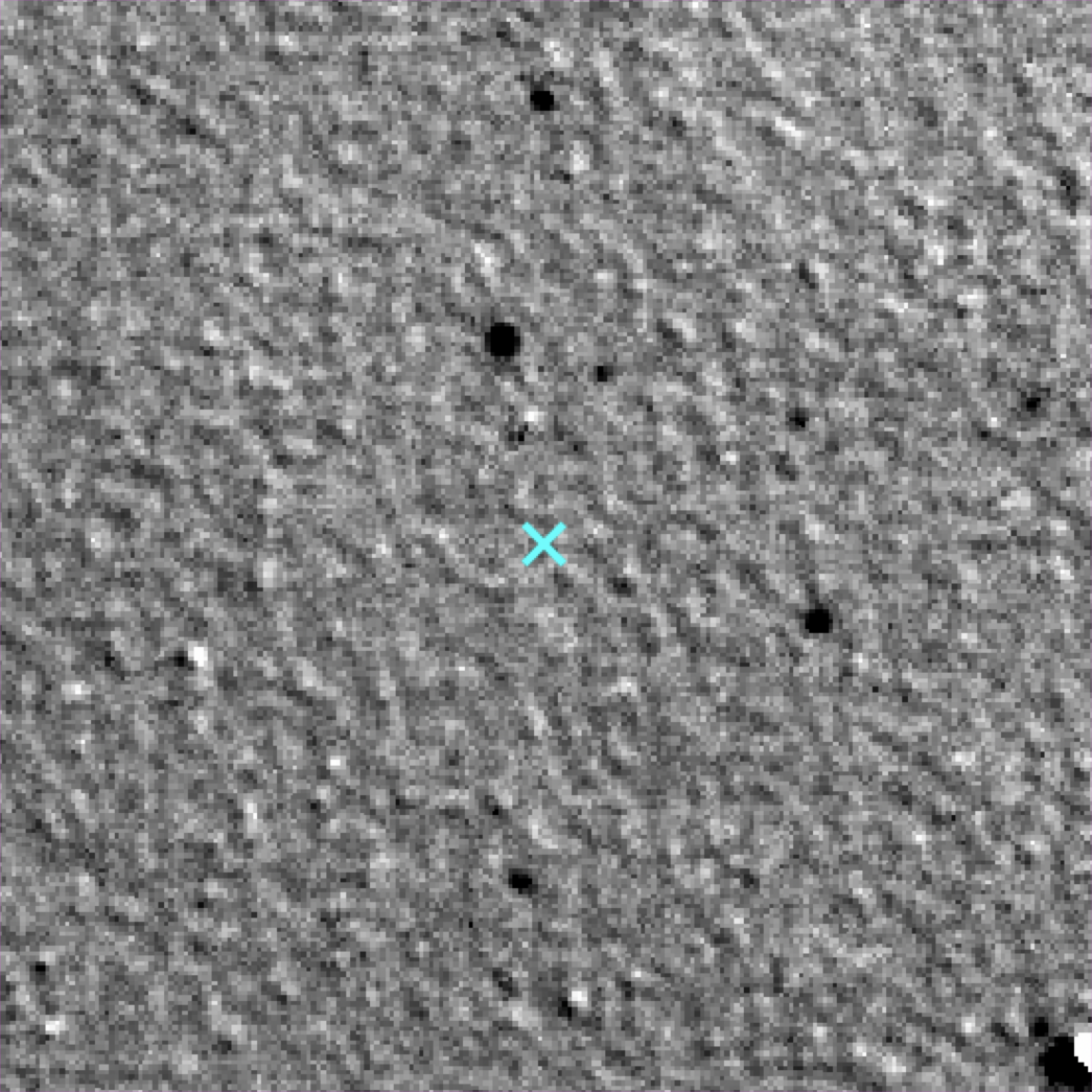}} \quad
\subfloat{\includegraphics[width=.32\textwidth]{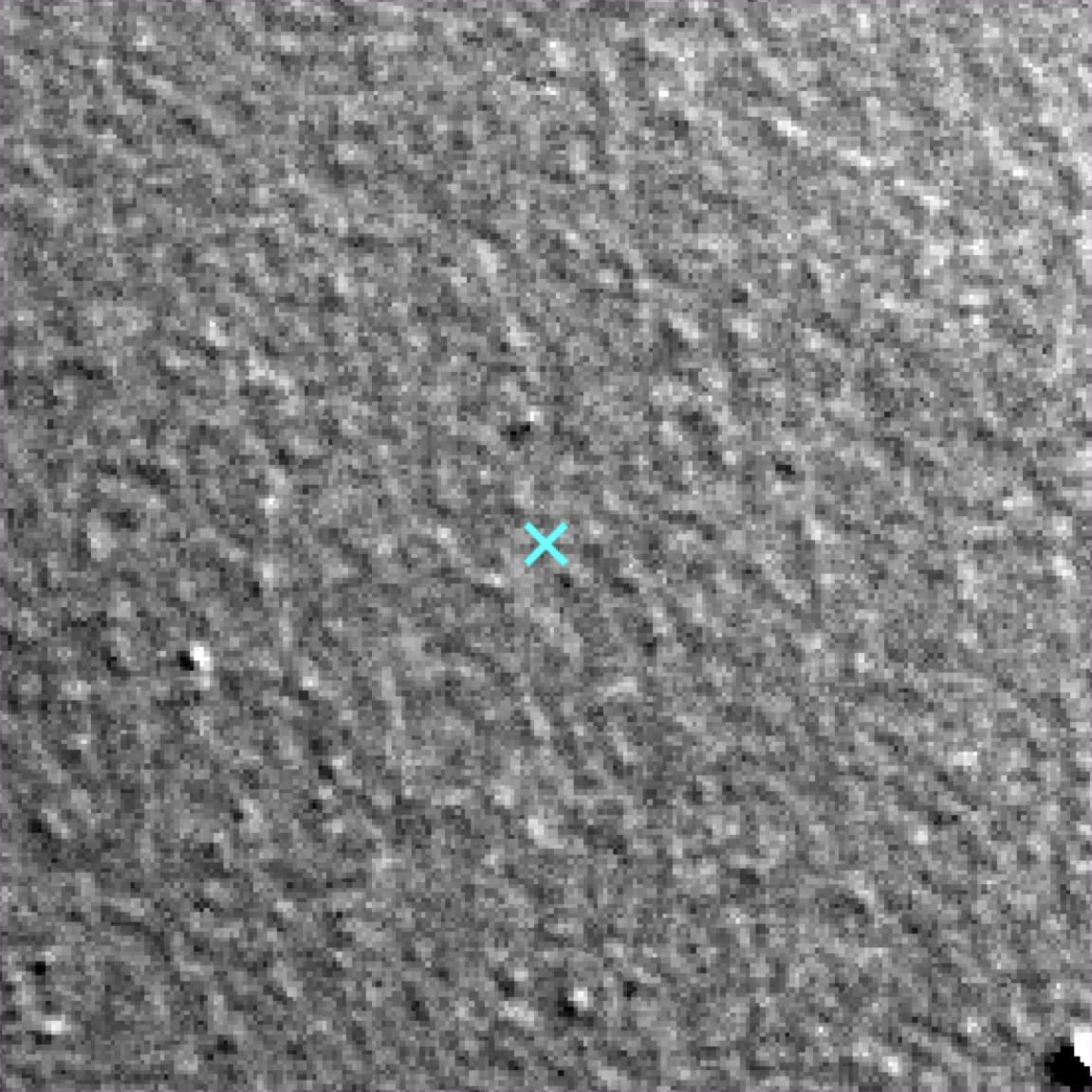}} \quad
\subfloat{\includegraphics[width=.32\textwidth]{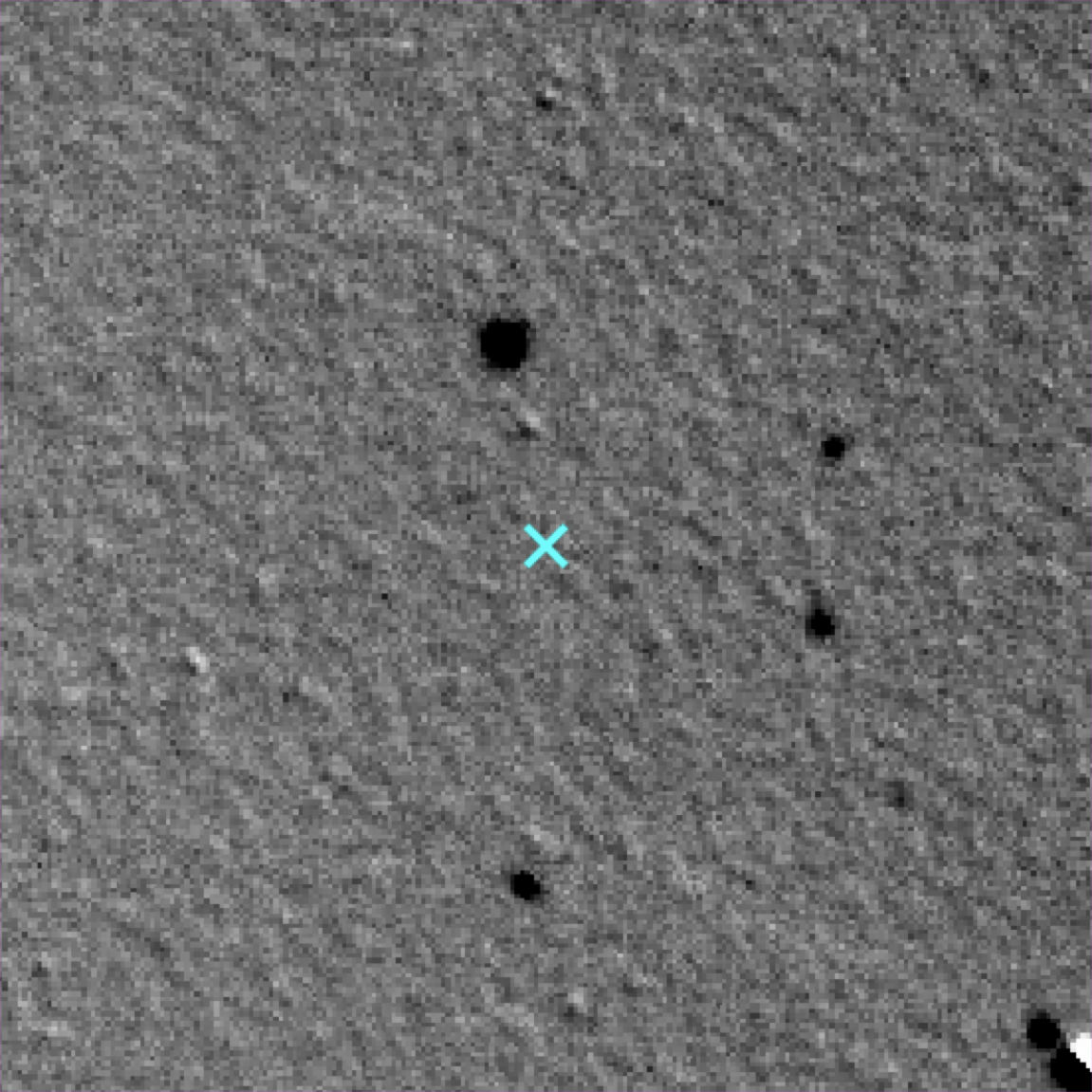}}
\caption{{\bf -- M\,31N 2013-10c}. The location of the nova (in field 6) is indicated by the blue cross. Left:\ Continuum subtracted LGGS H$\alpha$. Middle:\ Continuum subtracted LGGS [\ion{S}{ii}]. Right:\ Continuum subtracted LGGS [\ion{O}{iii}].}
\label{2013-10c surrounding sub images}
\end{figure*}

\begin{figure*}
\centering
\subfloat{\includegraphics[width=.32\textwidth]{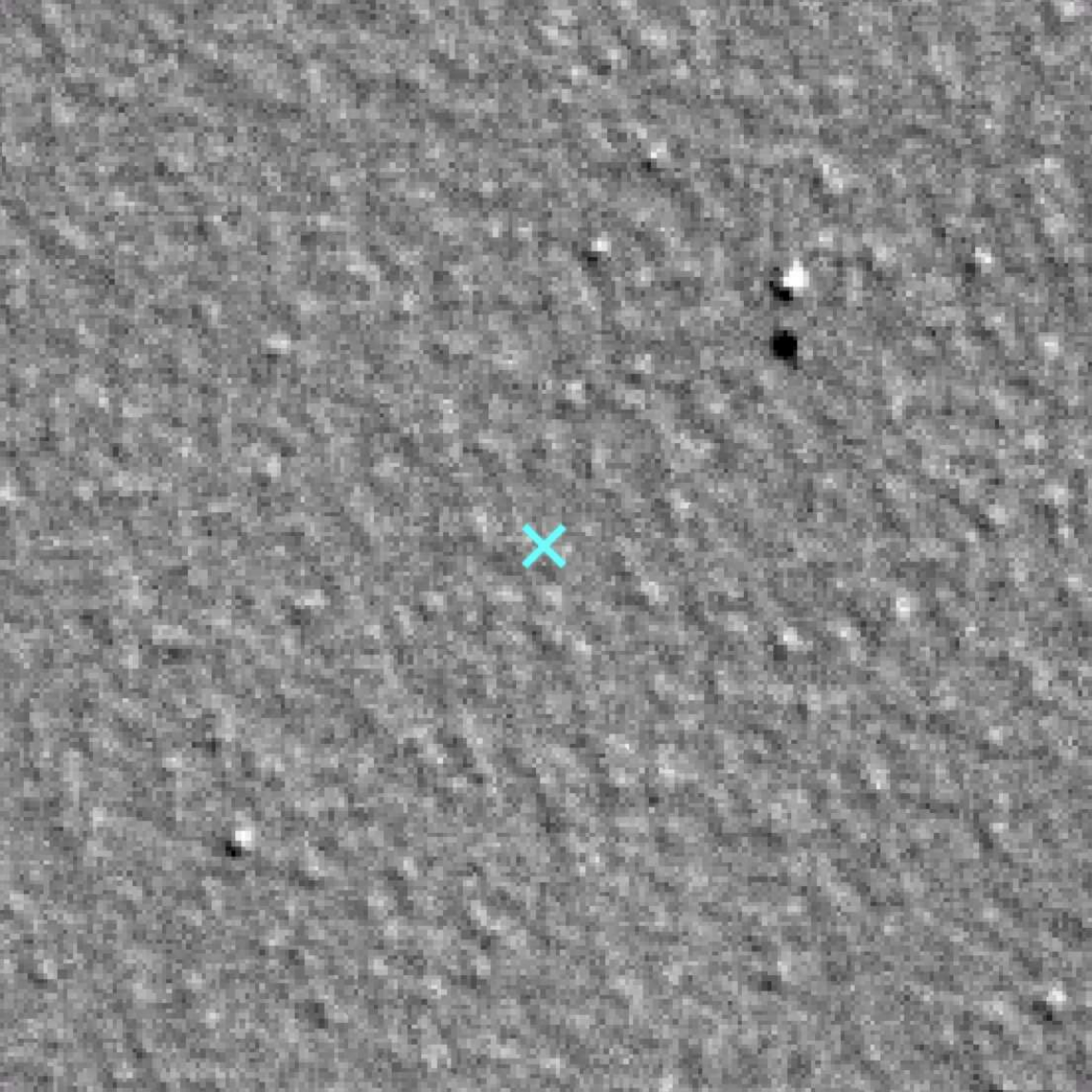}} \quad
\subfloat{\includegraphics[width=.32\textwidth]{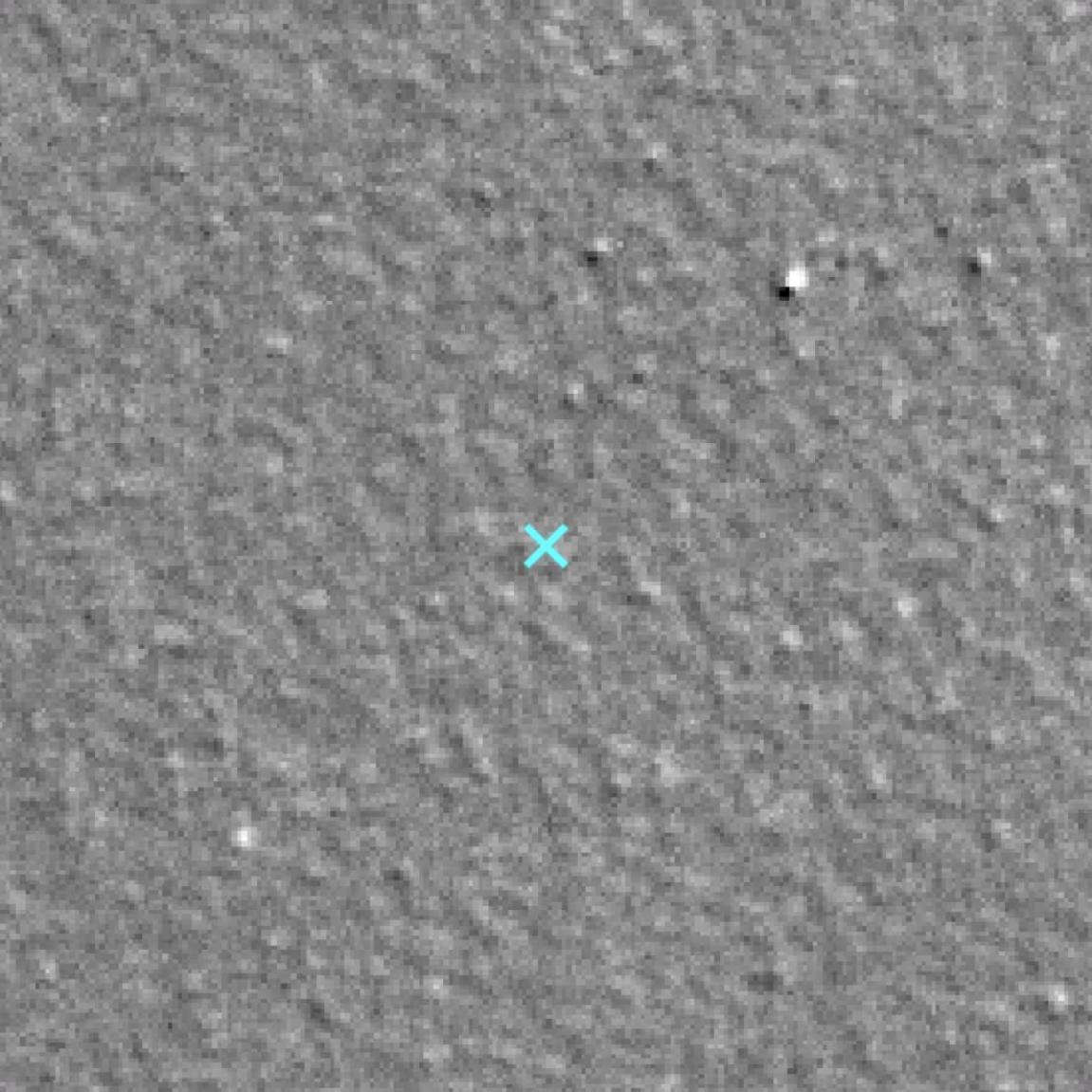}} \quad
\subfloat{\includegraphics[width=.32\textwidth]{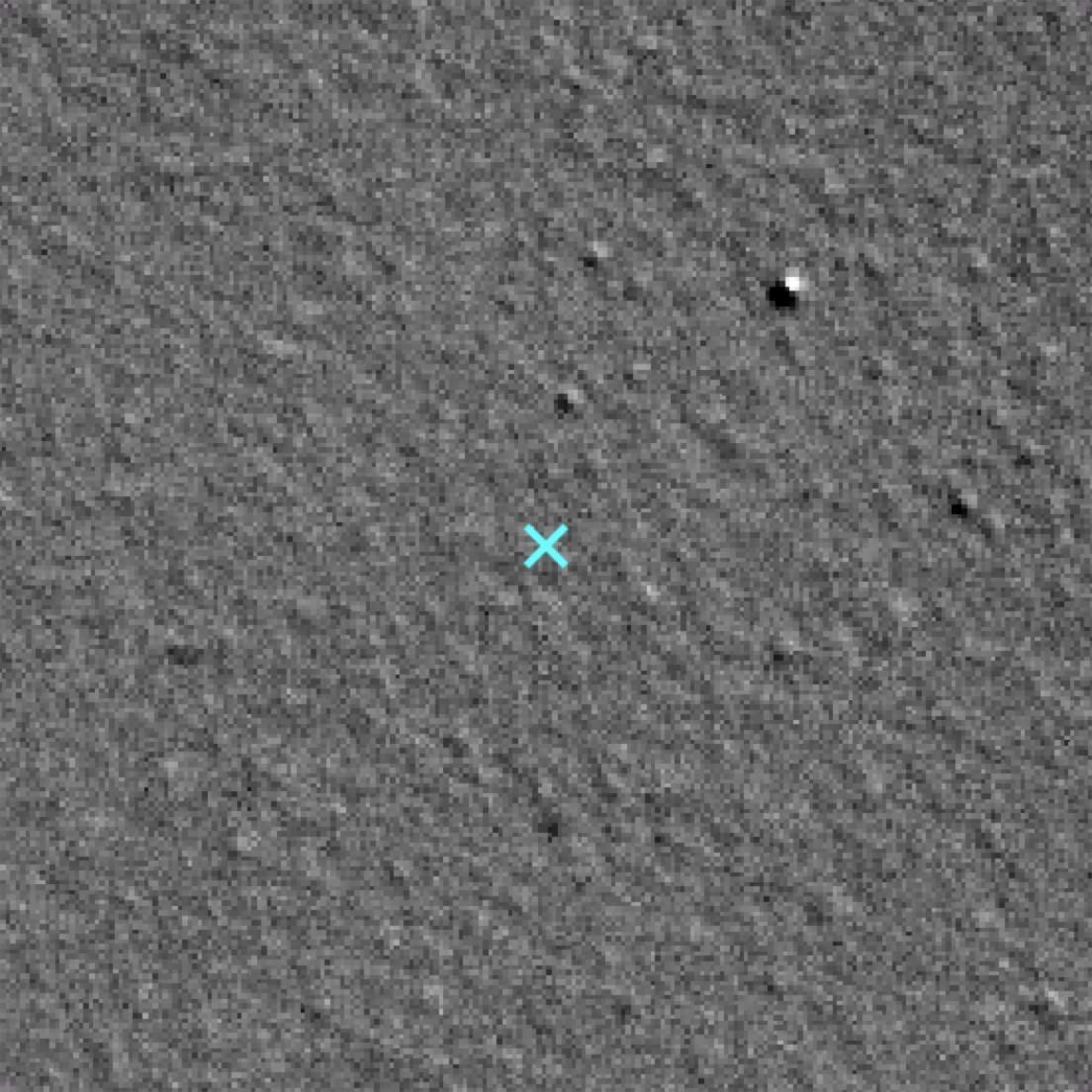}}
\caption{{\bf -- M\,31N 2007-10b}. The location of the nova (in field 6) is indicated by the blue cross. Left:\ Continuum subtracted LGGS H$\alpha$. Middle:\ Continuum subtracted LGGS [\ion{S}{ii}]. Right:\ Continuum subtracted LGGS [\ion{O}{iii}].}
\label{2007-10b surrounding sub images}
\end{figure*}

\begin{figure*}
\centering
\subfloat{\includegraphics[width=.32\textwidth]{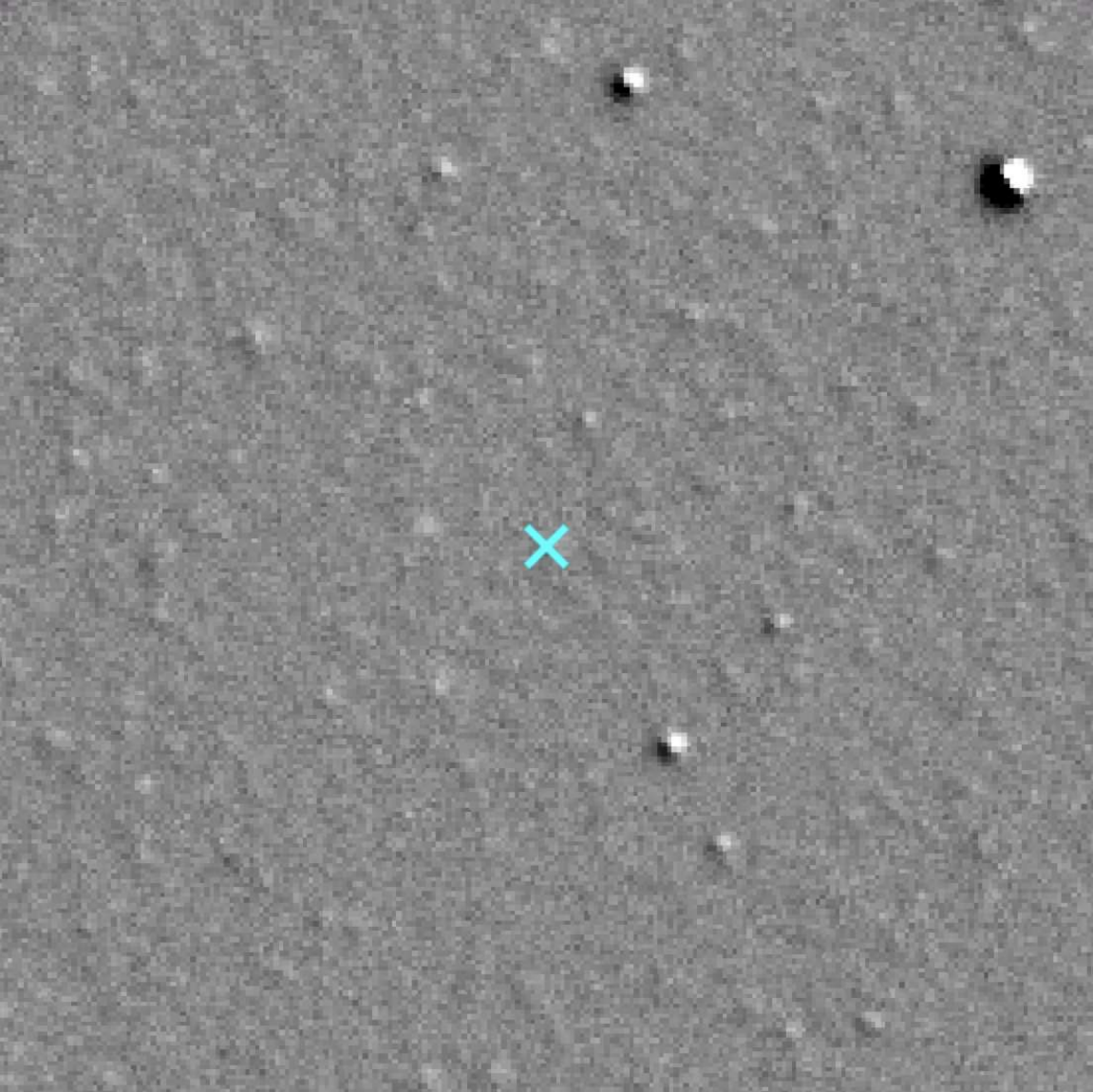}} \quad
\subfloat{\includegraphics[width=.32\textwidth]{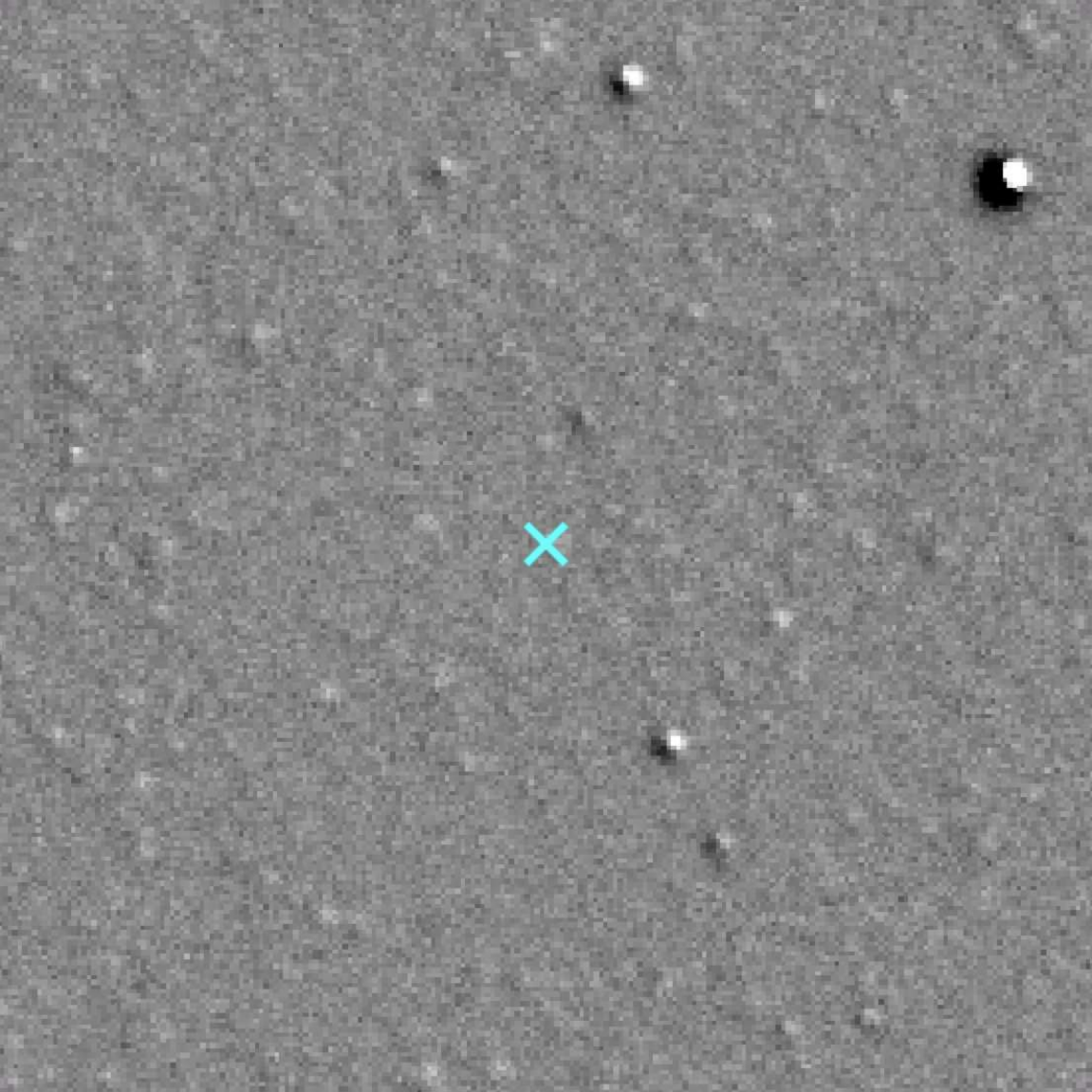}} \quad
\subfloat{\includegraphics[width=.32\textwidth]{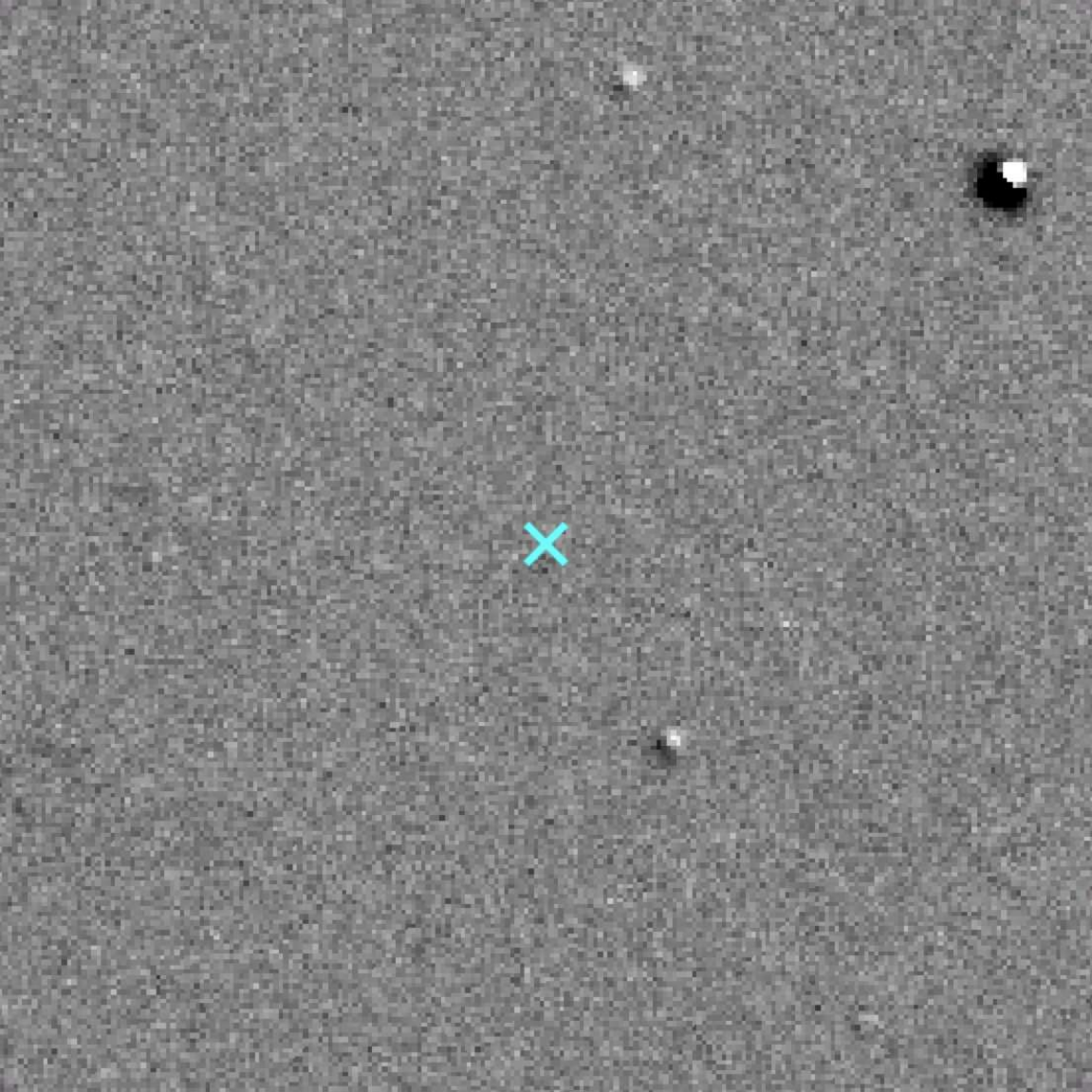}}
\caption{{\bf -- M\,31N 1982-08b}. The location of the nova (in field 3) is indicated by the blue cross. Left:\ Continuum subtracted LGGS H$\alpha$. Middle:\ Continuum subtracted LGGS [\ion{S}{ii}]. Right:\ Continuum subtracted LGGS [\ion{O}{iii}].}
\label{1982-08b surrounding sub images}
\end{figure*}

\begin{figure*}
\centering
\subfloat{\includegraphics[width=.32\textwidth]{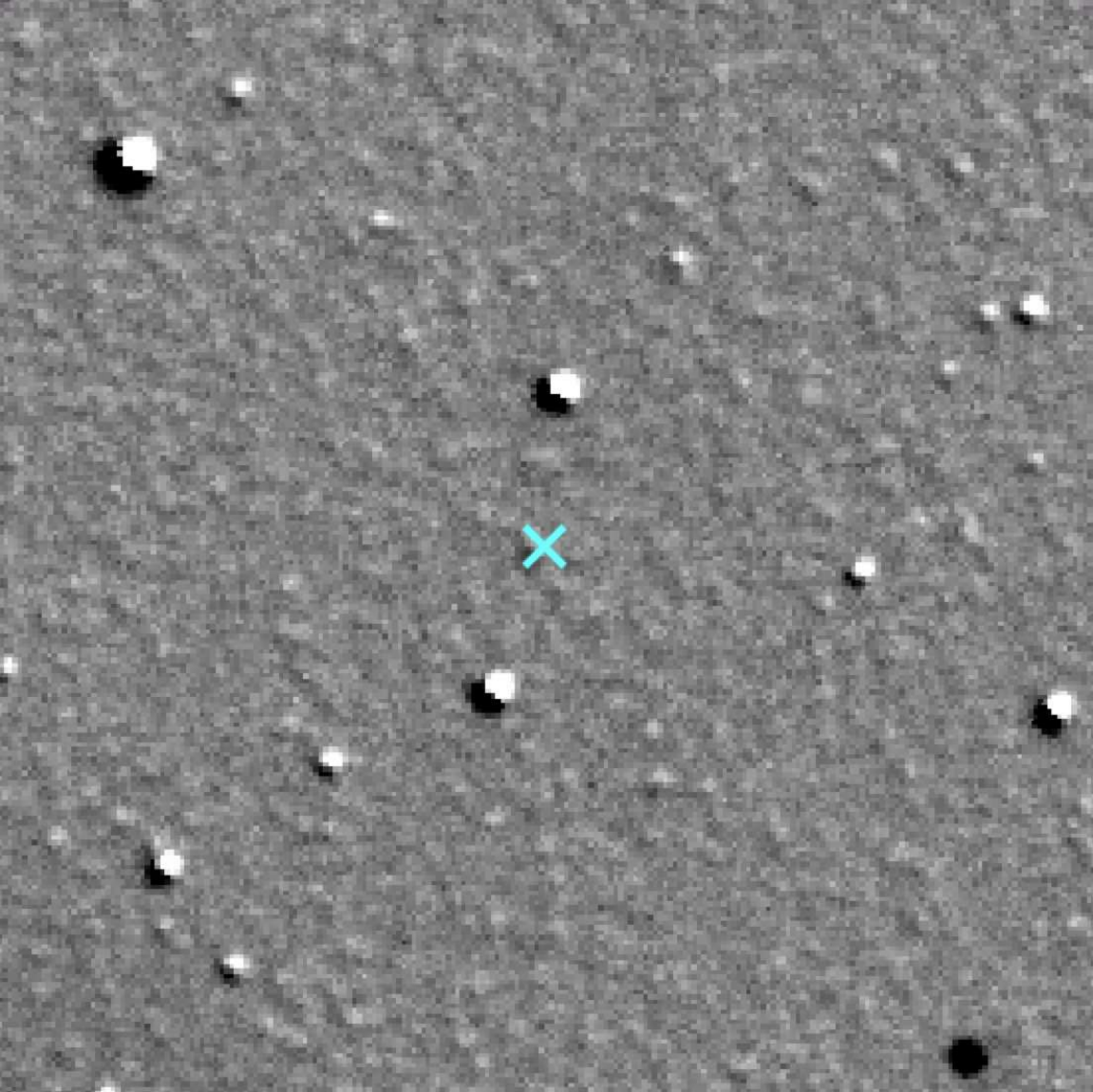}} \quad
\subfloat{\includegraphics[width=.32\textwidth]{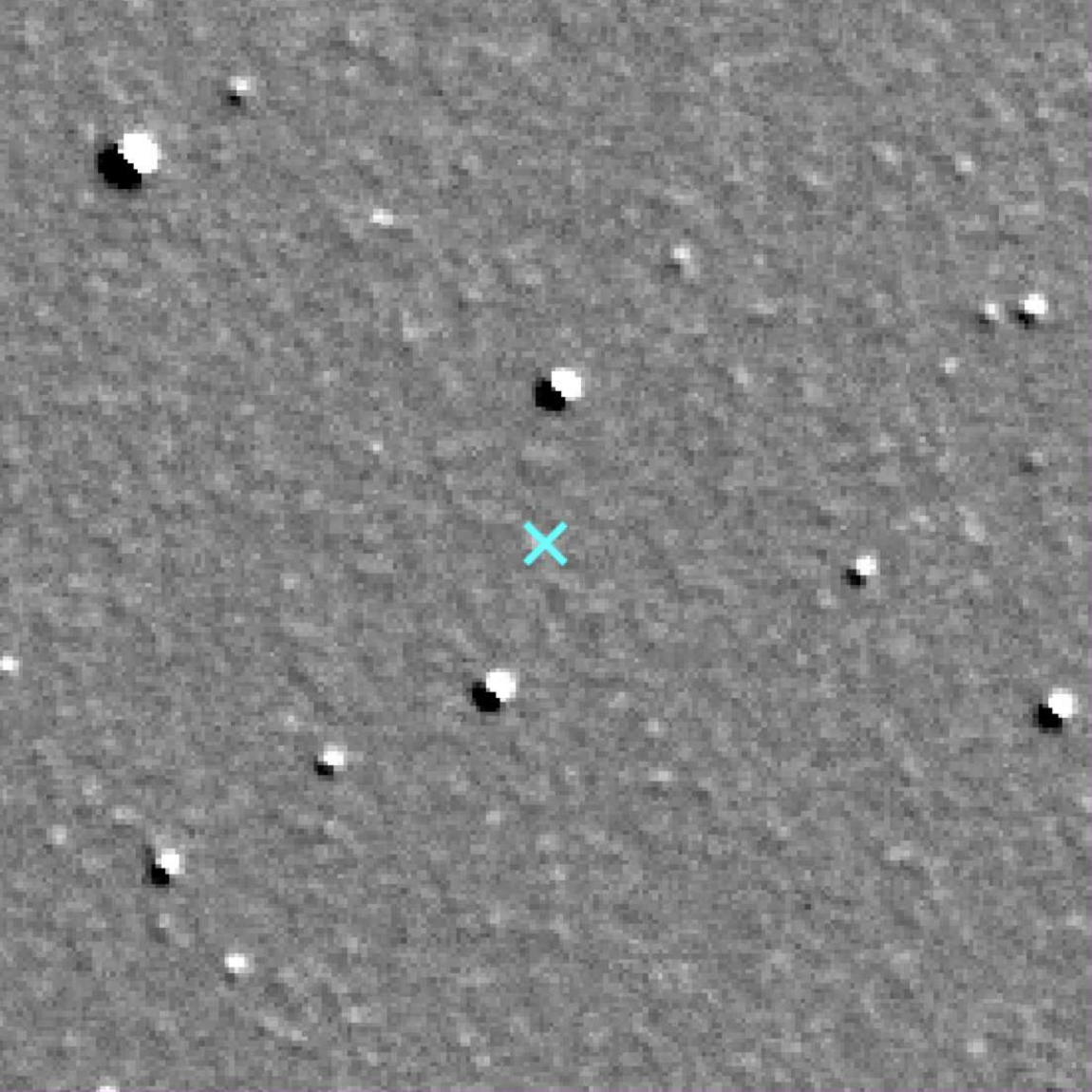}} \quad
\subfloat{\includegraphics[width=.32\textwidth]{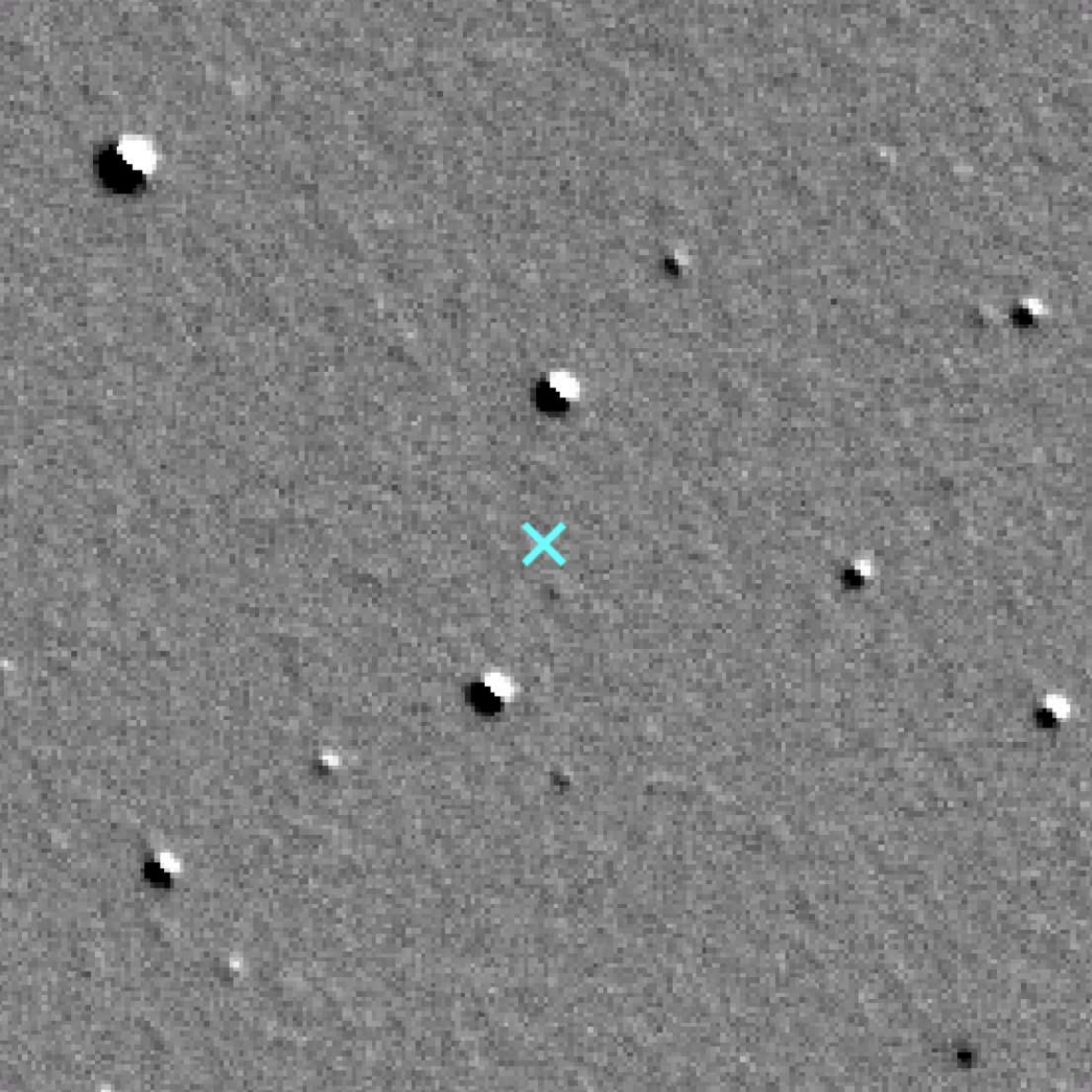}}
\caption{{\bf -- M\,31N 1945-09c}. The location of the nova (in field 6) is indicated by the blue cross. Left:\ Continuum subtracted LGGS H$\alpha$. Middle:\ Continuum subtracted LGGS [\ion{S}{ii}]. Right:\ Continuum subtracted LGGS [\ion{O}{iii}].}
\label{1945-09c surrounding sub images}
\end{figure*}

\begin{figure*}
\centering
\subfloat{\includegraphics[width=.32\textwidth]{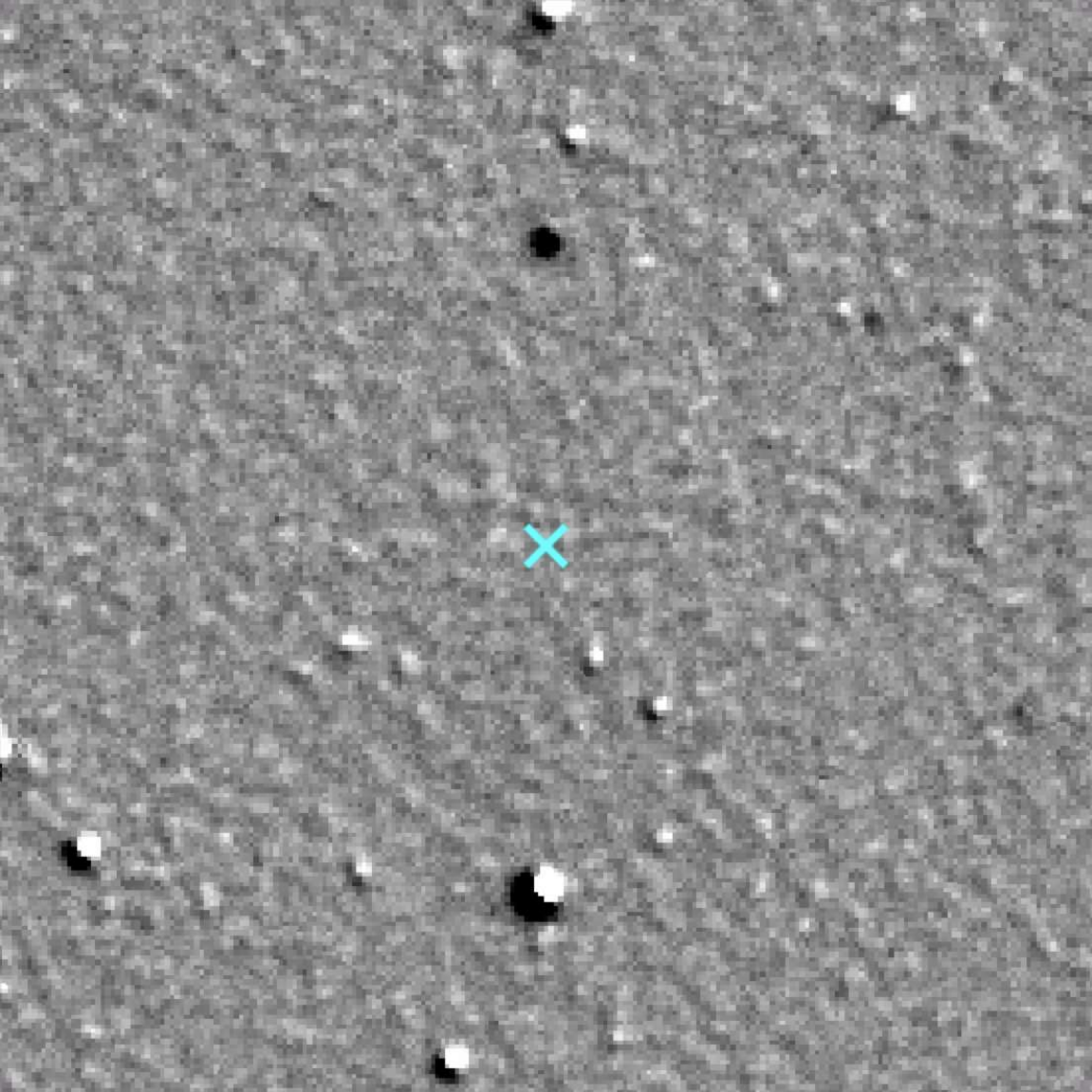}} \quad
\subfloat{\includegraphics[width=.32\textwidth]{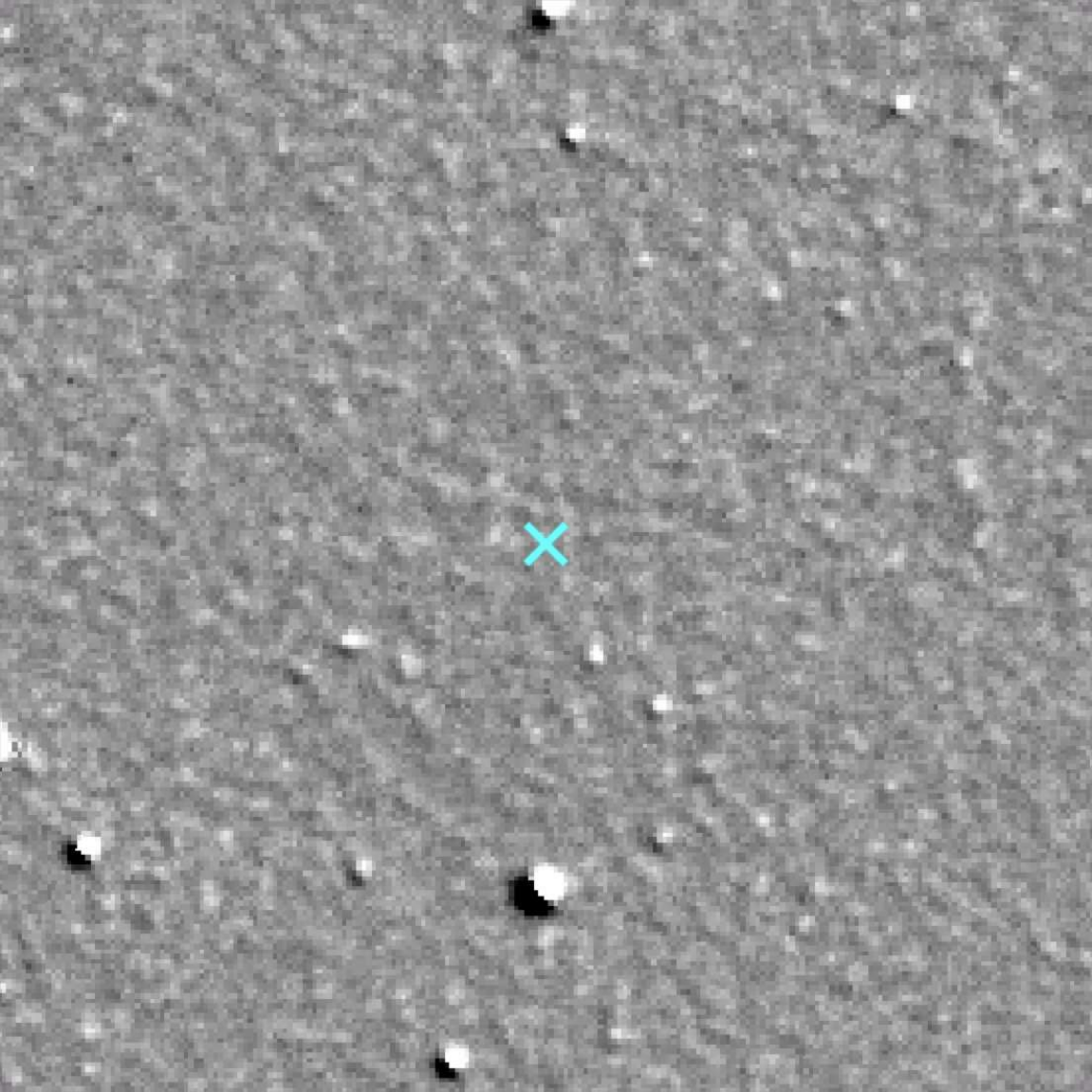}} \quad
\subfloat{\includegraphics[width=.32\textwidth]{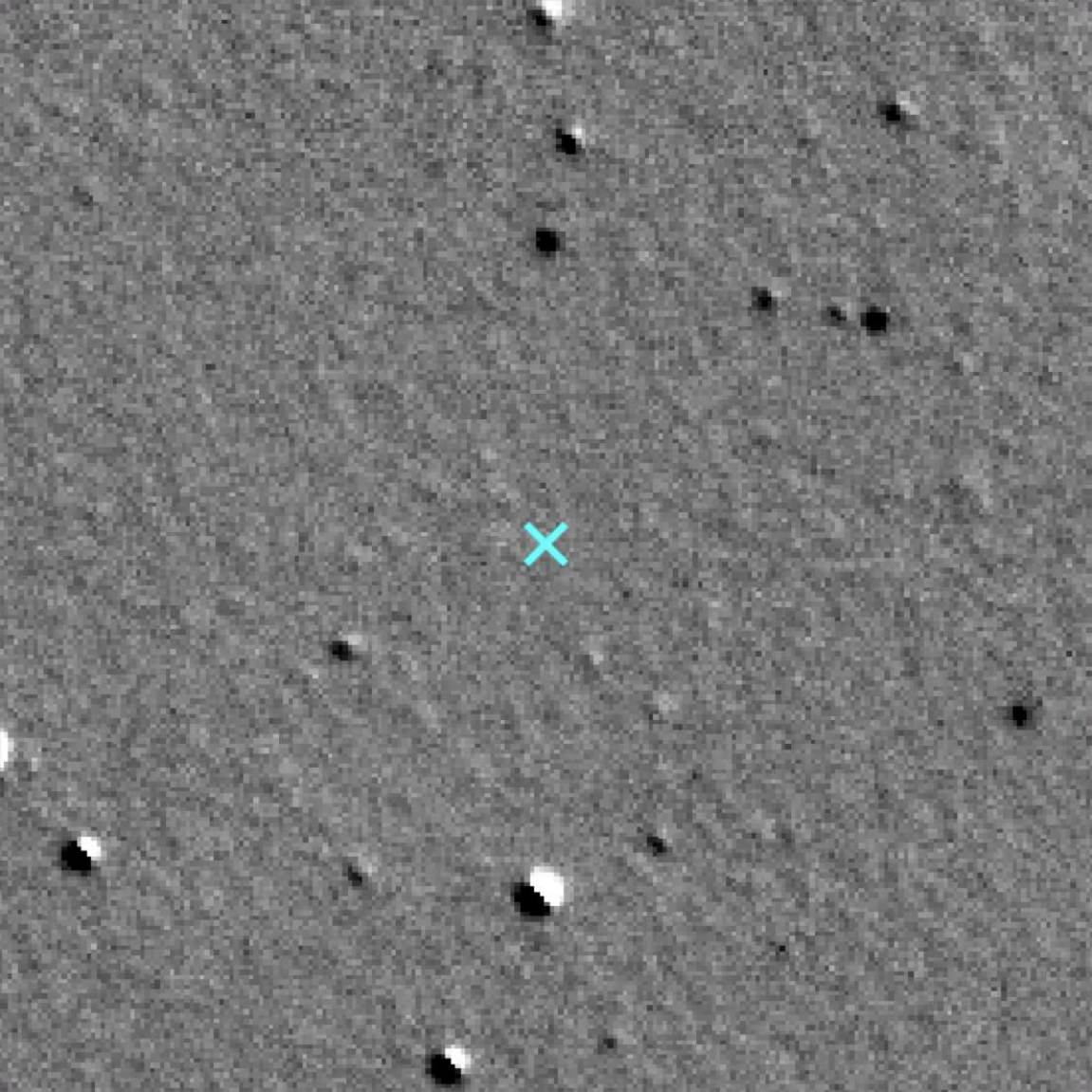}}
\caption{{\bf -- M\,31N 1926-06a}. The location of the nova (in field 6) is indicated by the blue cross. Left:\ Continuum subtracted LGGS H$\alpha$. Middle:\ Continuum subtracted LGGS [\ion{S}{ii}]. Right:\ Continuum subtracted LGGS [\ion{O}{iii}].}
\label{1926-06c surrounding sub images}
\end{figure*}

\begin{figure*}
\centering
\subfloat{\includegraphics[width=.32\textwidth]{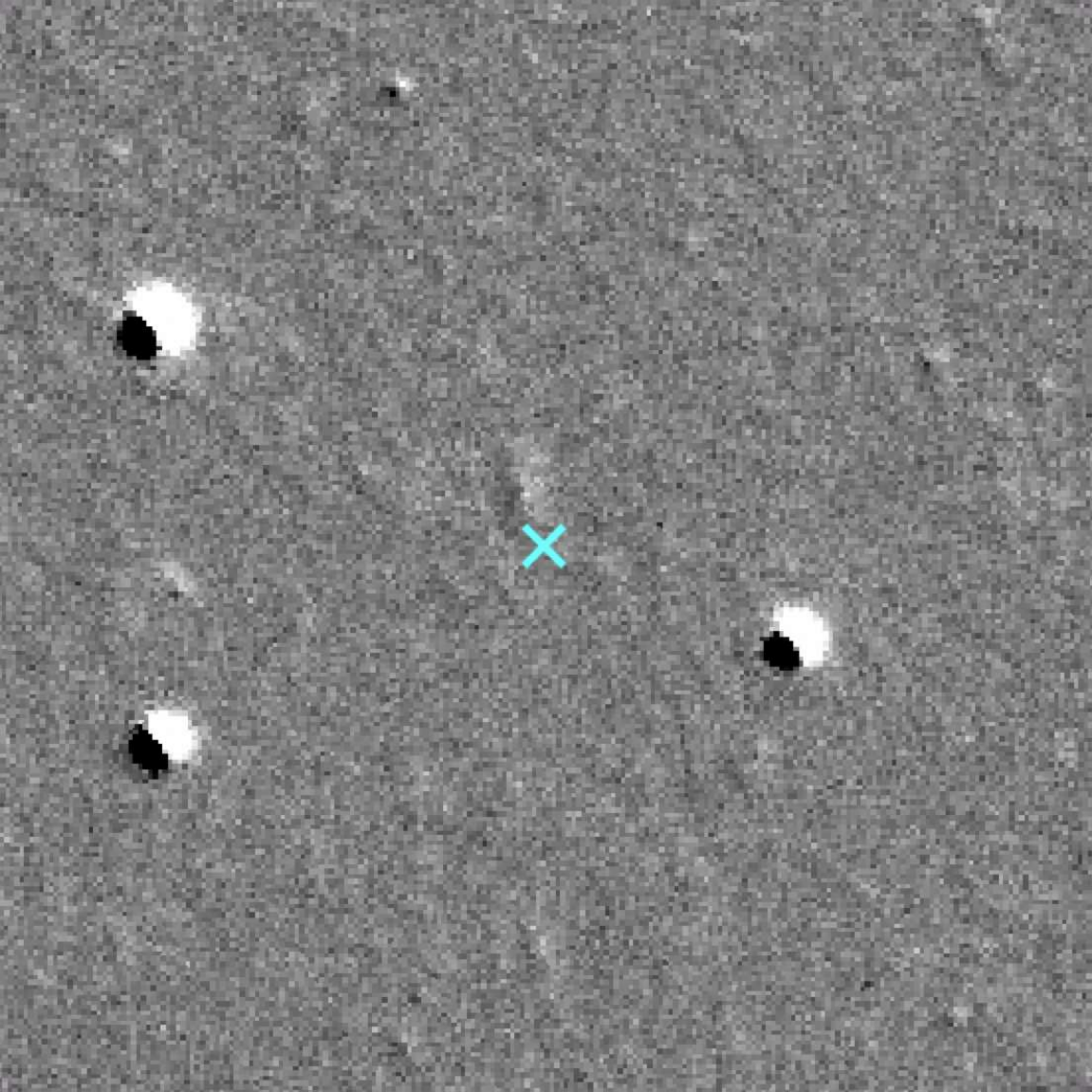}} \quad
\subfloat{\includegraphics[width=.32\textwidth]{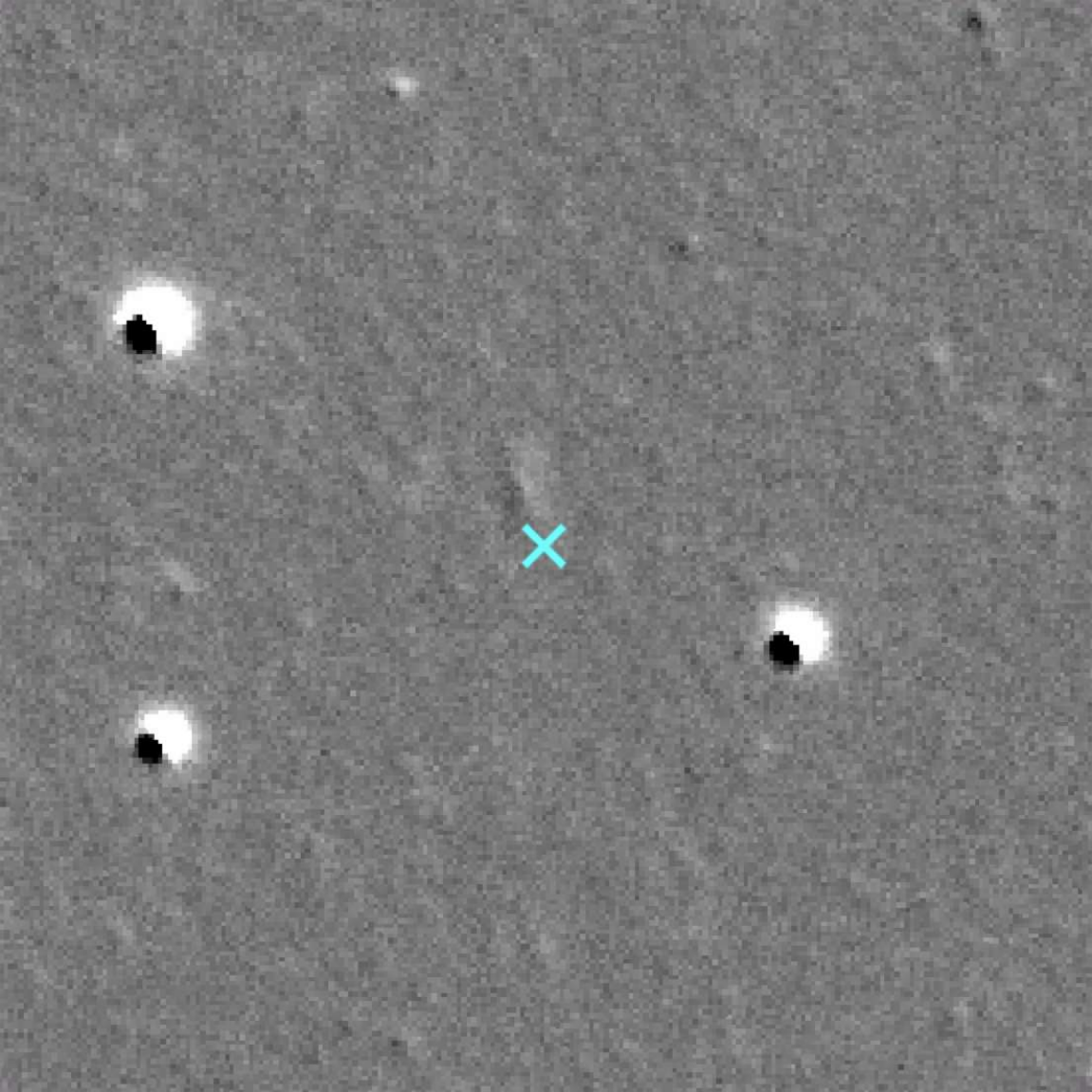}} \quad
\subfloat{\includegraphics[width=.32\textwidth]{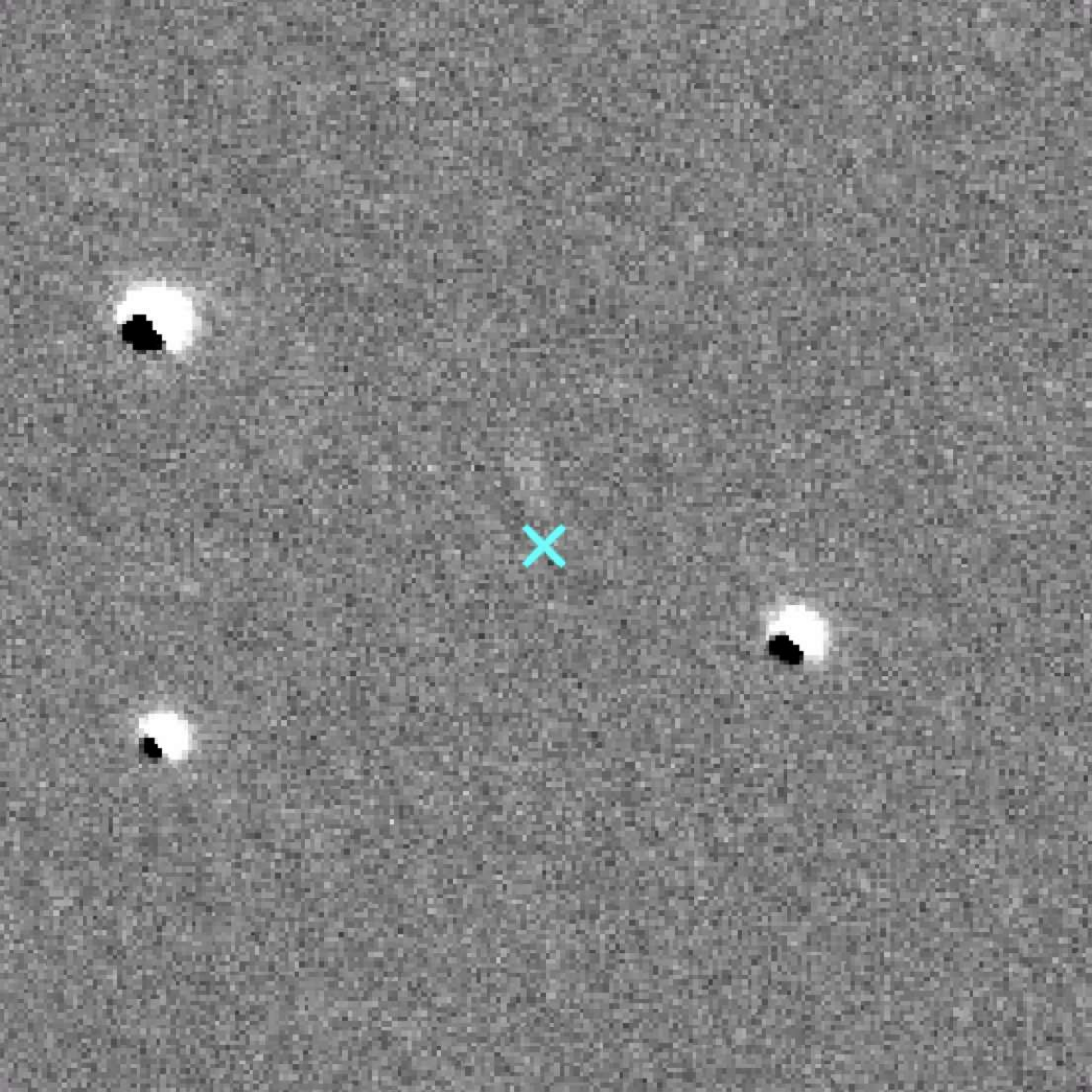}}
\caption{{\bf -- M\,31N 1966-09e}. The location of the nova (in field 8) is indicated by the blue cross. Left:\ Continuum subtracted LGGS H$\alpha$. Middle:\ Continuum subtracted LGGS [\ion{S}{ii}]. Right:\ Continuum subtracted LGGS [\ion{O}{iii}].}
\label{1966-09e surrounding sub images}
\end{figure*}

\begin{figure*}
\centering
\subfloat{\includegraphics[width=.32\textwidth]{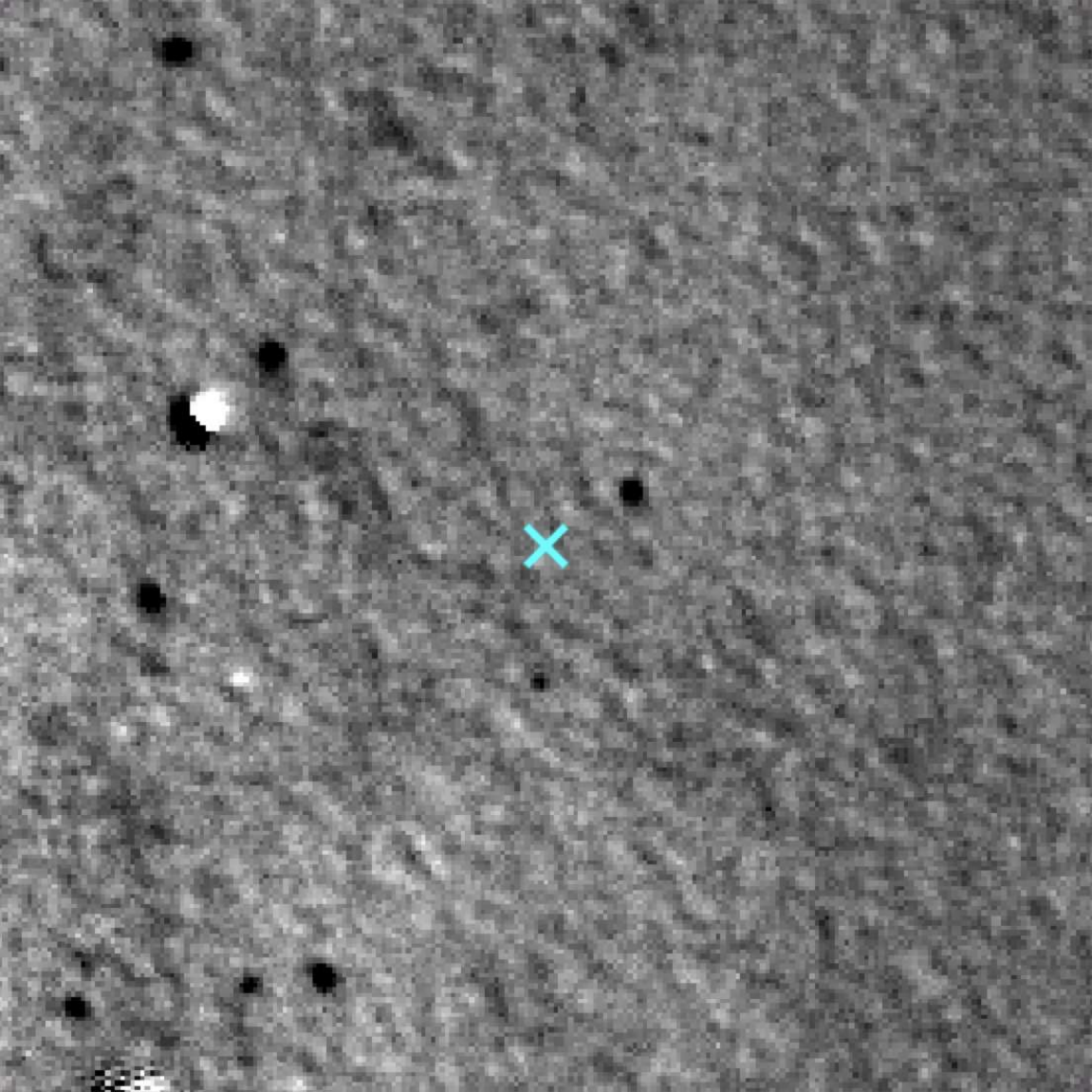}} \quad
\subfloat{\includegraphics[width=.32\textwidth]{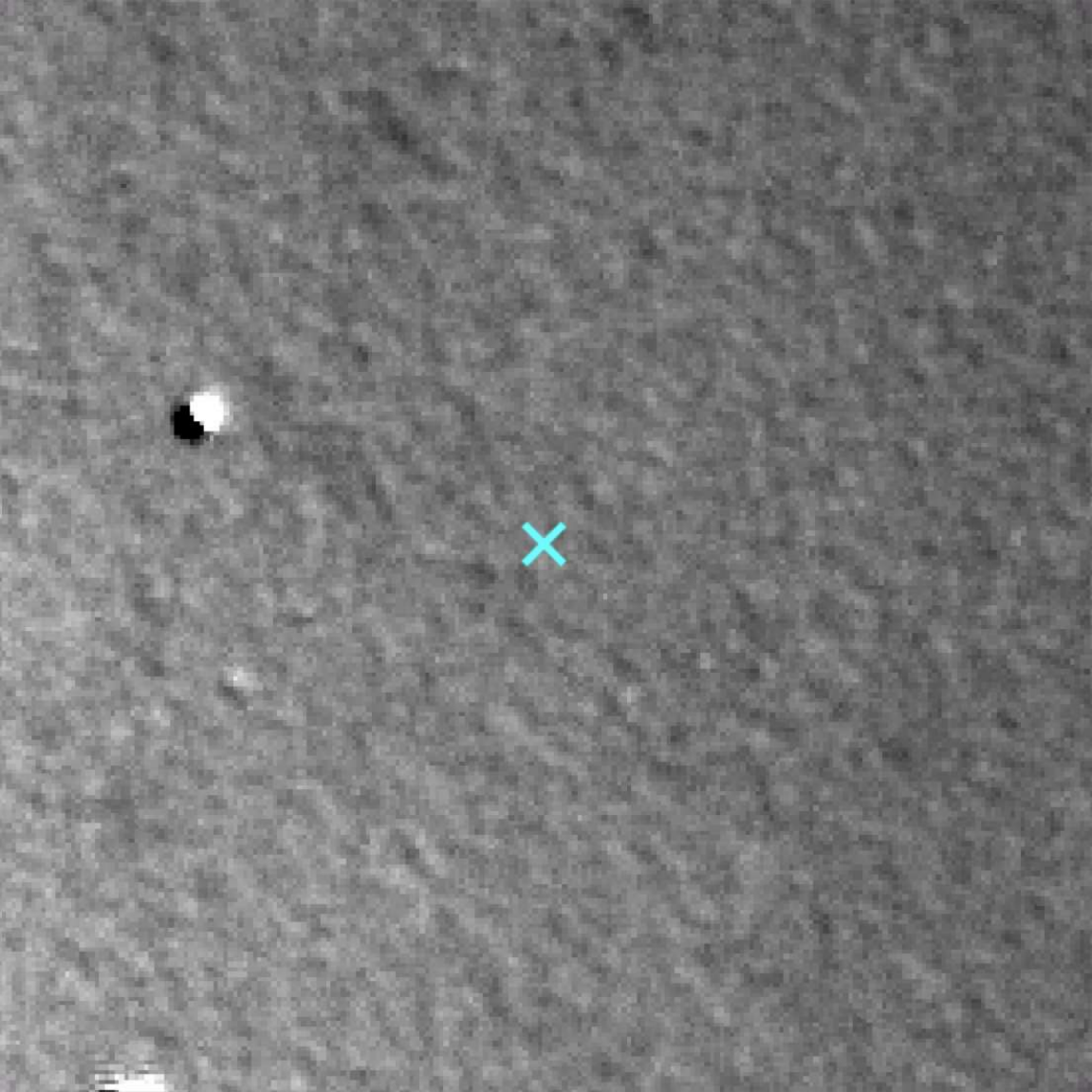}} \quad
\subfloat{\includegraphics[width=.32\textwidth]{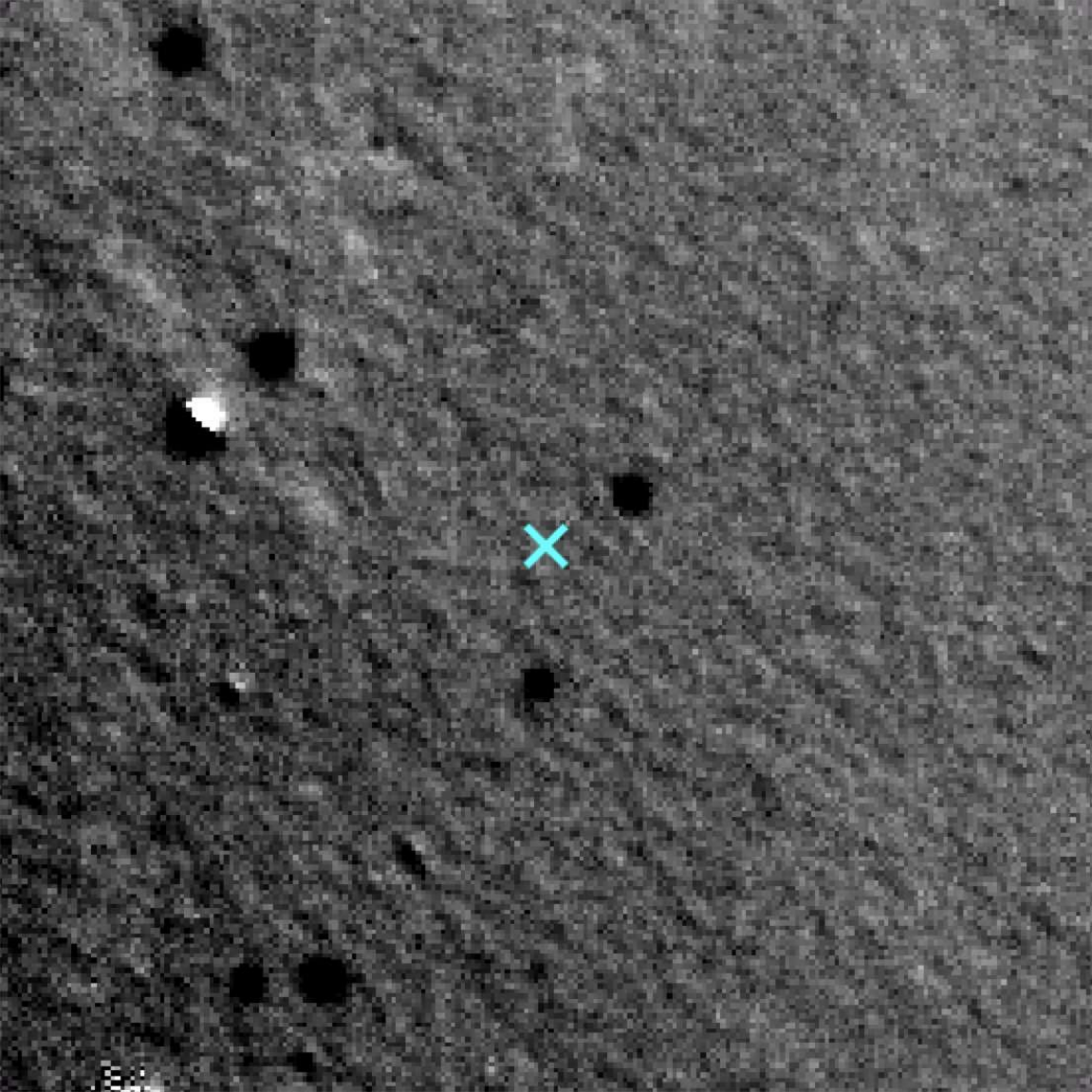}}
\caption{{\bf -- M\,31N 1961-11a}. The location of the nova (in field 6) is indicated by the blue cross. Left:\ Continuum subtracted LGGS H$\alpha$. Middle:\ Continuum subtracted LGGS [\ion{S}{ii}]. Right:\ Continuum subtracted LGGS [\ion{O}{iii}].}
\label{1961-11a surrounding sub images}
\end{figure*}

\begin{figure*}
\centering
\subfloat{\includegraphics[width=.32\textwidth]{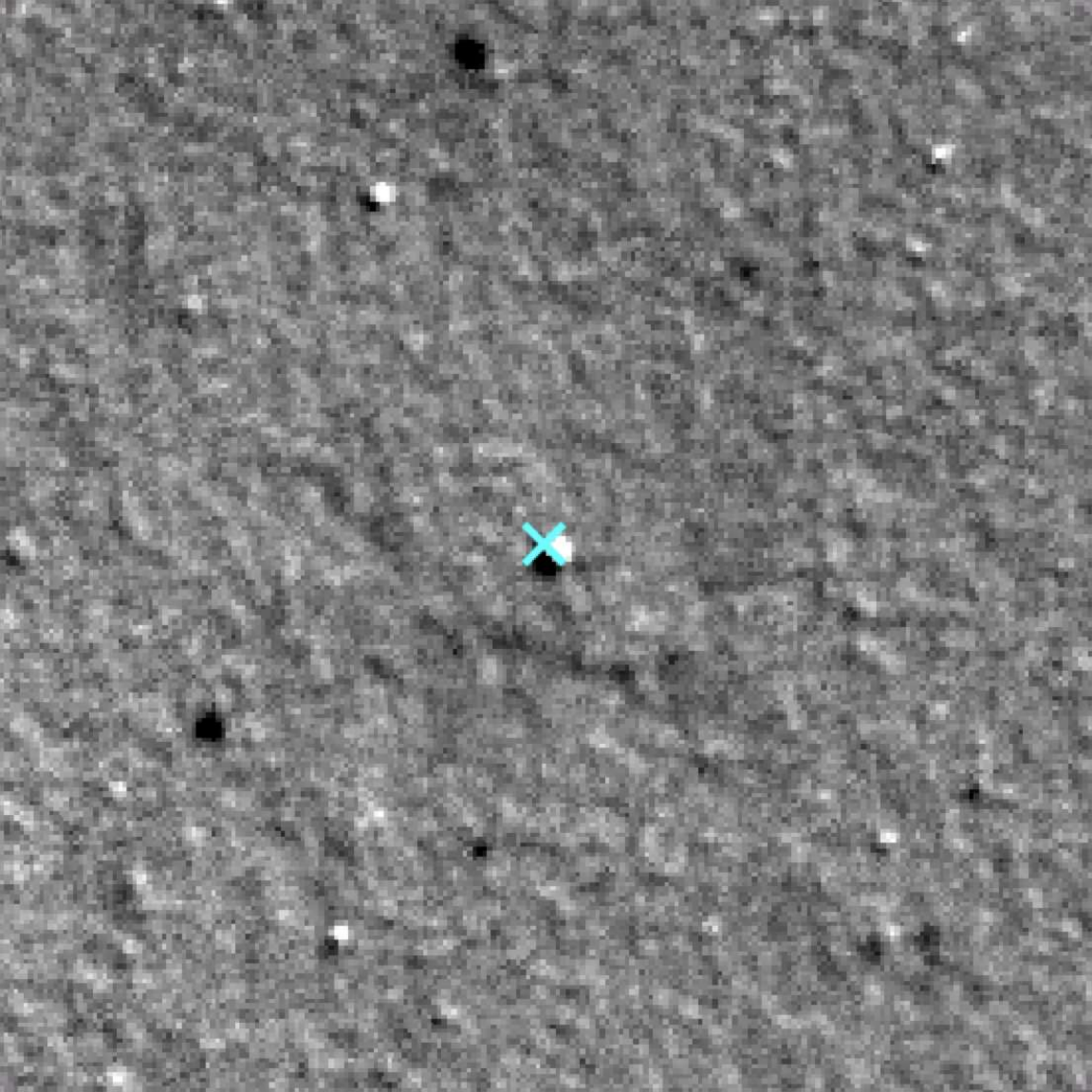}} \quad
\subfloat{\includegraphics[width=.32\textwidth]{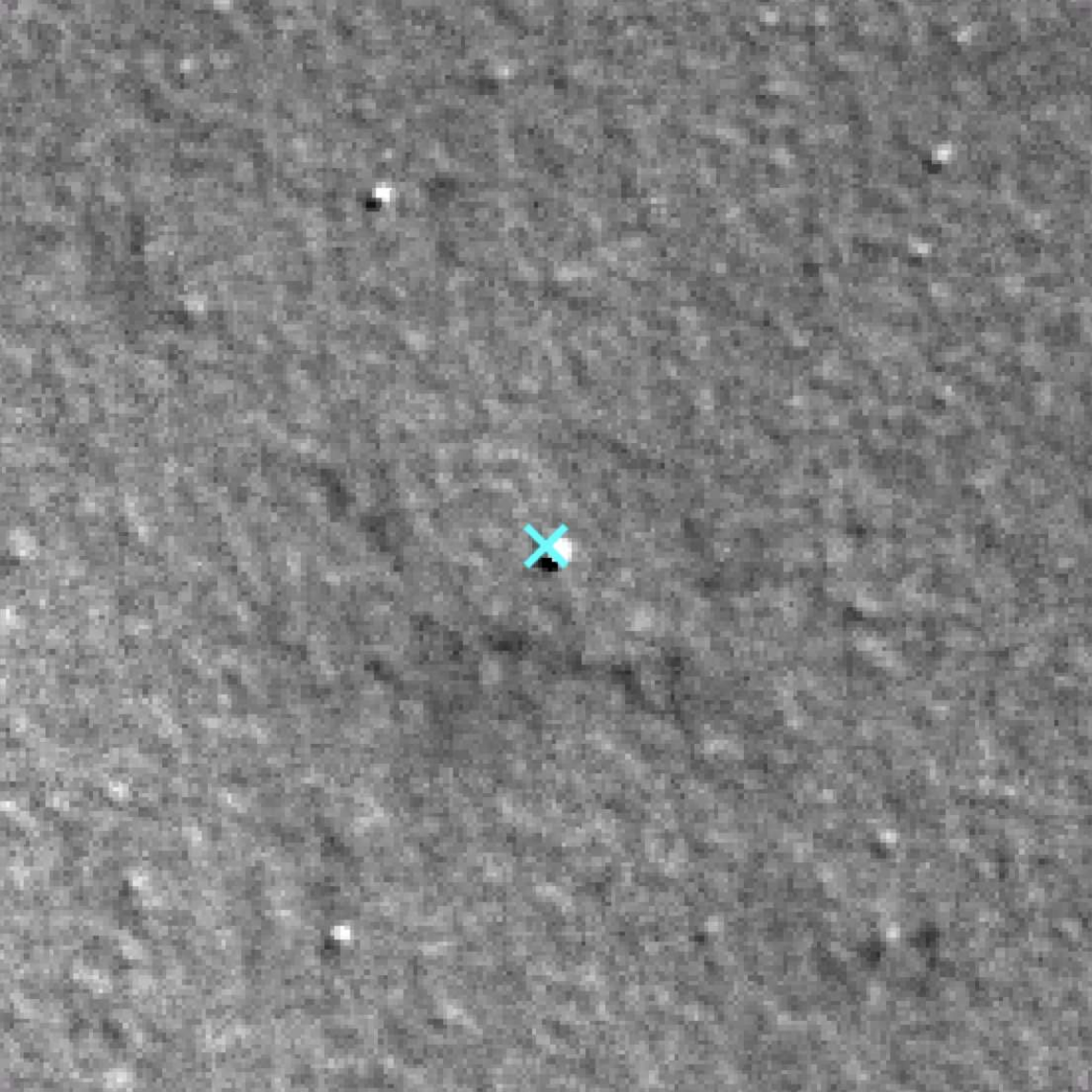}} \quad
\subfloat{\includegraphics[width=.32\textwidth]{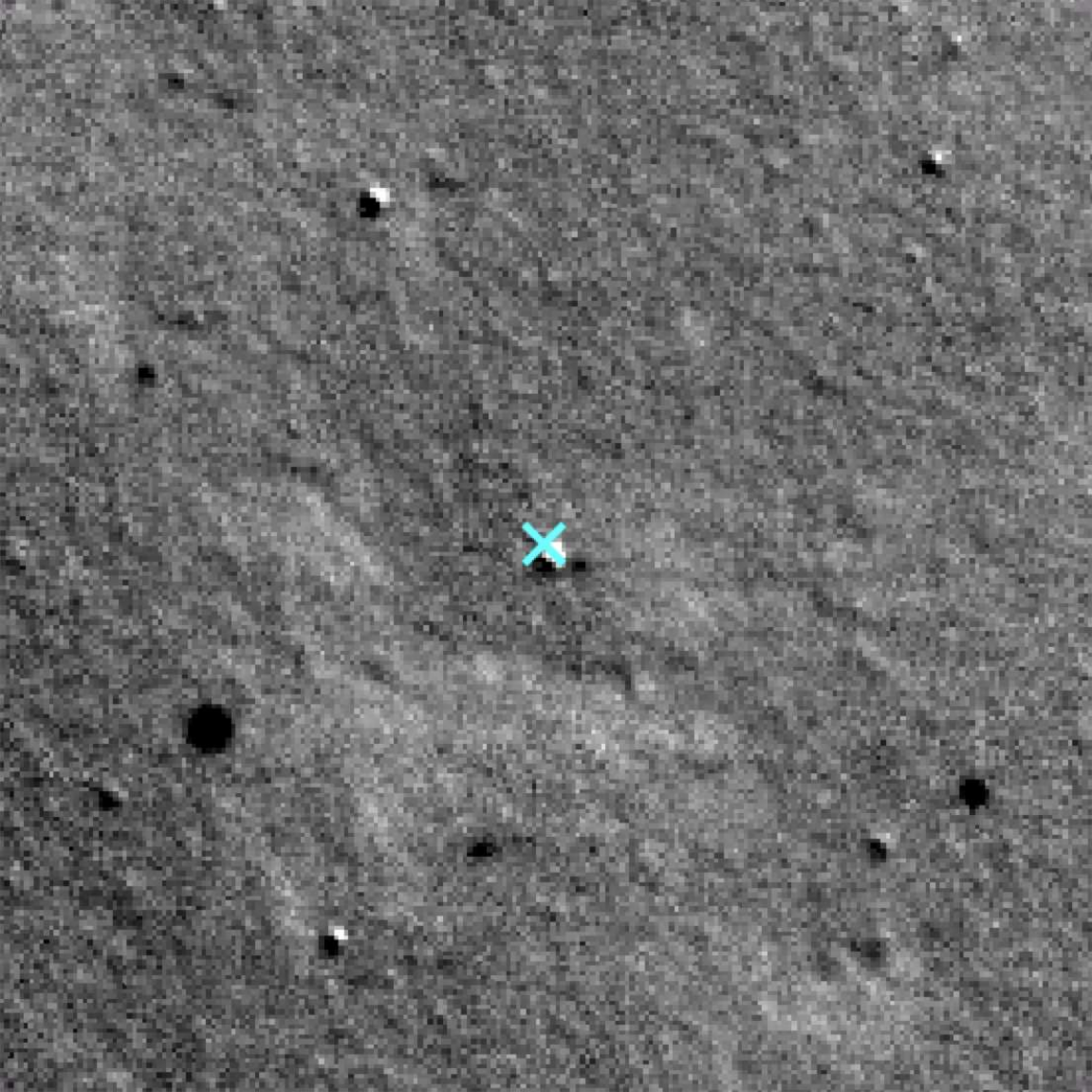}}
\caption{{\bf -- M\,31N 1953-09b}. The location of the nova (in field 6) is indicated by the blue cross. Left:\ Continuum subtracted LGGS H$\alpha$. Middle:\ Continuum subtracted LGGS [\ion{S}{ii}]. Right:\ Continuum subtracted LGGS [\ion{O}{iii}].}
\label{1953-09b surrounding sub images}
\end{figure*}

\begin{figure*}
\centering
\subfloat{\includegraphics[width=.32\textwidth]{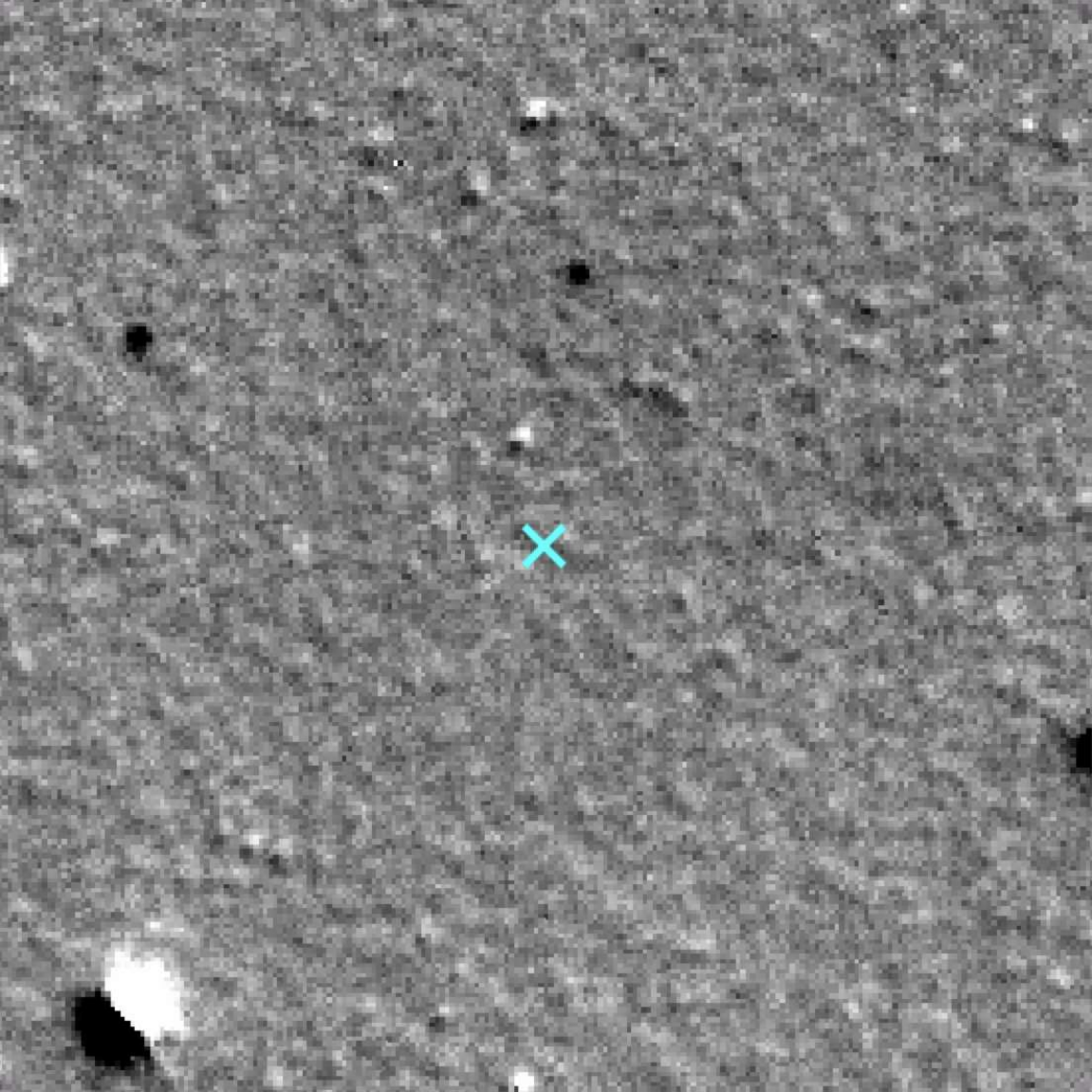}} \quad
\subfloat{\includegraphics[width=.32\textwidth]{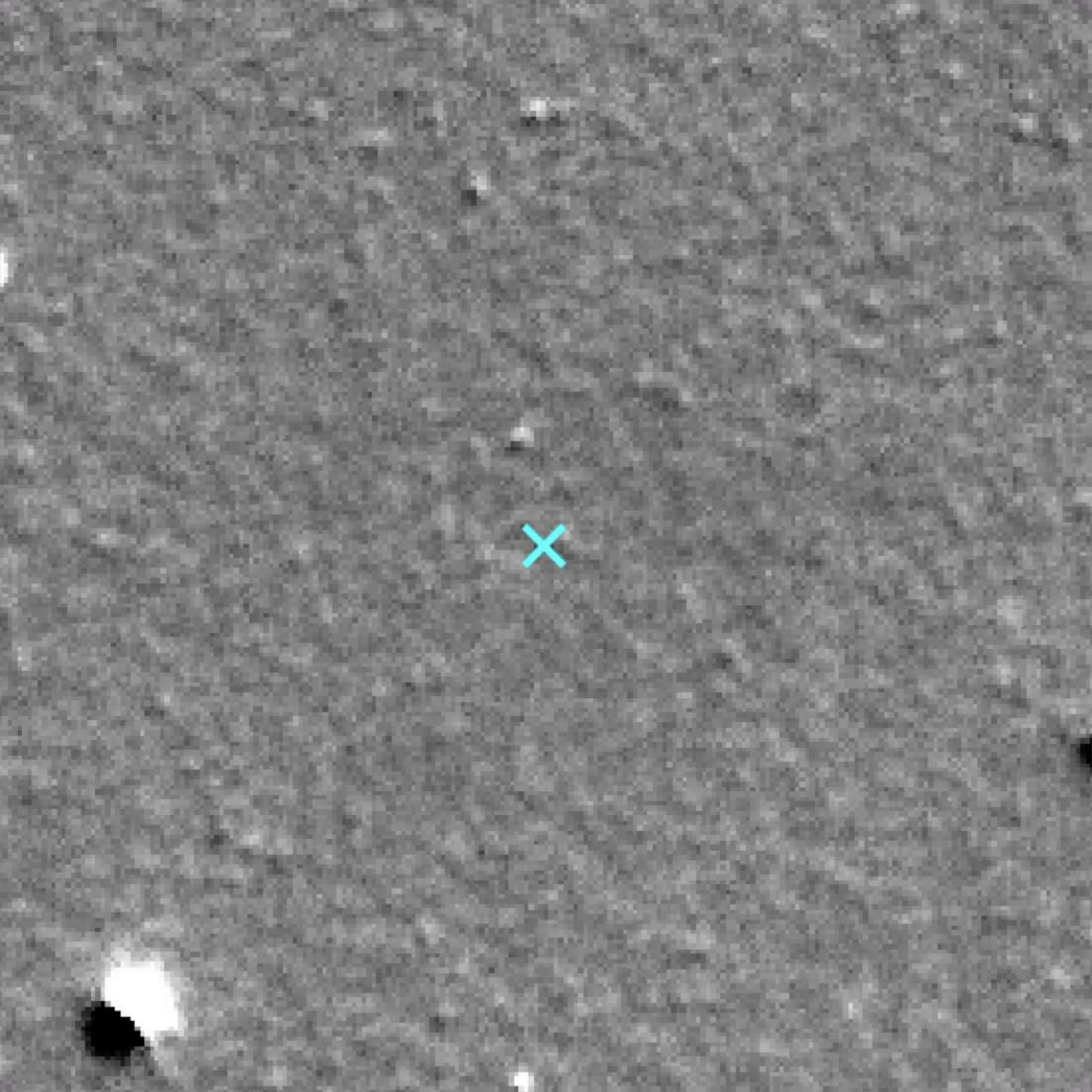}} \quad
\subfloat{\includegraphics[width=.32\textwidth]{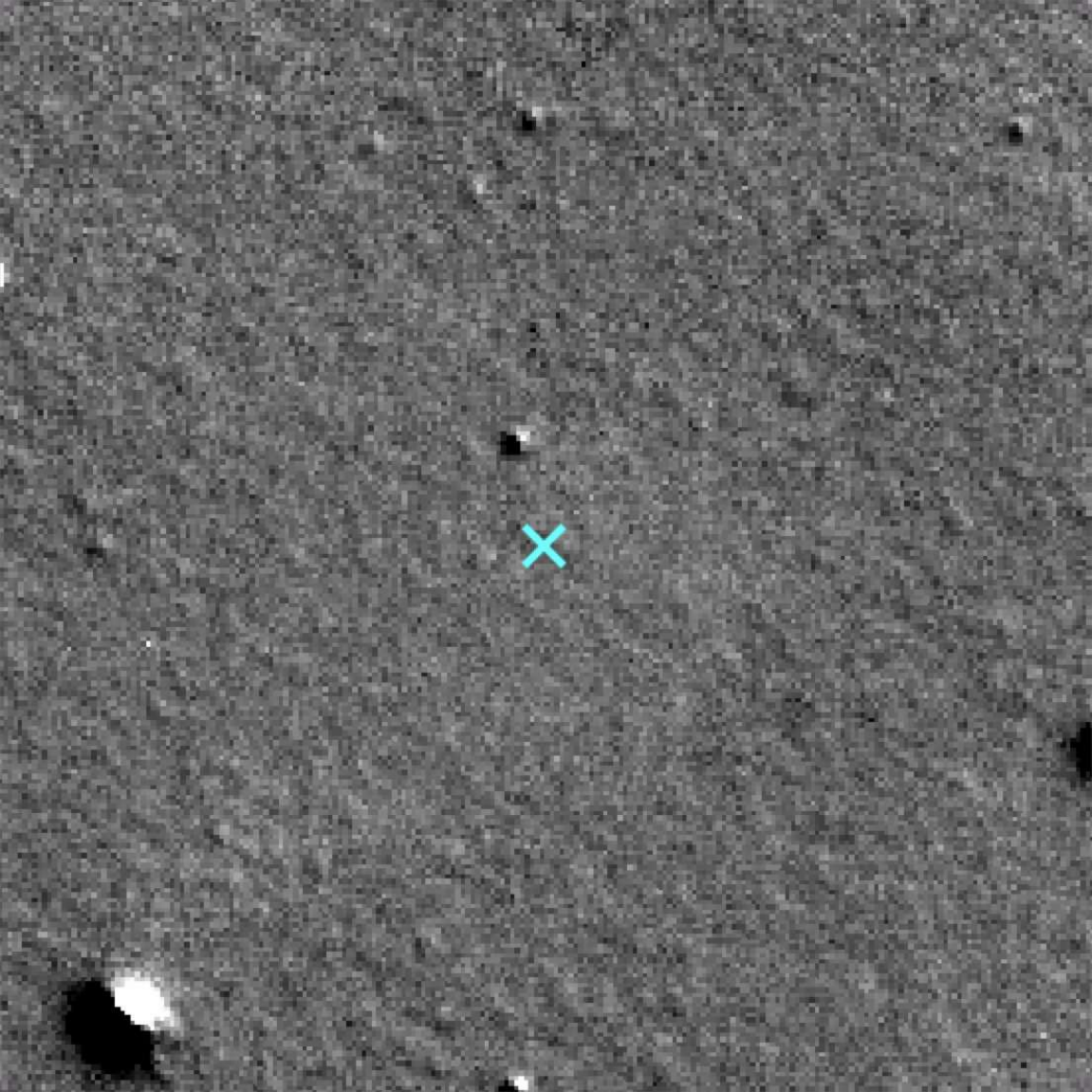}}
\caption{{\bf -- M\,31N 1919-09a}. The location of the nova (in field 6) is indicated by the blue cross. Left:\ Continuum subtracted LGGS H$\alpha$. Middle:\ Continuum subtracted LGGS [\ion{S}{ii}]. Right:\ Continuum subtracted LGGS [\ion{O}{iii}].}
\label{1919-09a surrounding sub images}
\end{figure*}

\begin{figure*}
\centering
\subfloat{\includegraphics[width=.32\textwidth]{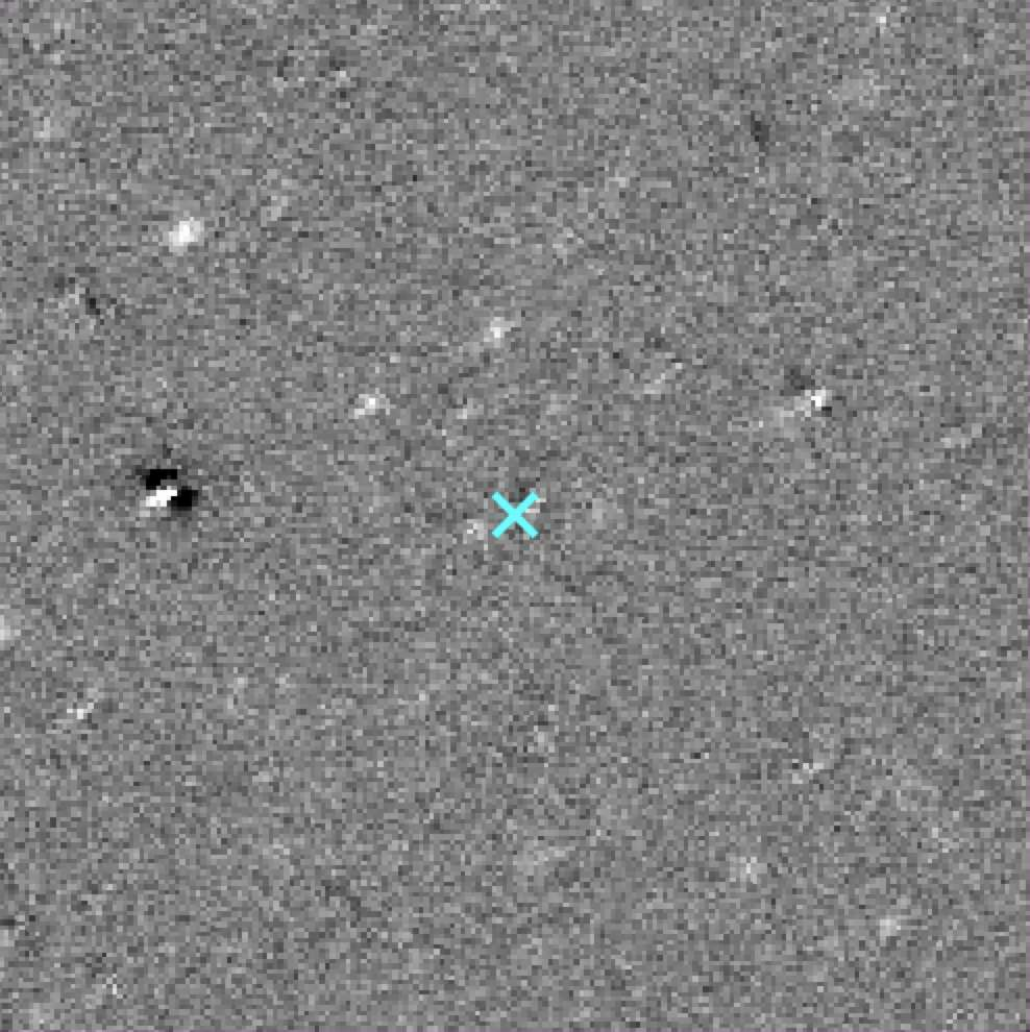}} \quad
\subfloat{\includegraphics[width=.32\textwidth]{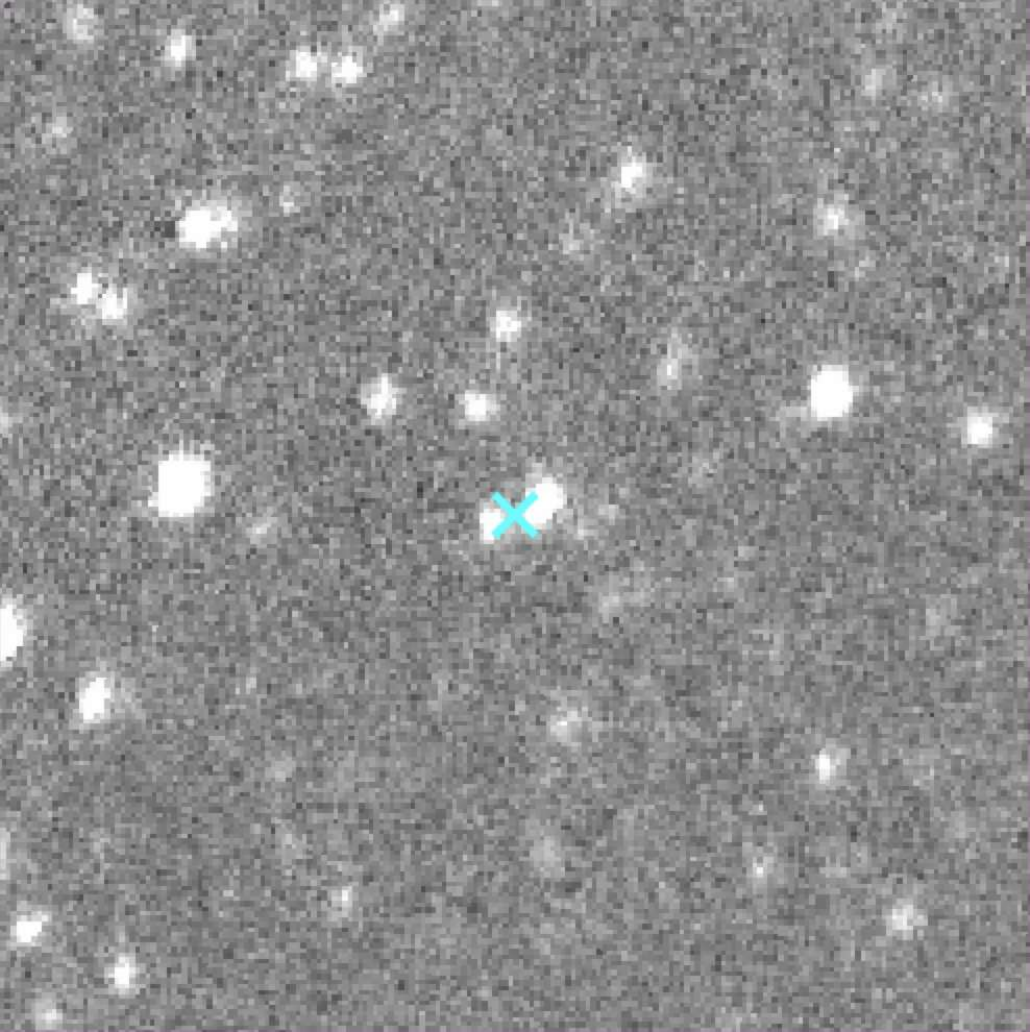}} \quad
\subfloat{\includegraphics[width=.32\textwidth]{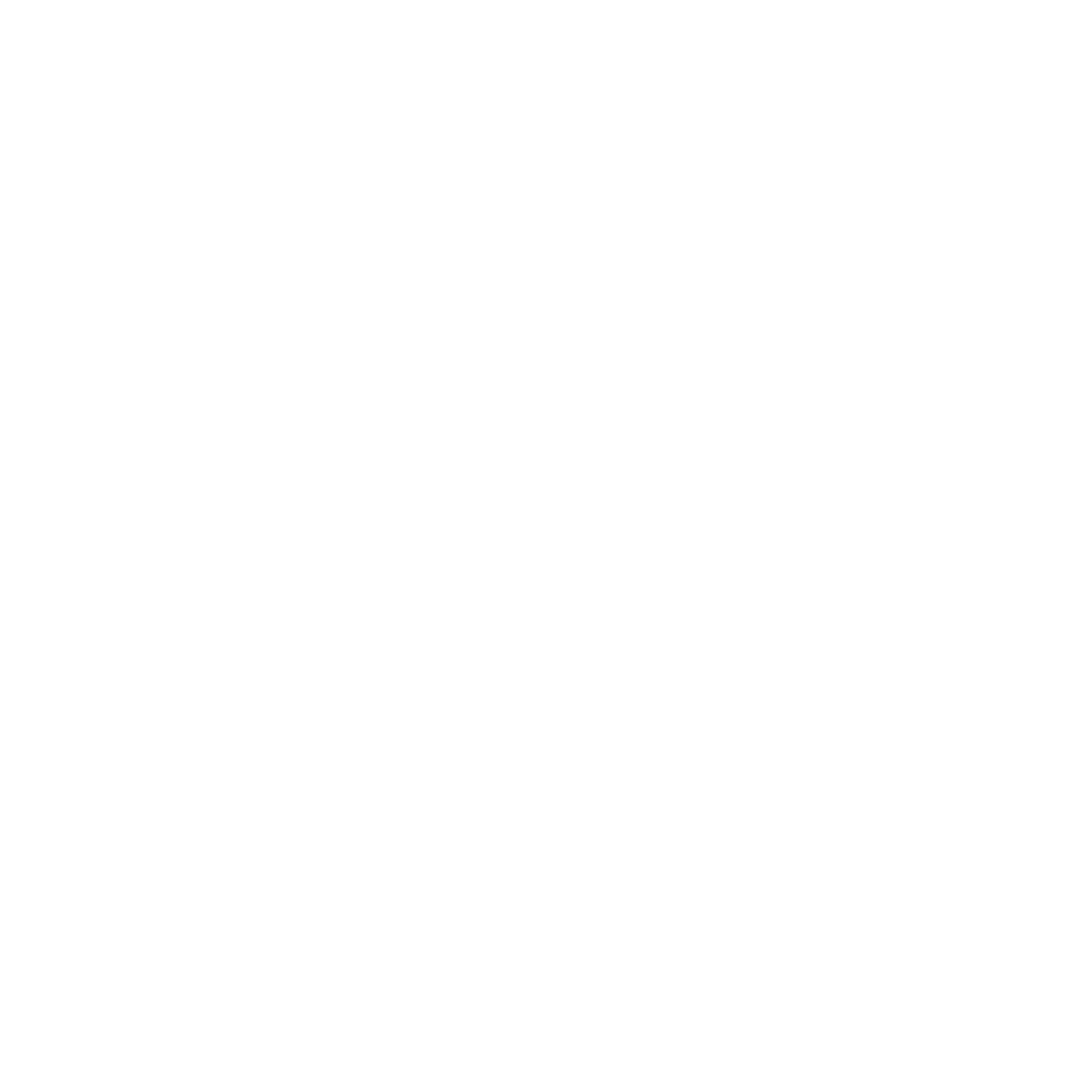}}
\caption{{\bf -- LMCN 1968-12a}. The location of the nova is indicated by the blue cross. Left:\ Continuum subtracted FTS H$\alpha$. Right:\ Continuum subtracted FTS [\ion{O}{iii}].}
\label{LMCN 1968-12a surrounding sub images}
\end{figure*}

\begin{figure*}
\centering
\subfloat{\includegraphics[width=.32\textwidth]{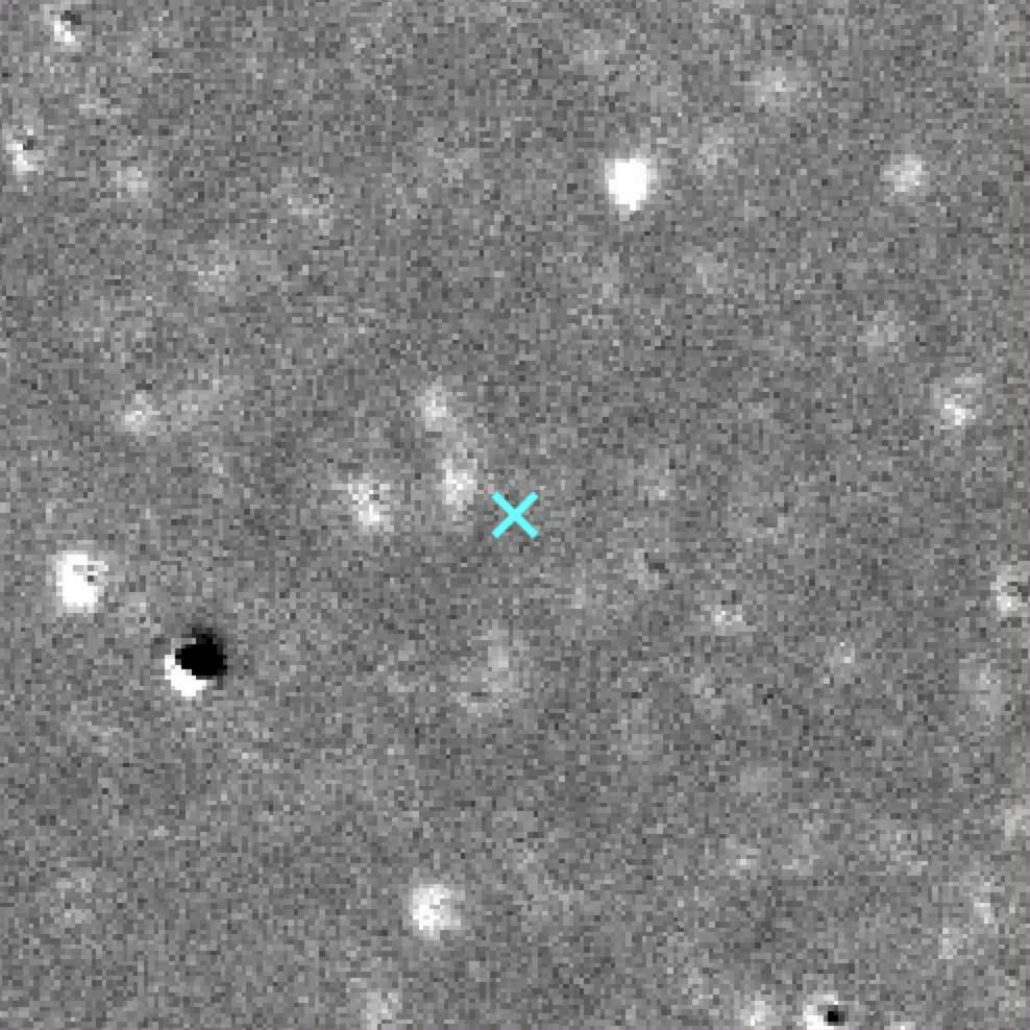}} \quad
\subfloat{\includegraphics[width=.32\textwidth]{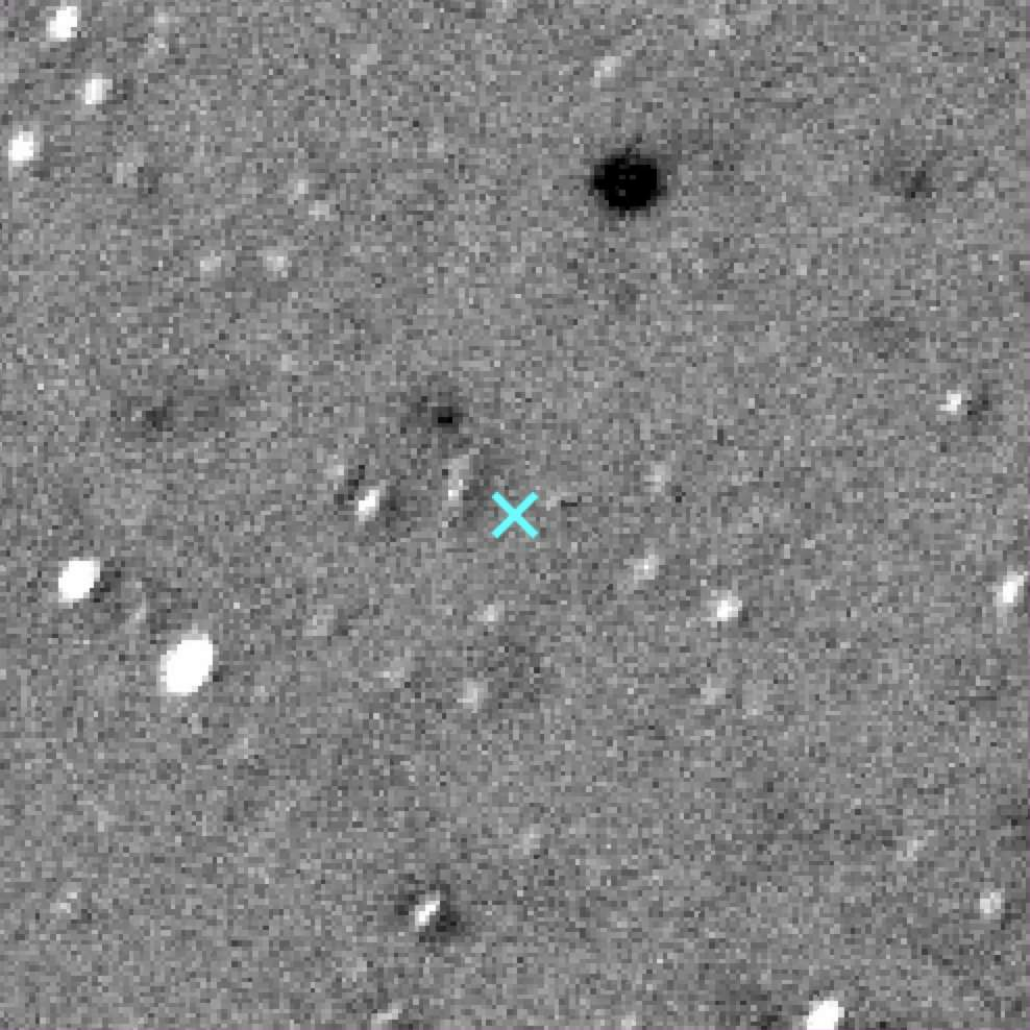}} \quad
\subfloat{\includegraphics[width=.32\textwidth]{Figures/Blank.pdf}}
\caption{{\bf -- LMCN 1996}. The location of the nova is indicated by the blue cross. Left:\ Continuum subtracted FTS H$\alpha$. Right:\ Continuum subtracted FTS [\ion{O}{iii}].}
\label{LMCN 1996 surrounding sub images}
\end{figure*}

\begin{figure*}
\centering
\subfloat{\includegraphics[width=.32\textwidth]{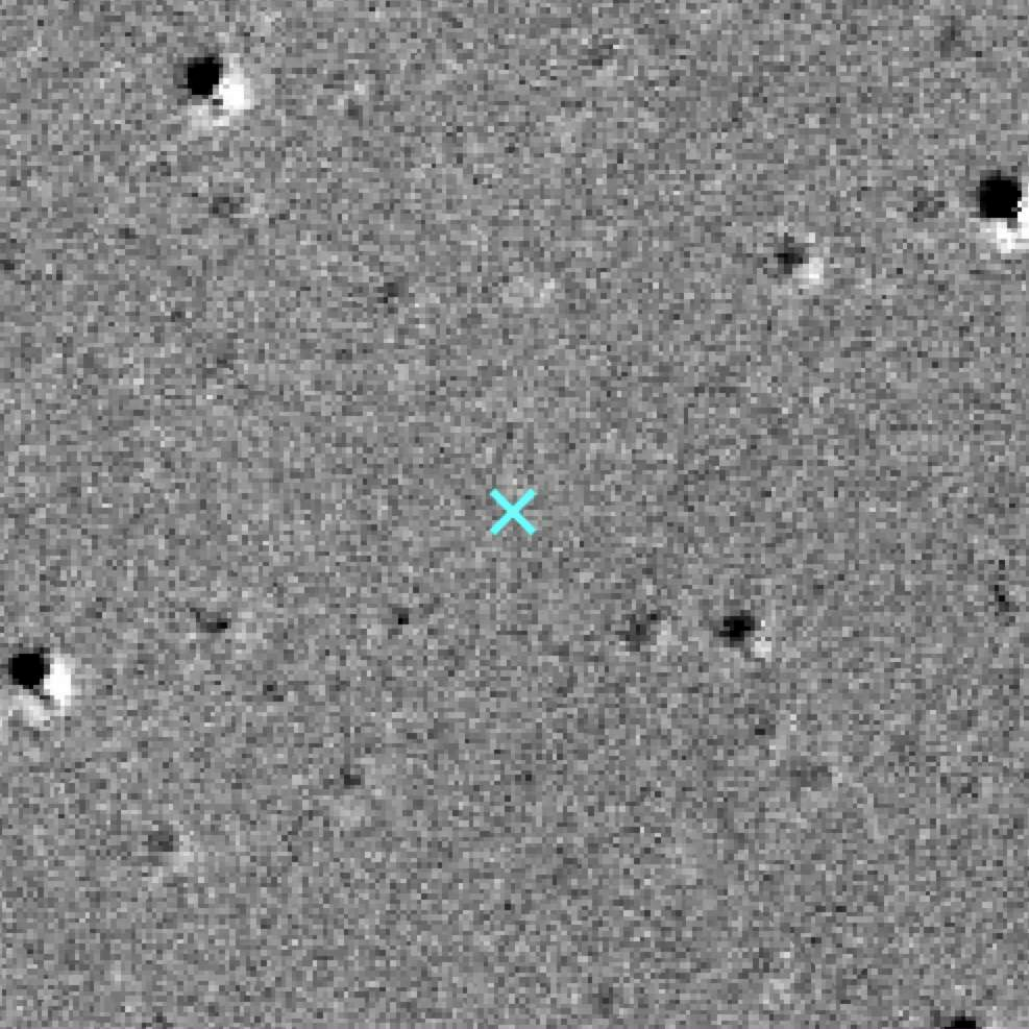}} \quad
\subfloat{\includegraphics[width=.32\textwidth]{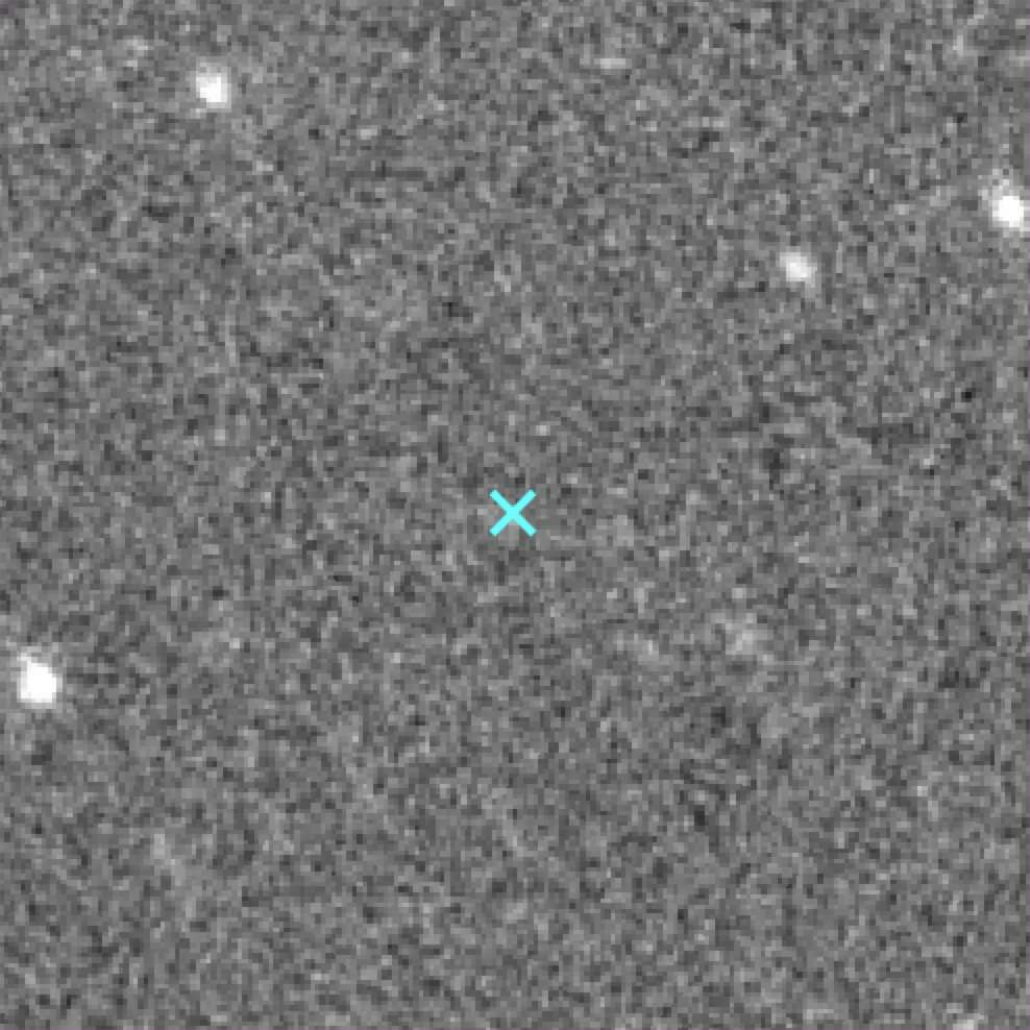}} \quad
\subfloat{\includegraphics[width=.32\textwidth]{Figures/Blank.pdf}}
\caption{{\bf -- LMCN 1971-08a}. The location of the nova is indicated by the blue cross. Left:\ Continuum subtracted FTS H$\alpha$. Right:\ Continuum subtracted FTS [\ion{O}{iii}].}
\label{LMCN 1971-08a surrounding sub images}
\end{figure*}

\begin{figure*}
\centering
\subfloat{\includegraphics[width=.32\textwidth]{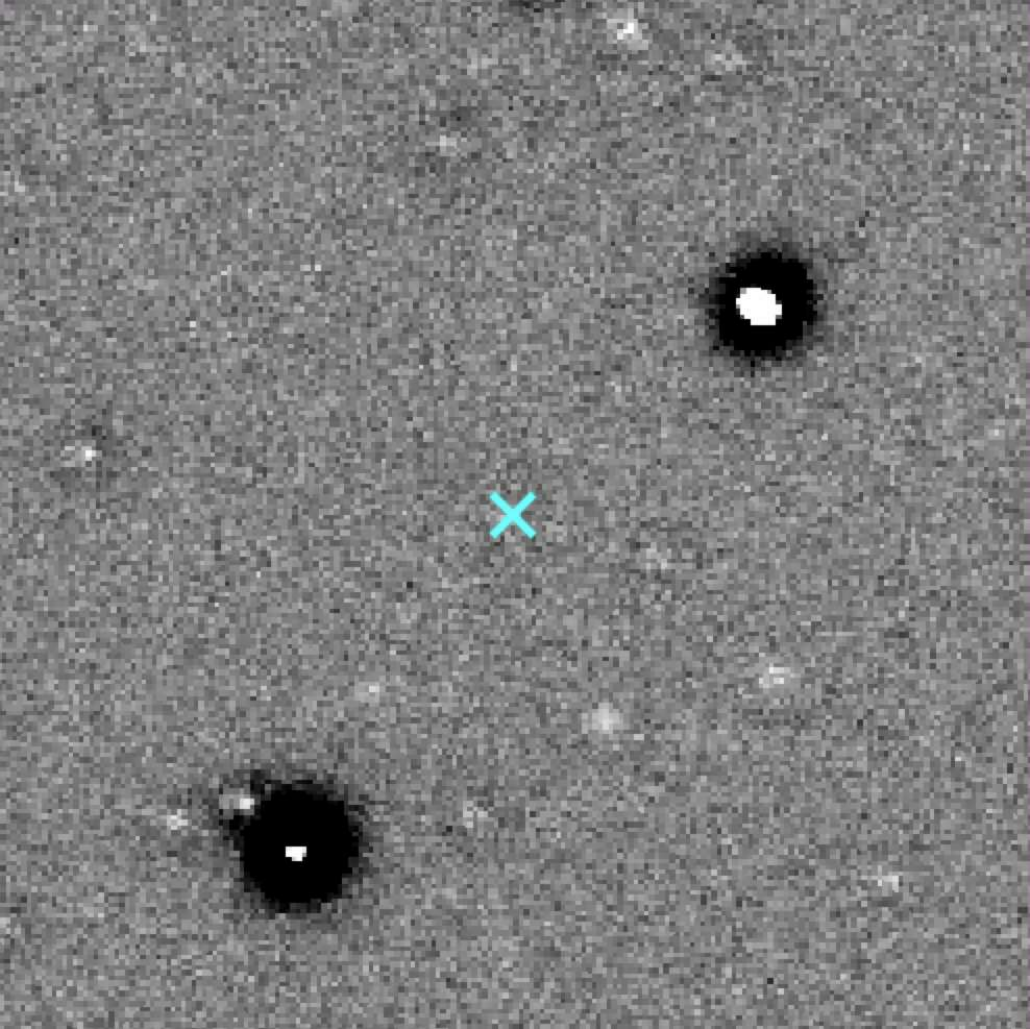}} \quad
\subfloat{\includegraphics[width=.32\textwidth]{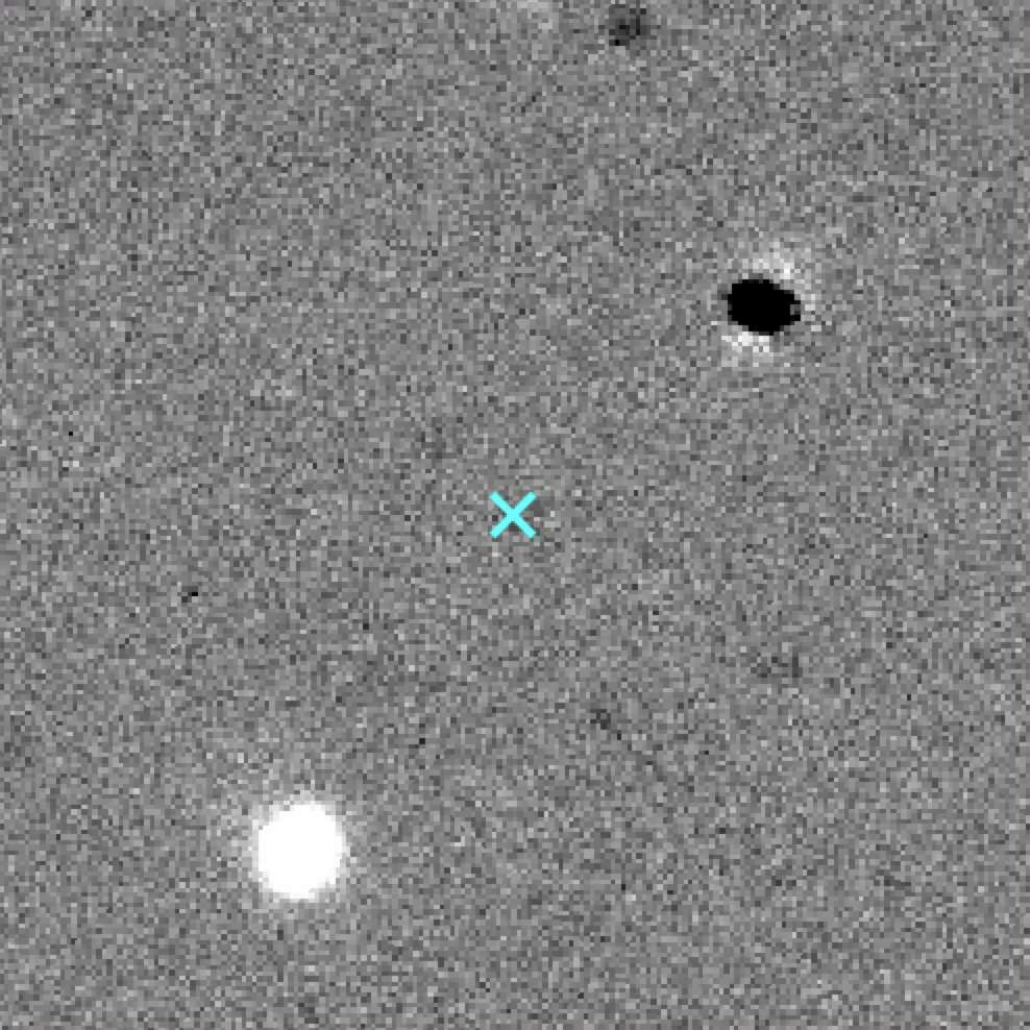}} \quad
\subfloat{\includegraphics[width=.32\textwidth]{Figures/Blank.pdf}}
\caption{{\bf -- YY Doradus}. The location of the nova is indicated by the blue cross. Left:\ Continuum subtracted FTS H$\alpha$. Right:\ Continuum subtracted FTS [\ion{O}{iii}].}
\label{YY Doradus surrounding sub images}
\end{figure*}

\bsp
\label{lastpage}
\end{document}